\documentclass[useAMS,usenatbib]{mn2e}

\usepackage[flushleft]{threeparttable}
\usepackage{graphicx}
\usepackage{xtab,afterpage}
\usepackage{amsmath}
\usepackage{epstopdf}
\usepackage{url}
\usepackage[flushleft]{threeparttable}
\usepackage{booktabs,fixltx2e}
\usepackage{amssymb}

\newcommand{\icarus}{Icarus}

\newcommand{\pasp}{{\it PASP~\/}}

\newcommand{\apj}{ApJ}
\newcommand{\apjl}{ApJ}

\newcommand{\aap}{A \& A}
\newcommand{\araa}{ARA\&A}
\newcommand{\aj}{AJ}
\newcommand{\mnras}{MNRAS}

\newcommand{\nat}{Nature}

\def\planss{\rmfamily{Planet.~Space~Sci.~}}% Planetary Space Science
\def\jgr{\rmfamily{J.~Geophys.~Res.~}}% Journal of Geophysics Research
\def\apss{\rmfamily{Ap\&SS~}}     % Astrophysics and Space Science

\renewcommand{\v}{\ensuremath{\mathbf{v}}}

\DeclareRobustCommand{\rchi}{{\mathpalette\irchi\relax}}
\newcommand{\irchi}[2]{\raisebox{\depth}{$#1\chi$}} % inner command, used by \rchi

\makeatletter
\newcommand*{\rom}[1]{\expandafter\@slowromancap\romannumeral #1@}
\makeatother

\title[Effects of impacts on the TRAPPIST-1 planets]{Cometary impactors on the TRAPPIST-1 planets can destroy all planetary atmospheres and rebuild secondary atmospheres on planets f, g, h}%Comets impacting on TRAPPIST-1 planets can destroy all primordial atmospheres and replenish that of planets f, g, h}
\author[Q. Kral et al.]{Quentin Kral,$^{1}$\thanks{E-mail: qkral@ast.cam.ac.uk} Mark C. Wyatt,$^{1}$ Amaury H.M.J. Triaud,$^{1,2}$ Sebastian Marino, $^{1}$ 
\newauthor{Philippe Th{\'e}bault, $^{3}$ Oliver Shorttle, $^{1,4}$}\\
$^{1}$Institute of Astronomy, University of Cambridge, Madingley Road, Cambridge CB3 0HA, UK\\
$^{2}$School of Physics \& Astronomy, University of Birmingham, Edgbaston, Birmingham B15 2TT, UK\\
$^{3}$LESIA-Observatoire de Paris, UPMC Univ. Paris 06, Univ. Paris-Diderot, France\\
$^{4}$Department of Earth Sciences, University of Cambridge, Downing Street, Cambridge CB2 3EQ, UK
}

\begin{document}

\date{Accepted 1928 December 15. Received 1928 December 14; in original form 1928 October 11}

\pagerange{\pageref{firstpage}--\pageref{lastpage}} \pubyear{2002}

\maketitle

\label{firstpage}

\begin{abstract}
The TRAPPIST-1 system is unique in that it has a chain of seven terrestrial Earth-like planets located close to or in its habitable zone. In this paper, we study the effect of potential cometary impacts on the TRAPPIST-1 planets and how they would affect the primordial atmospheres of these planets. We
consider both atmospheric mass loss and volatile delivery with a view to assessing whether any sort of life has a chance to develop. 
We ran N-body simulations to investigate the orbital evolution of potential impacting comets, to determine which planets are more likely to be impacted
and the distributions of impact velocities. We consider three scenarios that could potentially throw comets into the inner region (i.e within 0.1au where the seven planets are located) from an (as yet undetected) outer belt similar to the Kuiper belt or an Oort cloud:
Planet scattering, the Kozai-Lidov mechanism and Galactic tides. For the different scenarios, we quantify, for each planet, how much atmospheric mass is lost and what mass of volatiles can be delivered over the age of the system depending on the mass scattered out of the outer belt.
We find that the resulting high velocity impacts can easily destroy the primordial atmospheres of all seven planets, even if the mass scattered from the outer belt is as low as that of the Kuiper belt. However, we find that the atmospheres of the outermost planets f, g and h can also easily be replenished 
with cometary volatiles (e.g. $\sim$ an Earth ocean mass of water could be delivered). These scenarios would thus imply that the atmospheres of these outermost planets could be more massive than those of the innermost planets, and have volatiles-enriched composition. %We also discuss our results in the context of the most
%recent theories of abiogenesis, which involve impacts, UV irradiation and a hard planetary surface that may all be present in this system.

%Using a minimum mass solar nebula approach, we suggest that a potential belt of 20M$_\oplus$ may have formed around TRAPPIST-1, which would be the 
%cometary reservoir needed for impacts to happen. We also hypothesise that this belt would be located at $>10$au and observable in the sub-mm with ALMA. 
\end{abstract}


\section{Introduction}

The nearby (d=12pc) M8V ultra-cool dwarf star TRAPPIST-1 (2MASS J23062928-0502285) is now known to be surrounded
by at least seven terrestrial-like planets \citep{2016Natur.533..221G,2017Natur.542..456G,2017NatAs...1E.129L}. This old \citep[7.6$\pm$2.2 Gyr,][]{2017arXiv170602018B}, close-by, multi-planetary system 
may offer one of our best chances to study the chemistry, and structure of terrestrial planet atmospheres outside our Solar System \citep{2016Natur.537...69D,2017arXiv170804239M}. 
Moreover, several of the TRAPPIST-1 planets \citep[most likely planets e, f and g,][]{2017Natur.542..456G} lie within the liquid water habitable zone \citep[HZ, e.g.][]{2017MNRAS.469L..26O}.
However, the presence of liquid water and possible life strongly depends on the atmospheric content of these planets, the presence of oceans, the vegetation coverage, etc. \citep[e.g.][]{2017ApJ...839L...1W,2017arXiv170606005A,2016JGRE..121.1927E,2016MNRAS.461.1981C,2016A&A...592A..36G}.

This system being very close-by, we may soon be able to start characterising the atmospheres of the seven planets with new telescopes such as JWST \citep{2016MNRAS.461L..92B,2017MNRAS.469L..26O} and the E-ELT \citep[][]{2014ApJ...781...54R,2016A&A...596A.112T} and 
search for tracers of life. Such detailed spectral characterisation may eventually allow us to infer the presence of biological activity via the detection of 
gases such as ozone \citep{2016MNRAS.461L..92B}, or via the spectral signatures of pigmented micro-organisms \citep{2017AsBio..17..231P}.
Regardless, such observations will inform on the atmospheric compositions of these planets that is necessary to study the possibility that life may develop.

For now, little is known about the atmospheres of these seven planets. The two innermost planets b and c have been observed using transmission spectroscopy \citep{2016Natur.537...69D}.
This showed that the combined spectrum of both planets (obtained when transiting at the same time) is featureless, which favours atmospheres that are tenous (composed of a variety of chemical species), not hydrogen-dominated, dominated by aerosols or non-existent. Similar conclusions have been made for planets d, e, f (and potentially g) by \citet{2018arXiv180202250D}.
Also, from the combined measurement of planet radii (transit) and masses (transit timing variations), the derived planets' densities show that TRAPPIST-1 b, d, f, g, and h may require envelopes of volatiles in the form of thick atmospheres, oceans, or ice \citep{2018arXiv180201377G}.
We thus do not know yet information that would be important for considering the habitability of the planets such as whether these planets' atmospheres are primordial or created later, for instance by cometary impacts, 
although current observations suggest that current atmospheres may not be primordial due to a lack of hydrogen signatures in the observed spectra \citep{2018arXiv180202250D}. 

 %or even via detection of impact ejecta \citep{2017AsBio..17..721C}.
%Studying the habitability on these planets is also essential. 
%Learning more about the composition of the atmospheres of the seven planets (for instance, are they primordial or mainly made of volatiles delivered by comets?) is therefore key to be able to deduce the (thermal and pressure) conditions on the 
%surface of these planets and assess their potential habitability over long timescales \citep[e.g.][]{2016JGRE..121.1927E,2016MNRAS.461.1981C,2016A&A...592A..36G}.

Previous theoretical studies of the atmospheric composition of the TRAPPIST-1 planets have shown that they may vary with time and be affected by the early evolution of the star.
Indeed, ultra-cool dwarfs such as TRAPPIST-1 take up to 1Gyr to cool down \citep{2015A&A...577A..42B} and reach the main-sequence after the planets 
formed. This means that planets that are today in the HZ would have undergone a very hot pre-main-sequence era (with potentially a runaway greenhouse phase) and may have lost all (or part) of their initial water
content \citep{2017MNRAS.464.3728B}. Moreover, \citet{2017A&A...599L...3B} find that the total XUV emission from the star might be strong enough to entirely strip the primordial atmospheres of the planets over a few Gyr. One could then expect that a few of the TRAPPIST-1 
planets are devoid of atmospheres, or left with a gas layer too tenuous for life to persist over long timescales \citep{2017arXiv171102676R}.

Here we consider another process that can strongly influence the atmospheres, both positively and negatively for life: {\it exocomets}.
Impacting exocomets can influence planetary atmospheres in multiple ways: a) they can provide an energy source that depletes primordial atmospheres. b) They might also deliver volatiles that subsequently replenish a secondary atmosphere (i.e., dry, depleted atmospheres from impacts or XUV irradiation 
could be replenished by later impacts, and surviving primordial atmospheres could see their elemental abundances significantly transformed via exocomet impacts). c) Impacting exocomets may also act as catalysts 
for the development of life. Indeed, these impacts may initiate a cascade of chemical reactions, some of which can produce the necessary precursors to nucleobases on these planets \citep{2012PhLRv...9...84S,2015PNAS..112..657F,2015NatCh...7..301P,Suth17,2017ApJ...843..110R}.

For now, there is no evidence of exocomets in the TRAPPIST-1 system, however, this does not mean they are not present and part of the motivation of this work is to determine if evidence for such a population may be imprinted on the planets' atmospheres.

Many stars have large outer reservoirs of planetesimals that produce a detectable infrared excess due to collisional production of dust
\citep{2008ARA&A..46..339W,2013A&A...555A..11E}. Detections of CO gas in several systems are used to infer that these planetesimals are icy with a composition that is similar to Solar System comets \citep[e.g.][]{2016MNRAS.461..845K,2016MNRAS.460.2933M,2017ApJ...842....9M}. 
These planetesimal belts are harder to detect around low mass stars such as TRAPPIST-1 due to their low luminosity but this does not mean they are not present \citep{2009ApJ...698.1068P,2014ApJ...794..146T}. Some stars also have evidence that comets from these outer regions are being scattered into the inner regions. For example, CO detected at 20au in $\eta$ Corvi is inferred to originate in the sublimation of such an exocomet population\citep{2017MNRAS.465.2595M}. 
In addition, high-velocity metallic gas absorption lines in some systems \citep{2012PASP..124.1042M,2014Natur.514..462K,2016A&A...594L...1E} are inferred to originate in very eccentric comets passing very close to their host star \citep[called falling evaporating bodies, e.g.][]{1990A&A...236..202B}. 
Thus, it is not unreasonable that TRAPPIST-1 has (or indeed may have had) comets at some level.

In this study, we hypothesize that such comets exist in the TRAPPIST-1 system and use previous studies that looked at the effect of impacts onto planetary atmospheres \citep[e.g.][]{2012Icar..221..495D,2015Icar..247...81S}, and especially hydrodynamical simulations \citep{2009M&PS...44.1095S}
to derive some constraints on the TRAPPIST-1 planets' atmospheres in the presence of impacting comets. %Thanks to these studies, we can quantify how efficient these impacts are at destroying any primordial atmosphere in the TRAPPIST-1 system and
%evaluate how much cometary material or volatiles will end up in the atmospheres of these planets.

%The TRAPPIST-1 system is so far the best candidate for that type of studies. Indeed, the impact probability is high for very close-in planets \citep{2017MNRAS.464.3385W} meaning that most comets that make it to the inner regions of the system will likely collide with one of the planets 
%rather than being ejected. In addition, owing to the large number of planets in the systems, the innermost planets could be protected from being impacted if the outermost planets catch the comets first. This 
%potential shielding effect could however, counter-intuitively, be detrimental to the formation of life in the innermost planets by depriving them of essential volatiles. 
%This shielding effect would also give rise to different predictions for the atmospheric masses/volatile content of the seven planets, which we may be able to compare to observations in the near future.
 
We start by estimating the possible mass of a planetesimal belt that could have survived around TRAPPIST-1. In Sec.~\ref{nbody}, we then
study the dynamics of comets in the TRAPPIST-1 system that come close to the seven planets, i.e within 0.1au. 
Notably, we look into which planet will receive most impacts, at which velocity and derive the timescales on which impacts happen. In Sec.~\ref{appl}, we describe three plausible scenarios that can potentially scatter many exocomets over the lifetime of the system. In Sec.~\ref{results},
we show the results of our model, i.e how much atmospheric mass is removed from the primordial atmospheres of the seven planets by a continuous series of impacts and evaluate whether those impacts increase or reduce the amount of volatiles in the planets' atmospheres, 
and what kind of atmosphere each planet is likely to end up with. We then discuss our results in terms of their implications for the development of life in Sec.~\ref{discu} before concluding.

\section{The possible presence of a disc around TRAPPIST-1}\label{mmsn}
This paper is based on the potential presence of a yet undetected debris disc around TRAPPIST-1. 
To consider what this debris disc might look like, we construct a minimum mass extrasolar nebula for the TRAPPIST-1 system similar to \citet{1981PThPS..70...35H}, or \citet{2013MNRAS.431.3444C} who used 1925 extrasolar planets to constrain the minimum surface 
densities at different distances from the star assuming planets formed in situ. 

\begin{figure}
   \centering
  \includegraphics[width=9cm]{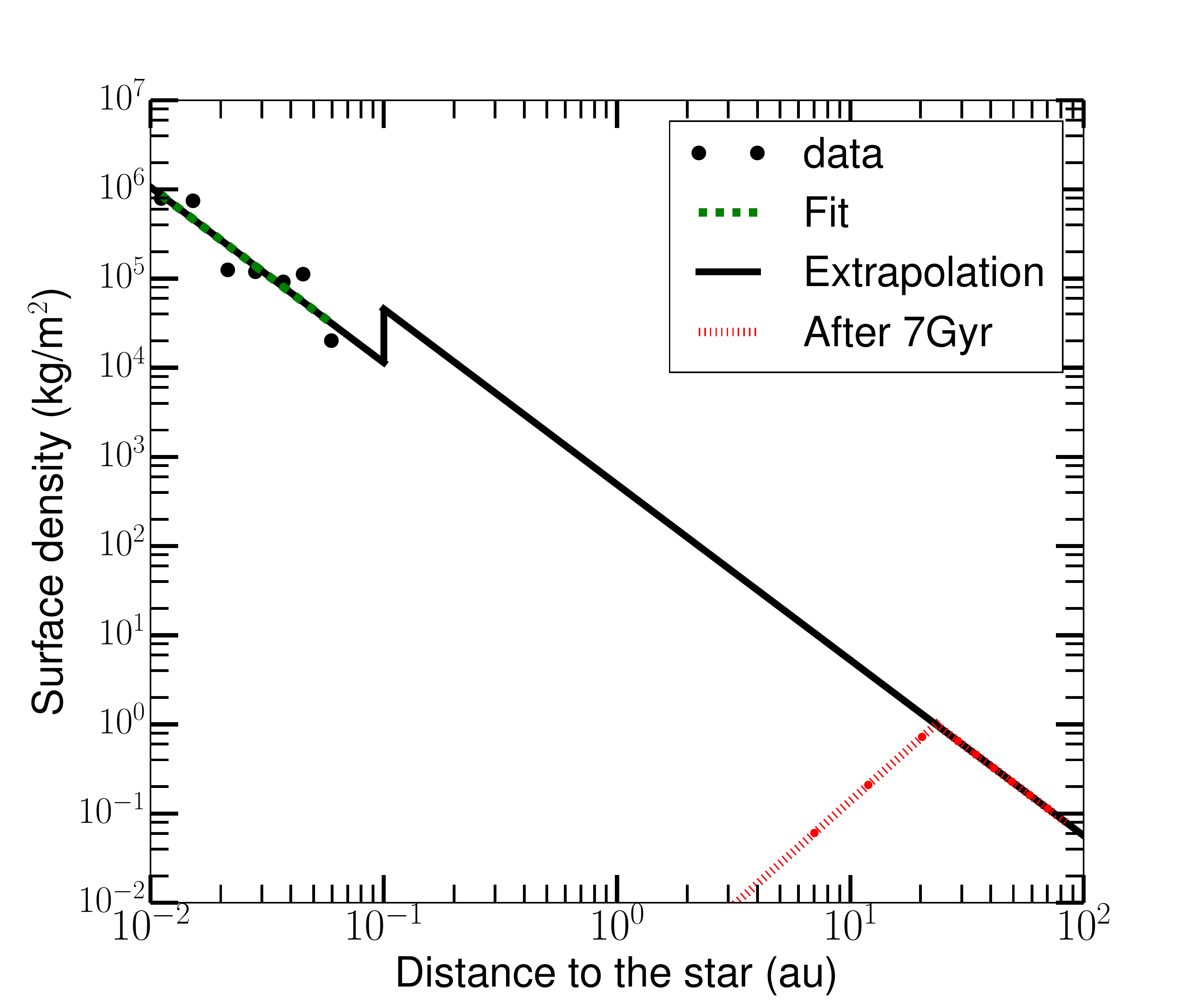}
   \caption{\label{fig1b} Surface density in the TRAPPIST-1 system assuming a minimum mass extrasolar nebula and extrapolating to tens of au to obtain a plausible mass that would be left in a potential, yet undetected, belt. In red, we show the predicted profile after 7Gyr of collisional evolution.}
\end{figure}

To get a surface density for each planet, we take the planet mass and divide it by the area of the annulus around the planet. For planets c to g, we define the annulus as being between the two midpoints to the neighbouring planets. 
For planets b and h, we work out the half width
using the centres between planets b and c and between planets g and h and multiply that width by two. This gives the following surface density (in solids) after fitting the data (see Fig.~\ref{fig1b}) %A,B -1.96907957124 2.08853855369

\begin{equation}\label{eqsig}
 \Sigma \sim 122 \left( \frac{r}{1{\rm au}}\right)^{-1.97} {\rm kg/m}^2,
\end{equation}

\noindent where $r$ is the distance to the star. Our fit of $\Sigma$ provides values a factor 4 smaller than \citet{2013MNRAS.431.3444C} at 1au (who used a large sample of Kepler planets around earlier-type stars) but steeper in $r$ and very close to the fit by \citet{2017MNRAS.470L...1G} who did it specifically for M-dwarf Kepler planets. It is
less than a factor 2 from the minimum mass solar nebula (MMSN) in solids for terrestrial planets at 1au \citep{1981PThPS..70...35H}.

The H$_2$O iceline during planetesimal formation is estimated to have been close to $\sim$0.1au in the TRAPPIST-1 system \citep{2017A&A...604A...1O}. It could as well have been slightly closer-in (by a factor $\sim$2) based on the (still not-well constrained) gradient of water compositions
 of the 7 planets \citep{2018NatAs...2..297U}.   
We assumed that beyond 0.1au, the solid \{rock+ice\} surface density is a factor 4 higher following \citet{1981PThPS..70...35H}.
We can now extrapolate the mass that may be present at several au and potentially form a disc of planetesimals rather than planets. 

A planetesimal belt at a radius $r$ with d$r/r \sim 0.5$ would have a mass of $\sim 12.6 (r/1{\rm au})^{0.03}$M$_\oplus$.
The collisional lifetime of the biggest planetesimals in such a belt is given by\footnote{We note that this formula can be used when the largest bodies from the collisional cascade have a large enough 
collision velocity that they can fragment after an impact. Depending on the level of stirring, using this formula for radii $\gtrsim$50au is therefore not accurate and only gives a lower limit on the timescale $t_c$.} $t_c=1.4 \times 10^{-3} \, r^{13/3} ({\rm d}r/r) D_c Q_D^{\star 5/6} e^{-5/3} M_\star^{-4/3} / M_{\rm tot} $ yr \citep{2008ARA&A..46..339W}. %Eq 16 of Wyatt+08, assume comme eux, Qd=150J/kg, dr/r=0.5, e=0.05, Dc=60km 
 This gives $t_c \sim 4 \times 10^3 (r/1{\rm au})^{4.3}$yr by assuming typical values \citep[as in][i.e $e=0.05$, $Q_D^{\star}=500$J/kg, $D_c$=100km]{2008ARA&A..46..339W,2017MNRAS.469..521K}. In other words, a belt at 1au would be significantly depleted after 7Gyr (the age of the system) of collisional evolution and we expect any belt this close in to have been significantly depleted.
However, a belt at $>$28au could survive over 7Gyr. At shorter radii, the mass that remains after collisional evolution for 7Gyr would be expected to have a radial profile that scales $\propto r^{7/3}$ \citep{2010MNRAS.405.1253K} as shown by the red dotted line in Fig.~\ref{fig1b}. 
While this formula depends on many uncertain parameters, it shows that we expect any potential surviving belt to be located at $\gtrsim10$au.

Using the extrapolation in Eq.~\ref{eqsig}, we expect such a leftover belt between 10 and 50au to have a mass of $\sim$20$M_\oplus$, which is compatible with the predicted large initial mass of the protoplanetary disc around TRAPPIST-1 required to have formed the seven planets \citep{2018MNRAS.475.5460H}. While this is at least two orders of magnitude more massive than 
the Kuiper belt \citep{2009AJ....137...72F,2010A&A...520A..32V}, note that the Kuiper belt is thought to have formed much more massive, with a solid mass of 20-40M$_\oplus$ compatible with the MMSN (\citealp[e.g.][]{1977Ap&SS..51..153W,1981PThPS..70...35H,2012AJ....144..117N}, but see \citealp{2016ApJ...818..175S} for a dissenting view).
%As a matter of comparison, we note that a belt between 42 and 47au similar to the Kuiper belt in the TRAPPIST-1 system would have a mass $\sim$1M$_\oplus$ using the surface density given by Eq.~\ref{eqsig}.
The left-over belt is not expected to extend much farther than 50-100au because protoplanetary discs around low-mass stars are less extended than around T-Tauri stars \citep{2017ApJ...841..116H}. One caveat to this estimate is that our approach is only accurate for an in situ formation of the seven planets. For planets that formed further out close to the water iceline as suggested by \citet{2017A&A...604A...1O}, the surface density would go down by a factor 10 at most and so would the belt mass
leading to an estimate of $\gtrsim$2M$_\oplus$.

The only observation of TRAPPIST-1 in the infrared is by WISE at 22$\mu$m \citep{2017AJ....153...54P}, which shows no signs of infrared excess. However, any belt that is warm enough to emit at 22$\mu$m would have to be inside 10au and so, as noted above, would be expected to be
collisionally depleted. The only region where significant mass is expected to remain at 7Gyr is beyond 10au, where such a belt would be $<15$K (assuming a black body) and therefore its emission would peak at $\lambda > 340\mu$m. 
This WISE observation is thus not constraining and observations at longer wavelengths are required to constrain such a cold belt, for instance using the ALMA interferometer.

\section{Dynamics of impacts for comets coming from an outer belt}\label{nbody}
There are many possible origins for the comets that may impact planets b to h. Rather than studying the details of the specific evolution for each scenario, we will start by assuming that very eccentric comets are produced and we will study their dynamics and look at their 
interactions with the seven planets. This framework is therefore general as soon as eccentric comets are produced and will be tied to specific scenarios (planet scattering, Kozai-Lidov mechanism or Galactic tides) in Sec.~\ref{appl}.

The pericentre $q$ of the eccentric comets we model can reach a few tenths to hundredths of au where they can collide with one of the seven detected planets around TRAPPIST-1. The apocentre Q can vary from a few au (for comets that originate in close-in belts) to $>100$s of au for comets coming from very cold outer belts or exo-Oort clouds. 
We perform N-body simulations of these very eccentric orbits 
assuming that the evolution is dominated by perturbations from the known TRAPPIST-1 planets to understand
how their fate depends on the comet's orbital parameters $q$ (pericentre) and $Q$ (apocentre). That is, for each of these different comet families (i.e for a given set of \{$q$,$Q$\}) we determine the fraction that is accreted onto the different 
planets and the fraction that is ejected. We also compute impact velocities
for each family of comets, which are used in Sec.~\ref{results} to assess if cometary impacts are able to destroy planetary atmospheres and if delivery of volatiles from these comets is possible.

\begin{table*}
  \centering
  \caption{Table describing the parameters used for the N-body simulations of the TRAPPIST-1 system \citep[from][]{2017Natur.542..456G,2017NatAs...1E.129L}.}

  \begin{threeparttable}
  \label{tab1}
  \begin{tabular}{|l|c|c|c|c|c|c|c|}
   \toprule
    Star mass ($M_\odot$) &\multicolumn{7}{c}{0.0802} \\
    Star radius ($R_\odot$) &\multicolumn{7}{c}{0.117} \\
    \midrule
    \midrule
    Planets & b & c & d & e & f & g & h\\
    \midrule
    Semi-major axis $a_{\rm pla}$ ($10^{-3}$ au) & 11.11 & 15.21 & 21.44 & 28.17 & 37.1 & 45.1 & 59.6\\
    Radius $R_{\rm pla}$ ($R_\oplus$) & 1.086 & 1.056 & 0.772 & 0.918 & 1.045 & 1.127 & 0.755\\
    Mass $M_{\rm pla}$ ($M_\oplus$) & 0.85 & 1.38 & 0.41 & 0.62 & 0.68 & 1.34 & 0.41$^1$\\
   \bottomrule
  \end{tabular}
\begin{tablenotes}
      \small
      \item $^1$ The mass of planet h is not well constrained \citep{2017NatAs...1E.129L} and as its radius is similar to that of planet d, we assume the same mass as planet d.
    \end{tablenotes}
  \end{threeparttable}
\end{table*}

\subsection{N-body simulations of impacts with the seven planets}\label{nbod}

The N-body simulations are run with REBOUND \citep{2012A&A...537A.128R} with the Hermes integrator which combines the IAS15 integrator for close encounters within a few Hill radii of the planets \citep{2015MNRAS.446.1424R} and uses the WHFast integrator otherwise \citep{2015MNRAS.452..376R}. 
The simulations include the seven planets orbiting around the central star TRAPPIST-1 (see Tab.~\ref{tab1} for the parameters used). We use a timestep of 5\% of planet b's orbital timescale.
We assumed zero eccentricities for the planets as the 2$\sigma$ upper limits are low \citep[$<0.09$ as implied by tidal forces 
and orbital stability,][]{2017Natur.542..456G,2017ApJ...840L..19T,2017ApJ...842L...5Q}. The planets gravitationally interact with each other, but their orbits do not evolve significantly over the course of all our simulations.

We start each simulation with 2000 test particles that all have a similar pericentre and apocentre \{$q$,$Q$\} spread in a narrow range defined by a grid (see Fig.~\ref{fig1}). We run the simulations until all test particles have either been ejected from the system (i.e., if their positions go beyond
100 times the initial comet's apocentre) or accreted onto the planets or the star. We note however that almost no particles collide with the central star. This is because for high-eccentricity orbits the pericentres will be locked and for low-eccentricity orbits, we notice that there are very few scattering events
that could potentially send the comets onto the star. Rather, the particles tend to be accreted or ejected by the planet close to their pericentres. 
We assumed zero inclination, which we expect to be unrealistic but leads to much faster simulations and can be scaled a posteriori to give results for a comet-like inclination distribution (see subsection \ref{incl}). Running the simulations assuming a 
zero inclination angle was necessary to allow the simulations to be performed in a reasonable timescale (i.e., not exceeding two months). The whole set of 900 simulations took $\sim$2 months on 20 CPUs, whereas inclined comets would have taken almost two years to compute. 
This is because we ran each simulation until there are no particles left. As the time to accrete/eject particles is much smaller in the zero inclination case, we gain a factor greater than 10 in overall computational time.
We note that of the results we derive in this section, the probability map as well 
as the accretion/ejection timescales are affected (in a quantifiable way) by a change in inclination but not the impact velocities (subsection~\ref{vel}).

We ran a grid of 900 N-body simulations for a wide range of \{$q$,$Q$\} values, with 90 logarithmically-spaced bins in pericentre covering $10^{-3}{\rm au}<q<10^{-1}$au and 10 logarithmically-spaced bins in apocentre covering $10^{-1}{\rm au}<Q<10^{2}$au, which
form the grid seen in Fig.~\ref{fig1}. The grid is defined by the pericentres and apocentres at the start of the simulations.
The TRAPPIST-1 planets are located between 0.01 and 0.06au (white vertical lines in Fig.~\ref{fig1}) so that the chosen range of pericentres is large enough to follow what happens when the comets' orbits cross those of the planets.

\subsection{Probability to impact the different planets or to be ejected for comet-like orbits}\label{probaimp}

\begin{figure*}
   \centering
  \includegraphics[width=19cm]{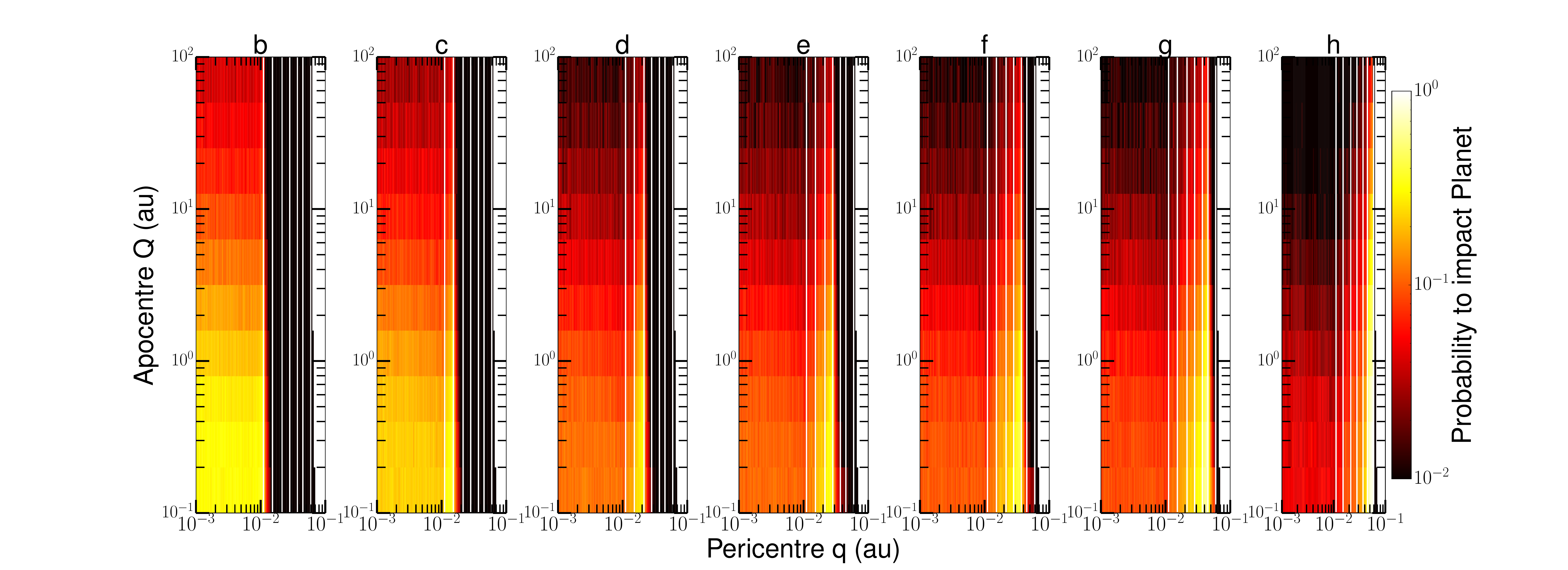}
   \caption{\label{fig1} Map showing the probability to impact (from left to right) planets b to h for each family of orbits defined by a given pericentre $q$ and apocentre $Q$. The vertical white lines show the positions of planets b to h.
The white background colour is used to show the part of the parameter space that have pericentres too far from the planets to either collide or be ejected. On the contrary, the black colour is for 
orbits that collide with a low probability $\leq 10^{-2}$.}
\end{figure*}

\begin{figure}
   \centering
  \includegraphics[width=9cm]{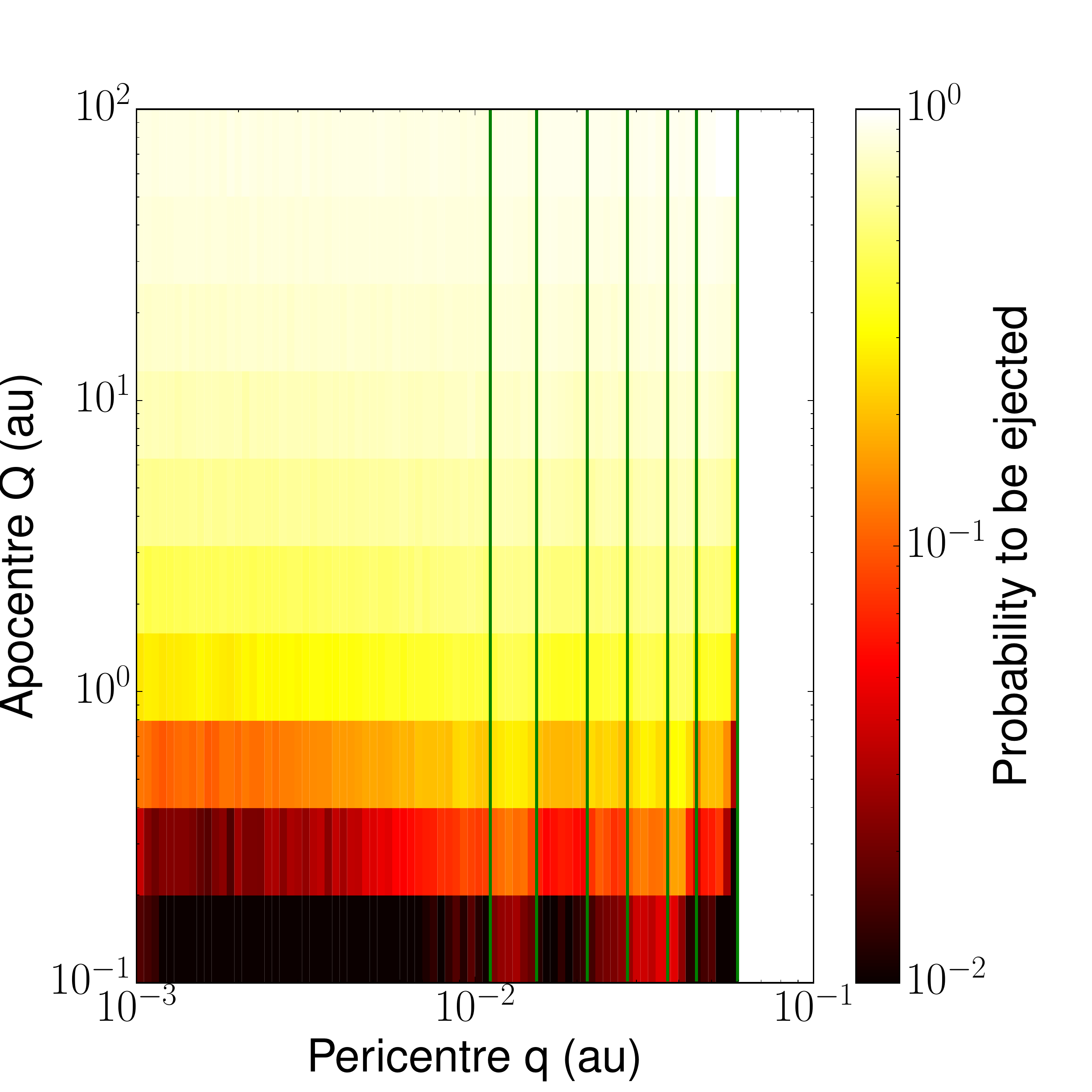}
   \caption{\label{fig2} Map showing the probability to be ejected for an orbit with a given pericentre and apocentre \{$q$,$Q$\}. The vertical green lines show the positions of planets b to h.
The white colour is used to show the part of the parameter space that have pericentres too far from the planets to either collide or be ejected. The black colour is for 
orbits that collide and eject particles but with a very low ejection probability $\leq 10^{-2}$.}
\end{figure}

Fig.~\ref{fig1} shows a map of the probability to impact the different planets (each inset is for a given planet, planet b to h from left to right), while Fig.~\ref{fig2} shows the probability to be ejected for each given \{$q$,$Q$\} of our parameter space. 
Some of the large scale features in these figures can be readily understood. 

For example, the extended black regions in Fig.~\ref{fig1} at large pericentres are because in order for a comet to collide with a given planet, the comet's pericentre must be
smaller than the planet's semi-major axis $a_{\rm pla}$. Since the pericentre and apocentre of comets do not evolve significantly  from their starting values, this means that the region of the parameter space with $q>a_{\rm pla}$
appears in black. %This is why, for instance for collision with planet b (leftmost inset in Fig.~\ref{fig1}), all simulations to the right of the leftmost green line (showing planet b's position) have a very small or zero probability to collide with the planet. 
Comets with such pericentres collide with the more distant planets.% as can be seen looking at the other insets.

Another large scale feature is that the probability to impact one of the planets is higher for smaller cometary apocentres. This can be explained by looking at Fig.~\ref{fig2} which shows that the ejection rate goes up with increasing $Q$, 
noting that the sum of the impact probabilities over the seven planets and the ejection probability equals 1. 

%After each close-encounter with a planet, a comet gets a kick of the order of the escape velocity $v_{\rm esc}$, i.e close to 10km/s for Earth-like planets, which could potentially eject it from the system.
%Fig.~\ref{fig1} shows that the probability to impact one of the planets is higher for smaller cometary apocentres. Fig.~\ref{fig2} shows the ejection map, and we conclude that for those small apocentres, the most likely outcome after a close encounter with a planet is accretion rather than ejection, 
%which is similar to what is found by \citet[][see their Fig.~4]{2017MNRAS.464.3385W}. For the case of higher $Q$s, the comet's energy is lower and it is easier to get ejected after a close encounter as shown in Fig.~\ref{fig2}.
%We describe this more thorougly in subsection \ref{anproba}.

The increased ejection probability seen in Fig.~\ref{fig2} with $Q$ (for all pericentres) is because the 
comet's energy ($\sim GM_\star / Q$) is lower for these larger apocentres and so a comet is ejected by a smaller kick when passing by a planet. The biggest kick in velocity that the comet can receive from a planet (without colliding onto it) is roughly
equal to $v_{\rm esc}$ \citep{2017MNRAS.464.3385W}, where $v_{\rm esc}$ is the escape velocity of the planet. The resulting increase in the comet's orbital energy can be enough to unbind the comet if $v_{\rm orb} v_{\rm esc} > GM_\star/Q$, where the comet's orbital velocity $v_{\rm orb}$ close to a planet is

\begin{equation}\label{vcom}
 v_{\rm com} \sim \sqrt{ \frac{2 GM_\star}{a_{\rm pla}}},
\end{equation}

\noindent which leads to the apocentre value $Q$ where ejections start becoming dominant

\begin{multline}\label{Qeq}
Q \gg 0.2 \left( \frac{M_\star}{0.08M_\odot} \right)^{1/2} \left(\frac{M_{\rm pla}}{1 M_\oplus}\right)^{-1/2} \left(\frac{R_{\rm pla}}{1 R_\oplus}\right)^{1/2} \\ \left(\frac{a_{\rm pla}}{0.03 {\rm au}}\right)^{1/2}  {\rm au},
\end{multline}

\noindent where $M_{\rm pla}$, $R_{\rm pla}$ and $a_{\rm pla}$ are the planet mass, radius and semi-major axis, respectively. This calculation explains why Fig.~\ref{fig2} shows that for $Q \gtrsim 1$au, ejection 
is the more likely outcome. %Of course, for the same given kick, a comet with a smaller energy (or larger $Q$) at the same radial location will be easier to eject, which is why we see that the ejection rate increases gradually when the apocentre $Q$ becomes larger.

Another feature in Fig.~\ref{fig1} is that for pericentres inside planet b, the accretion probability is higher for planets closer to the star. 
In fact, the accretion probability decreases as $a_{\rm pla}^{-1}$ from planet b to h for a fixed \{$q,Q$\} in this regime. This can be explained by the different
accretion timescales of each planet as showed in Sec.~\ref{times}.

Finally, another noticeable feature is that the highest probabilities of impacts (the narrow yellow regions) are for comet orbits that have pericentres close to but slightly smaller than the positions of the planets. For instance, on the planet d inset, we see that the yellow region is concentrated 
in a narrow region of the parameter space between $0.015-0.021$au (planets c and d positions). This can be readily explained because comets with such a pericentre cannot collide with planets b, c so increasing the rate of collisions with planets d, e, f, g, h. The most extreme case
is for comets that have pericentres just below planet h, thus ensuring that they can only collide with planet h and explaining the very narrow yellow region in the planet h inset.

%However, collision timescales are going to change depending on the region of the parameter space and it could be that the timescales to accrete are actually longer in some regions of the parameter space 
%even if the probability to be accreted after an infinite amount of time (as shown in Fig.~\ref{fig1}) is the same for two given cells (see Sec.~\ref{times}).

%The yellow regions can also be understood analytically because for these regions the collision between the comet and planet is at maximum velocity, thus increasing the collision rate $n \sigma v$. We describe that in more details in Sec.~\ref{vel}.

%Of course, the ejection map (Fig.~\ref{fig2}) can be understood analytically using that a cell in the ejection map 
%is just $p_e=1-\sum\limits_{i=1}^7 p_a(i)$, where we recall that $p_a$ is the probability to be accreted on to planet $i$.

\subsection{Accretion/ejection timescales}\label{times}
It is important to consider the timescale on which particles are accreted (or ejected) in the simulations, because we will later be considering how these outcomes compete with other processes that may be acting to modify 
the particles' orbits (such as the processes that brought them onto comet-like orbits in the first place).

%The accretion timescales appear
%to be very similar from one planet to another (which is expected because the particle loss is acting for the seven planets all together in the simulations and may be dominated by one only) and there is little dependence 
%on the pericentre of the comets' orbits (because the comet velocity is almost independent of $q$). The most significant dependence is in the comet's apocentre. For this reason, 

\begin{figure}
   \centering
  \includegraphics[width=9cm]{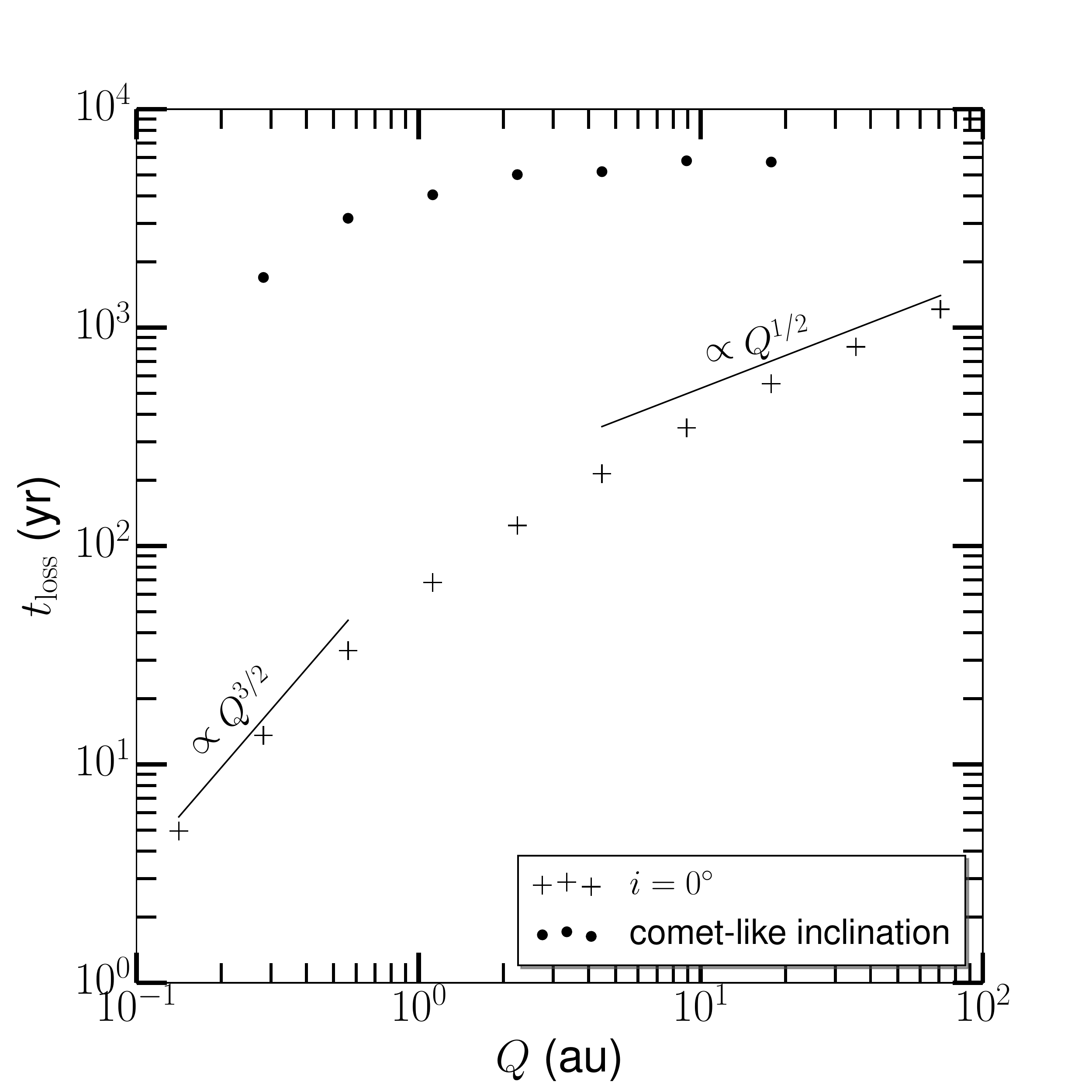}
   \caption{\label{fig3}Plot showing the removal timescale (in yr) for a comet as a function of the apocentre $Q$ (in au) for an $i=0^\circ$ (crosses) and a realistic comet-like inclination distribution (filled dots, see subsection~\ref{incl}).}
\end{figure}

\begin{figure}
   \centering
  \includegraphics[width=9cm]{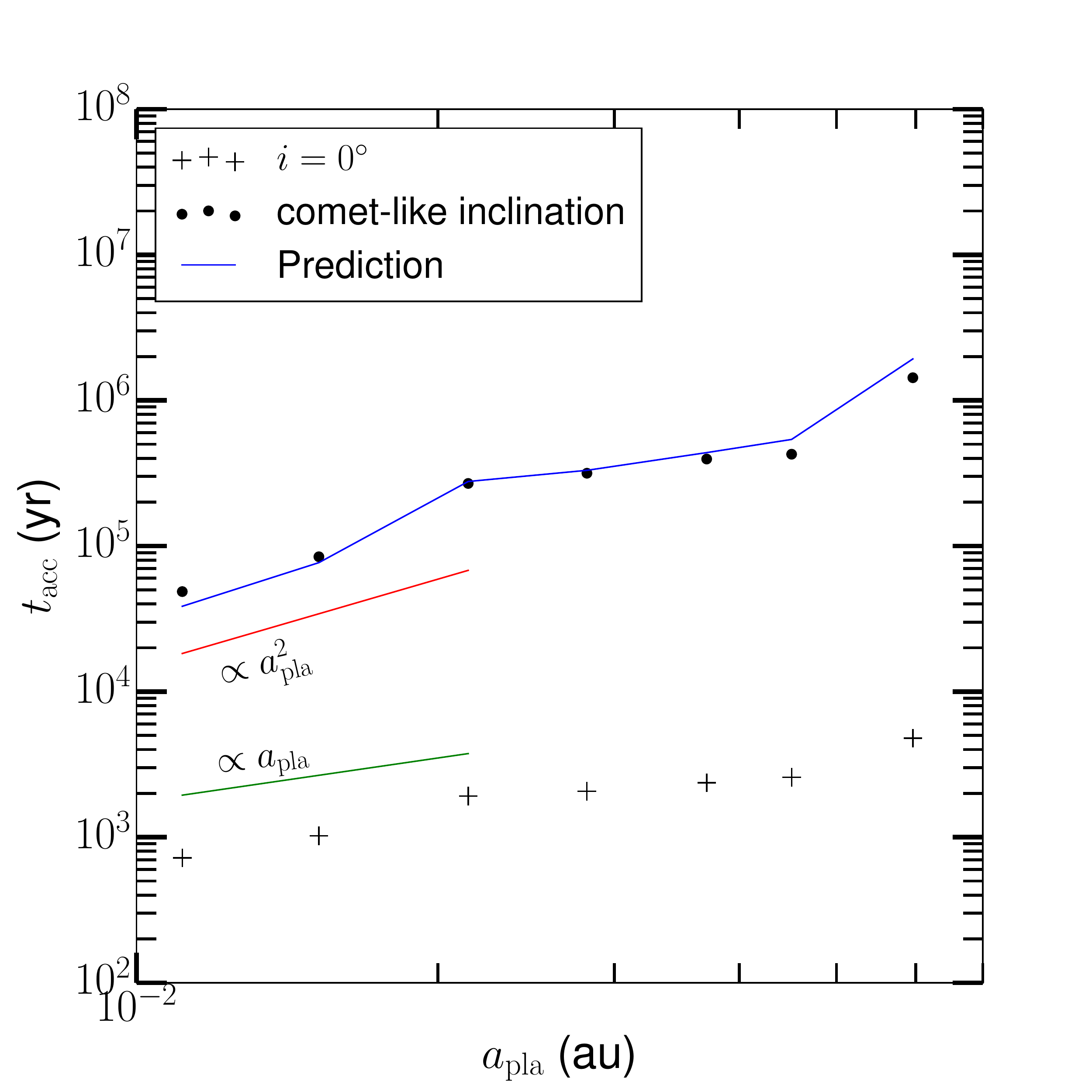}
   \caption{\label{fig3c} Accretion timescale $t_{\rm acc}=R^{-1}_{\rm col}$ as a function of the semi-major axes of the planets $a_{\rm pla}$ for an $i=0^\circ$ (crosses) and a realistic comet-like inclination distribution (filled dots).
We show the analytically predicted $t_{\rm acc}$ (blue line) for an inclined distribution of comets from the $i=0^\circ$ case (see subsection~\ref{incl}). Here, $t_{\rm acc}$ is plotted for planet b for $Q \sim 1$au and it scales as $Q^{3/2}$.}
\end{figure}

In Fig.~\ref{fig3}, we plot the loss timescale $t_{\rm loss}$ which is the timescale for half of the 2000 test particles to be lost from the simulation (through accretion or ejection) as a function of the apocentre $Q$. Since there is little dependence 
on the pericentre of the comets' orbits (because the comet velocity is almost independent of $q$), this shows that
for $Q \ll 1$au, the loss timescale scales as $Q^{3/2}$ and as $Q^{1/2}$ for larger $Q$. For $Q \ll 1$au, the loss of particles is dominated by accretion onto the planets. For a 2D geometry, the rate of collisions between a given comet and planet is proportional to $R_{\rm col}=n_\sigma \sigma_{2D} v_{\rm rel}$, where $n_\sigma$ is the fraction of the comet's orbit per unit cross-section spent in a region d$r$ around the planet's orbit, 
$v_{\rm rel}$ is the relative velocity at encounter and $\sigma_{2D}$ is the collisional cross section. 
Considering the fraction of the orbit spent in an annulus d$r$ around the planet's orbit we find that $n_\sigma$ (per cross-sectional area) is $\propto Q^{-3/2} a_{\rm pla}^{-1/2}$.
In practice, the velocity at encounter $v_{\rm rel}$ is the same as the impact velocity $v_{\rm imp}$, and
we show in Sec.~\ref{vel} that $v_{\rm imp}$ is close to the comet's velocity (see Eq.~\ref{vcom}), which is large enough for gravitational focusing to be ignored such that $\sigma_{2D}=2 R_{\rm pla}$. 
Therefore, we find that $R_{\rm col} \propto Q^{-3/2} a_{\rm pla}^{-1}$, so that the accretion timescale is $t_{\rm acc} \propto R^{-1}_{\rm col} \propto Q^{3/2} a_{\rm pla}$, explaining why the loss timescale 
scales as $Q^{3/2}$ for small $Q$. It also shows that the accretion timescale scales as $a_{\rm pla}$ as shown in Fig.~\ref{fig3c}, where we plot $t_{\rm acc}$ for planet $i$ by computing $t_{\rm loss}/p_i$, where $p_i$ is 
the probability to be accreted on planet $i$ (that we have for every $\{q,Q\}$ cell in Fig.~\ref{fig1}). This also explains why the accretion probability ($\propto t_{\rm acc}^{-1}$) decreases as $a_{\rm pla}^{-1}$ from planet b to h as noted in Sec.~\ref{probaimp}.

For $Q \gg 1$au, the loss is dominated by ejections. In that case, the cross section $\sigma_{\rm ej}$ used to calculate the rate of ejection is proportional to the impact parameter $b_{\rm ej}$ at which encounters are just 
strong enough to cause ejection. The kick $\Delta v$ that the comet receives from a planet after a close encounter scales with $1/b$, and for the ejection to happen $v_{\rm com} \Delta v > GM_\star/Q$ (see Sec.~\ref{probaimp}). This means 
that for a flat geometry $\sigma_{\rm ej} \propto Q$ and so $t_{\rm ej} \propto (n_{\rm \sigma} \sigma_{\rm ej} v_{\rm rel})^{-1} \propto Q^{1/2}$, explaining the dependencies.

%For a flat inclination case (such as our simulations), the collisional cross section close to the planet's orbit equals $\sigma_{\rm flat}=2 R_{\rm pla} H_{\rm max}$, where $H_{\rm max}=2 a_{\rm pla} I_{\rm max}$
%so that $R_{\rm col} \propto a_{\rm pla}^{-1} Q^{-3/2}$, explaining the dependency. We note that for comets with a distribution of inclinations, $\sigma=\sigma_{\rm inc}=\pi R^2_{\rm pla}$ and $R_{\rm col} \propto I_{\rm max}^{-1} \, a_{\rm pla}^{-2} Q^{-3/2} $ (see Sec.~\ref{incl}). This can be seen in Fig.~\ref{fig3c} showing $t_{\rm acc}=R^{-1}_{\rm col}$.

%As for the eccentricity, they are very high as $e=(Q/q-1)/(Q/q+1)$ and $Q \gg q$ for most simulations. Only
%in the lowermost right part of the planet 7 inset in Fig.~\ref{fig3} can we start looking at eccentricities close to 0.4, which explains why $T$ goes up by a small amount in this lowermost right part of the inset.

\subsection{Impact velocities for the different planets}\label{vel}

An important parameter to determine the effects of a cometary impact onto the atmosphere of a planet is the impact velocity. 

\begin{figure}
   \centering
  \includegraphics[width=9cm]{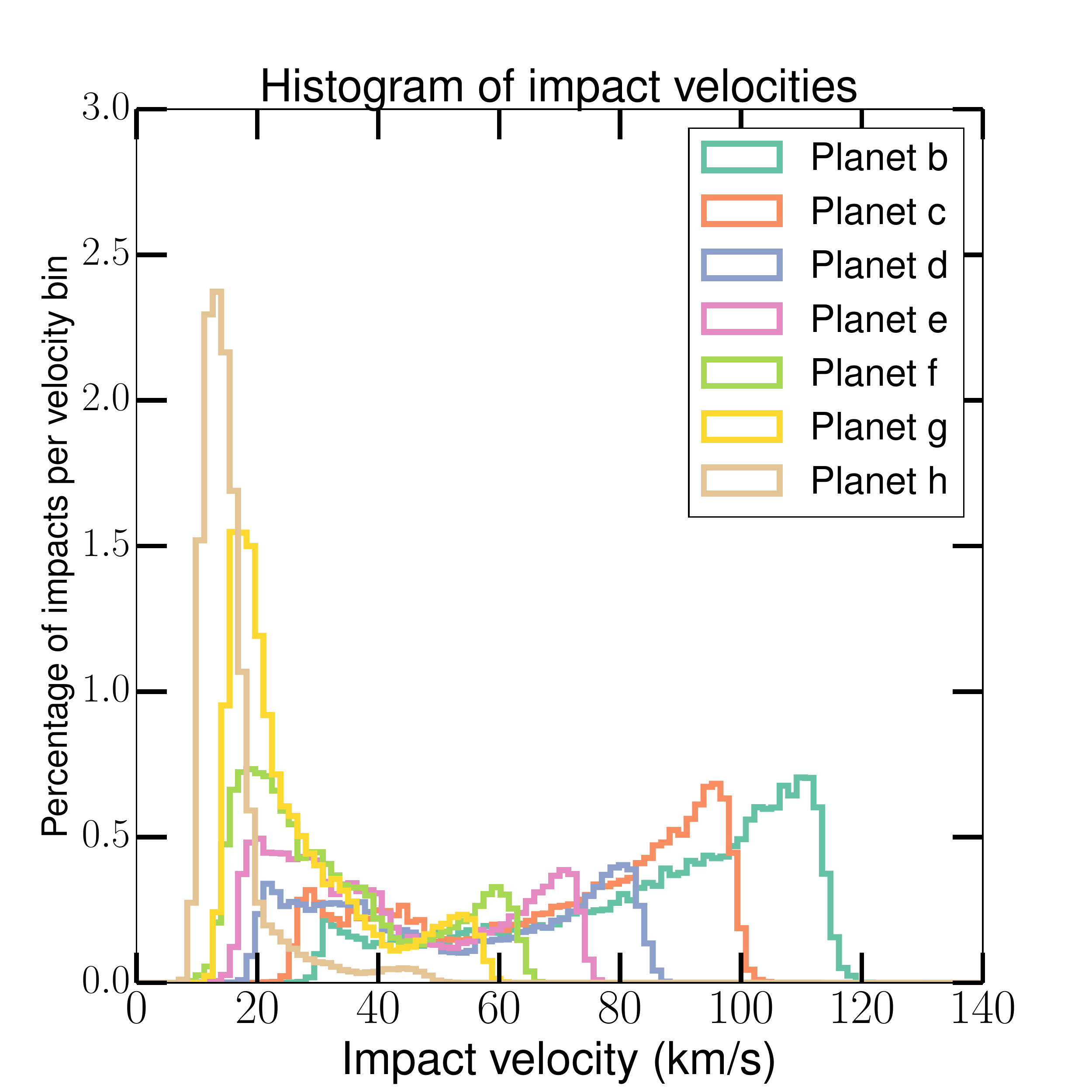}
   \caption{\label{fig4} Histogram of impact velocities (cut in 100 bins) for each planet 
weighted by the impact probability from Fig.~\ref{fig1}. The y-axis shows the percentage of impacts per velocity bin.}
\end{figure}

\begin{figure}
   \centering
  \includegraphics[width=9cm]{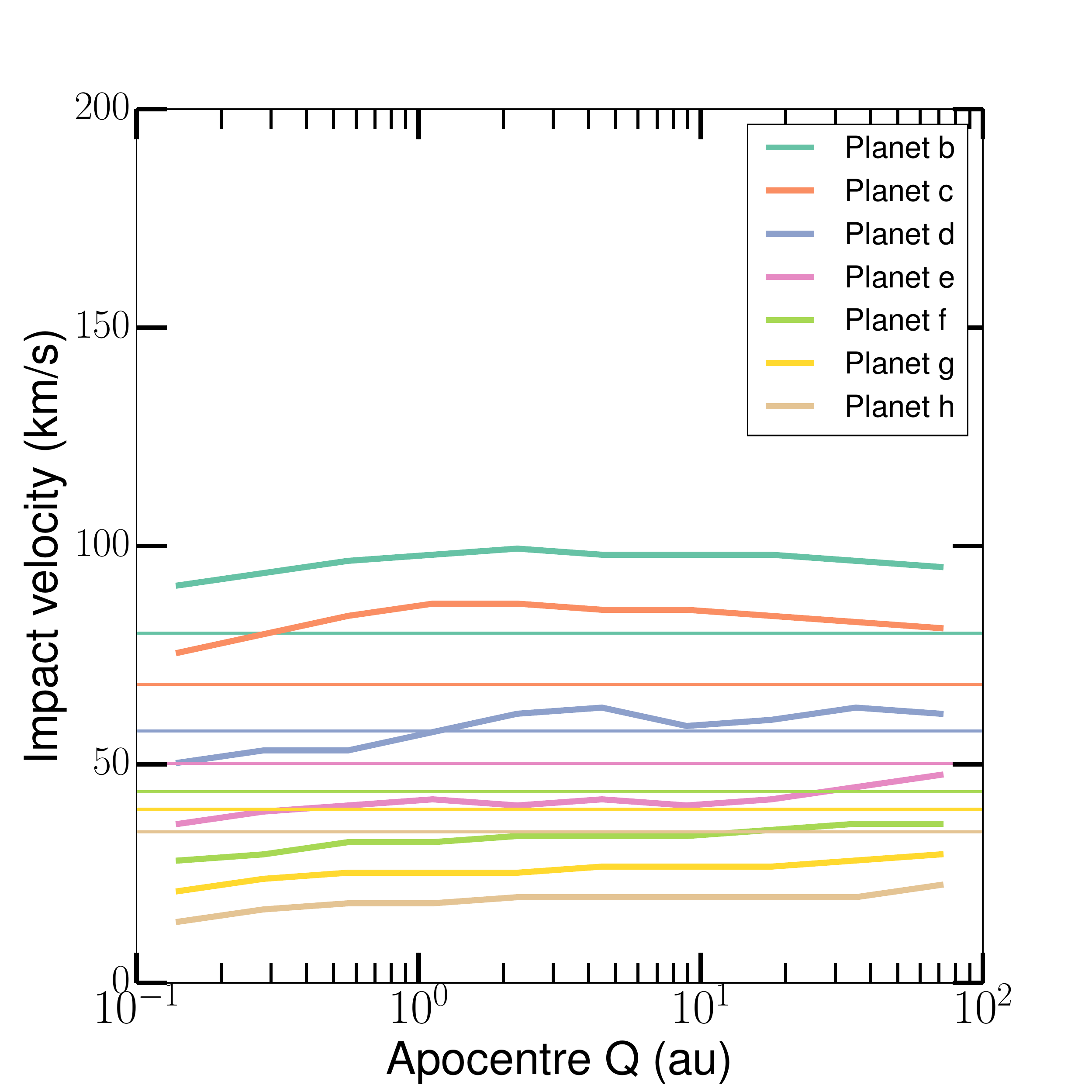}
   \caption{\label{fig4b} Median velocity of the distributions shown in Fig.~\ref{fig4} for different apocentres Q. The thin horizontal lines show the Keplerian velocity of each planet for comparison.}
\end{figure}

In Fig.~\ref{fig4},
we show histograms of impact velocities for the different planets. We computed the impact velocities for each simulation (i.e. for specific pericentres and apocentres), but
find that the distributions of impact velocity do not depend significantly on the comet's pericentre. To get Fig.~\ref{fig4}, we therefore average the $v_{\rm imp}$ distributions over the pericentres in the grid, assuming that
comet orbits are uniform in log $q$ (keeping a fixed apocentre). Averaging in this way results in more accurate histograms of impact velocities. To do so, the impact velocities from the different simulations are weighted by the probability to impact the different planets (using Fig.~\ref{fig1}).
Furthermore, Fig.~\ref{fig4b} shows that the medians of the impact velocity distributions for each planet also do not depend significantly on apocentre. Thus while Fig.~\ref{fig4} shows the distributions for an apocentre of $\sim$1au, these distributions are also representative of that of a large range of apocentres.

We see that the impact velocity distribution is peaked at a different location for each planet from $\sim 15$ to $\sim$110km/s from planet h to b. A much smaller
secondary peak can also be seen for each planet. This is because there are two extreme types of impacts. Collisions can occur when the comet is on a near radial orbit approaching or receding from pericentre. They may also occur when the planet and comet velocities are parallel (i.e when the comet encounters the planet near its pericentre). 
As shown by Eq.~\ref{vcom}, the comet velocity at impact is $\sim \sqrt{ 2 GM_\star/a_{\rm pla}}$
(for $a_{\rm pla} \ll a$), which is thus always higher than the planet's Keplerian velocity of $\sqrt{ GM_\star/a_{\rm pla}}$ (which varies from $\sim$35km/s for the farthest to $\sim$80km/s for the closest planet).
Therefore, we find that the impact velocity distributions should peak at $\sqrt{GM_\star/a_{\rm pla}} (\sqrt{2}-1)$ 
(33, 29, 24, 21, 18, 16, 14km/s, for planet b to h) for parallel orbits at impact and would be maximal for radial encounters at $\sqrt{ 3 GM_\star/a_{\rm pla}}$. %Perpendicular encounters are an extreme case
%but the velocity should peak close to 45 degree encounters equal to $1.31\sqrt{GM_\star/a_{\rm pla}}$ (105, 90, 76, 66, 57, 52, 45km/s, for planet 1 to 7), which is what is observed in Fig.~\ref{fig4}. 
We note that the impact velocities are much greater than the escape velocities of the planets ($\sim$10km/s) and therefore gravitational focusing is not important.

Thus the high velocity peaks correspond to comets colliding on radial orbits and the low velocity peaks to comets falling on the planets at their pericentres (i.e., parallel collision). By looking at
Fig.~\ref{fig1}, we see that for planet h, the highest impact probability region (the yellow region) is very narrow and restricted to comets
whose pericentres are close to planet h's position so that most collisions are going to be parallel. This explains why the low velocity peak is higher for this planet. For planet b, however,
the yellow region is large and not peaked close to planet b's semi-major axis. Therefore, most impacts will happen with comets on nearly radial orbits and the
high velocity peak is therefore higher than the low velocity peak. Histograms for the other planets can be understood following the same procedure.

The non-dependence of impact velocities on apocentres shown in Fig.~\ref{fig4b} also derives from the velocity at impact, which, as
shown by Eq.~\ref{vcom}, only depends on $a_{\rm pla}$ and not $a$.
We notice that the median velocities are close to the Keplerian velocities of the corresponding planets (Fig.~\ref{fig4b}).

\subsection{Simulations for realistic inclinations}\label{incl}
The simulations assumed comets with zero inclination. To check how our results change for different inclinations, we ran a set of 30 additional simulations (spread across the $\{q,Q\}$ parameter space) with more realistic comet-like inclinations. 
The chosen inclination distribution follows a Rayleigh distribution peaking at 10 degrees, i.e
close to the distribution of JFC\footnote{The Jupiter-family comets are short period comets that orbit part of the time in inner regions of our Solar System and whose orbits are primarily influenced by Jupiter's gravity.} comets \citep{1997Icar..127...13L,2009Icar..203..140D}.
We find that the loss timescale (see Fig.~\ref{fig3}, filled dots) and the timescale for accretion onto the different planets are affected (see Fig.~\ref{fig3c}, filled dots), but that the impact velocities are unaffected. 

The difference in $t_{\rm acc}$ between the inclined and flat cases in Fig.~\ref{fig3c} can be explained by generalising the analytics in subsection \ref{times}. The ratio of the rates at which comets collide with a planet is expected to be $(n_v \sigma_{3D})/(n_\sigma \sigma_{2D})=\pi R_{\rm pla}/(4 I_{\rm max} a_{\rm pla})$, where $n_v$ is now number per unit volume in the vicinity of the planet and $I_{\rm max}$ the median inclination of the comets in the 3D case. 
We plot this analytical prediction (blue line) together with some numerical simulations for a distribution of inclinations (filled dots) in Fig.~\ref{fig3c}. A similar comparison shows that the dependence in $Q$ remains the same for $t_{\rm acc}$ for the two types of simulations. We thus conclude that the zero inclination simulation 
collisional rates can be scaled to account for the inclined case.

To recover the probability map shown in Fig.~\ref{fig1} for the case of an inclined distribution, we also need to rescale the loss timescale $t_{\rm loss}$ shown in Fig.~\ref{fig3} because the probability $p_i$ to be accreted on a given planet $i$ is equal to $t_{\rm loss}/t_{\rm acc}$.
The results for $t_{\rm loss}$ from the inclined numerical simulations are shown in Fig.~\ref{fig3} (filled dots). We can predict the change in ejection timescale (which dominates $t_{\rm loss}$ at $Q \gg 0.2$au) for the 3D case from the 2D simulations in the same way as for the timescales for accretion onto the planets. This prediction is that ejection timescale should be longer by $0.8 (M_\star/M_{\rm pla}) (a_{\rm pla} I_{\rm max}) / Q$. This is reasonably accurate but we prefer to use the numerical ratio of $t_{\rm loss}$ for the inclined and zero-inclination cases which is best fit by a power law equal to $63 \, (I_{\rm max}/10^\circ) \, Q^{-0.61}$.

\begin{figure*}
   \centering
  \includegraphics[width=18cm]{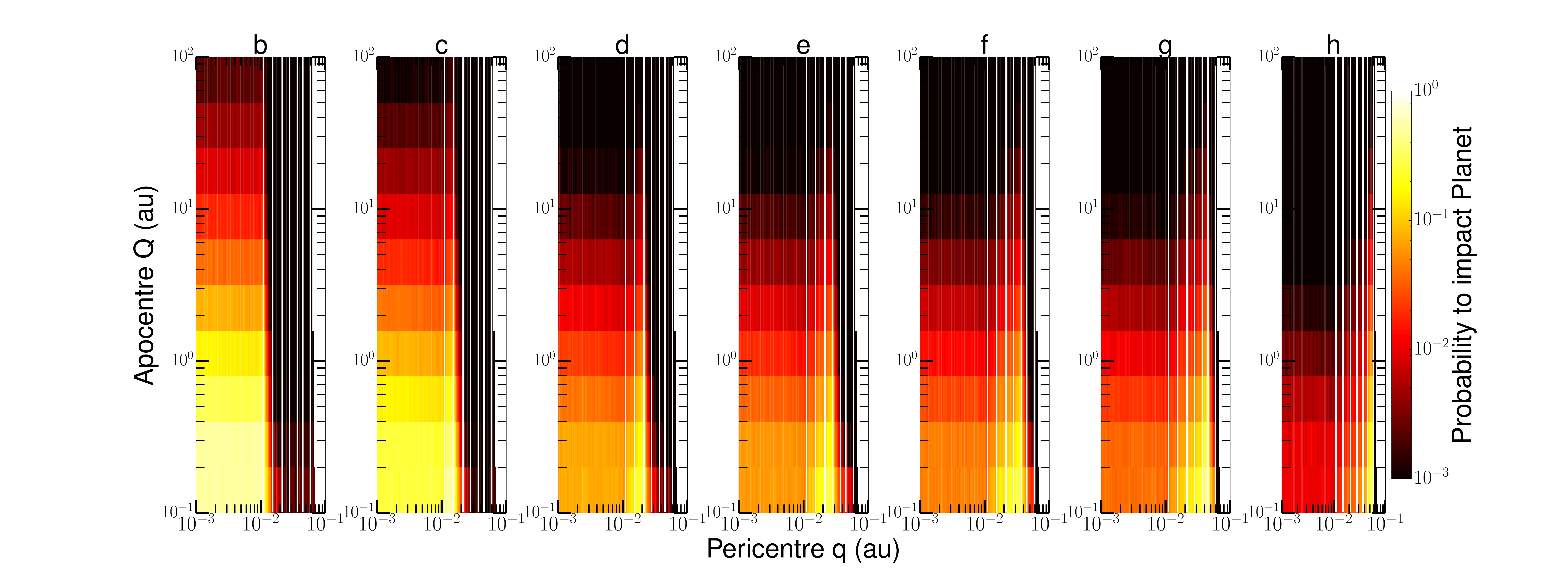}
   \caption{\label{fig3d} New probability map: similar to Fig.~\ref{fig1} but for an inclined distribution of comets (see Sec.~\ref{incl} for more details).}
\end{figure*}

\begin{figure}
   \centering
  \includegraphics[width=9cm]{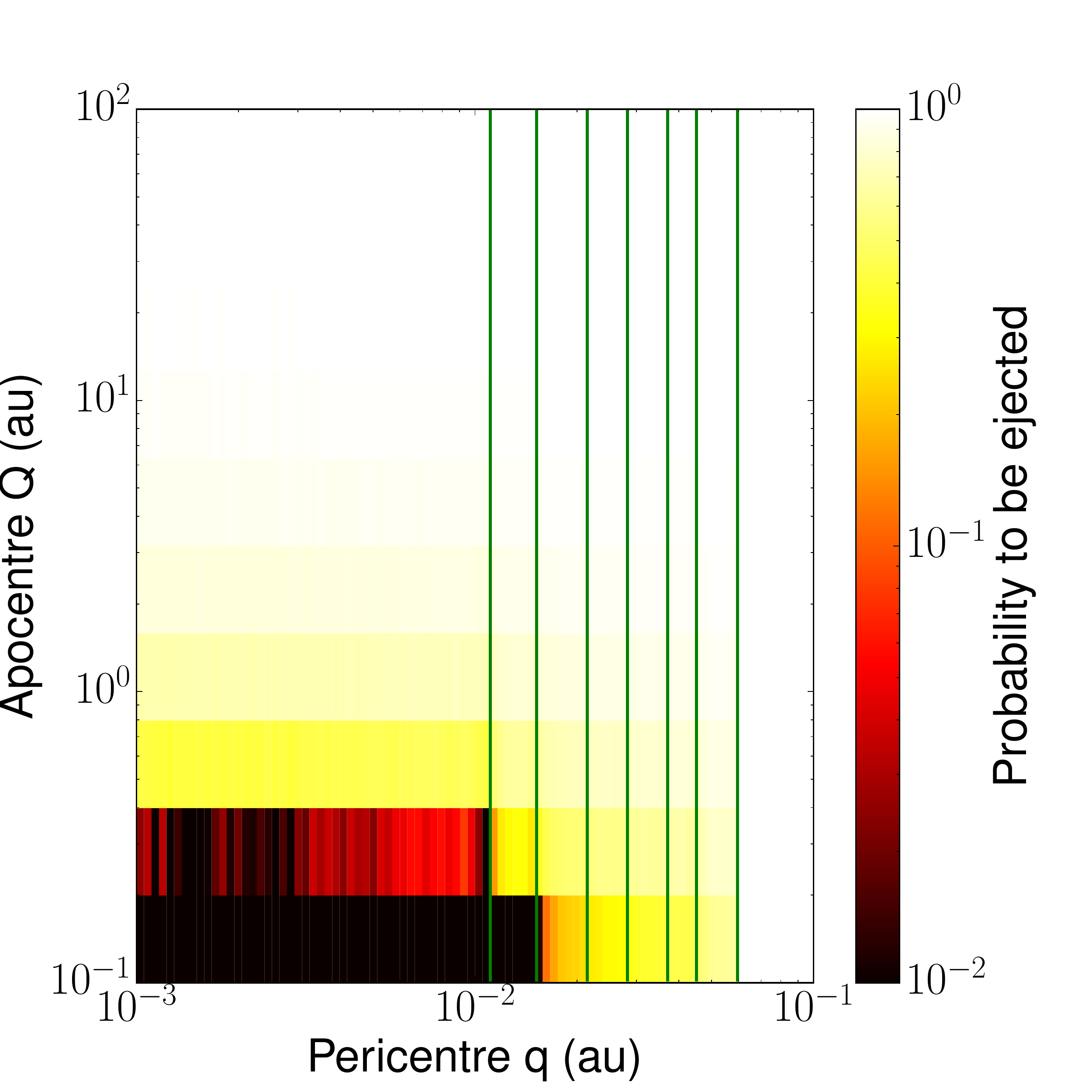}
   \caption{\label{fig3e} New ejection map: similar to Fig.~\ref{fig2} but for an inclined distribution of comets (see Sec.~\ref{incl} for more details).}
\end{figure}

Using these different scalings, we calculate the new probability map (see Fig.~\ref{fig3d}) and use that the sum of the probabilities to be accreted onto each of the planets and to be ejected equals 1 to compute a new ejection map (see Fig.~\ref{fig3e}).
Comparing the predictions from our scalings to the different results from the inclined distribution simulations, we find that we are accurate within a factor 2.

\section{Different scenarios to make eccentric comets}\label{appl}
\begin{figure}
   \centering
  \includegraphics[width=9cm]{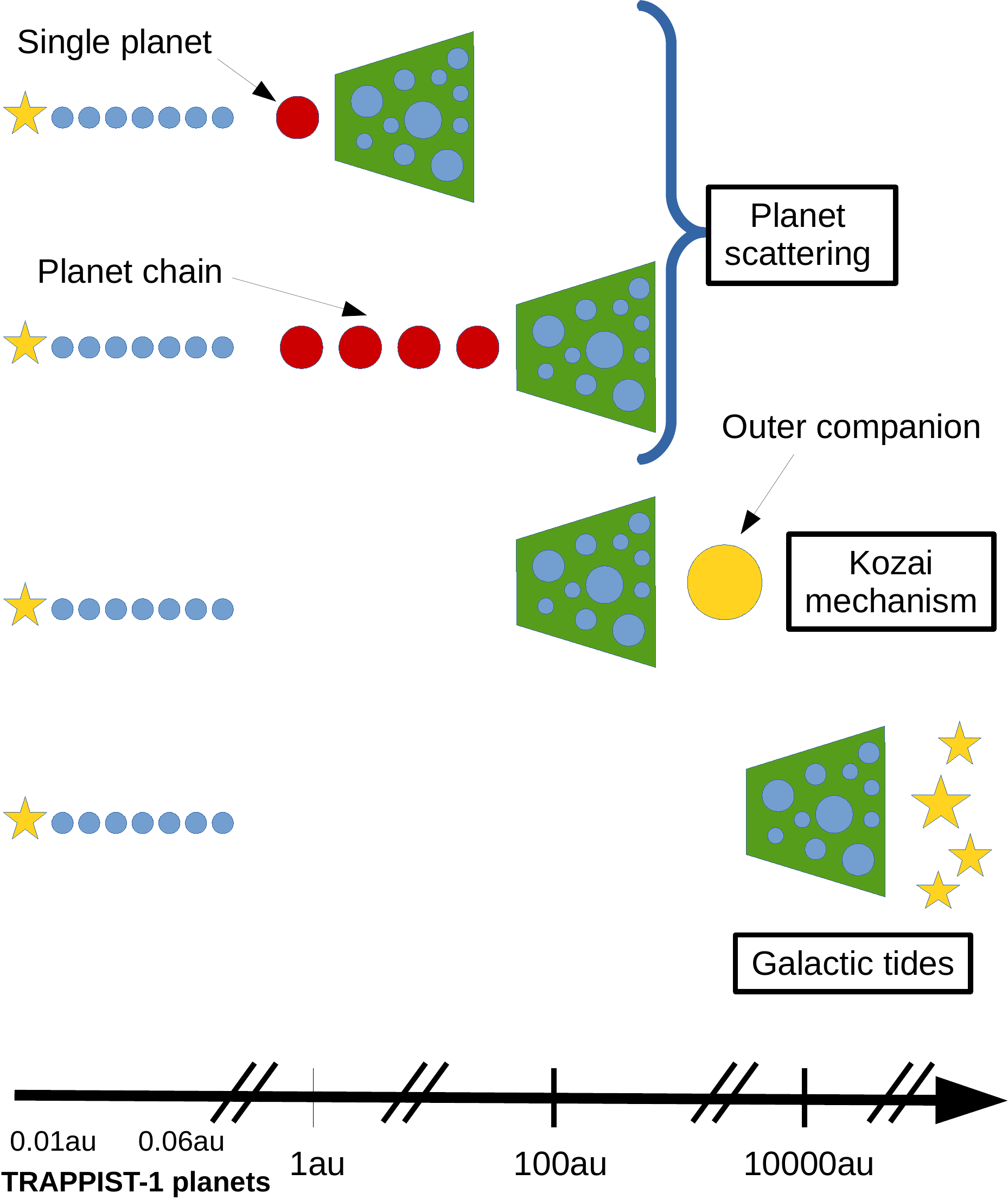}
   \caption{\label{schem}Schematic of the different scenarios tested that could potentially scatter comets in the inner regions of the TRAPPIST-1 system.}
\end{figure}

%\setcounter{subsection}{-1}
%\subsection{A general model for impact scenarios: constant $\dot{q}$}
We have studied the dynamics of highly eccentric comets in the presence of the seven TRAPPIST-1 planets in the previous section. Here, we consider three different scenarios that can send comets from the outer regions of the TRAPPIST-1 system onto such eccentric orbits (see Fig.~\ref{schem});
1) A planetesimal disc is perturbed by a nearby planet and comets are scattered inwards
by this single planet or through a chain of planets \citep[similar to comets scattered in our Solar System, e.g.][]{1997Sci...276.1670D,2012A&A...548A.104B,Marino17}. 2) A distant companion to TRAPPIST-1 forces comets in a Kuiper belt-like disc to undergo Kozai-Lidov oscillations \citep[e.g.][]{2016ApJ...826...19N}, which
can bring the comets to very close pericentres. 3) Galatic tides perturb a far away exo-Oort-cloud and send comets to decreasing pericentres.

We assume that the evolution of comets' orbits in these three scenarios can be approximated as an evolution in which their apocentres $Q$ remain constant and their pericentres $q$ decrease at a constant rate $\dot{q}$. This approximation 
allows us to use the results of Sec.~\ref{nbody} to consider the
outcome for comets scattered into the inner regions without having to consider the detailed dynamics of the comets' origin. The simplified dynamics allows us to study a wide range of different possible scenarios. 
Owing to this simplification, the results are expected to give order of magnitude correct estimates, which is justified by the uncertainties on the presence of a belt in this system and its yet unknown properties.
We explore expectations for the different $\dot{q}$ values for each of these three scenarios below.

\subsection{Impacts from comets scattered by a single or a series of planets}\label{sca}

In our Solar System, comets from the Kuiper belt are thrown into the inner Solar System thanks to a series of scattering by different planets. This planet scattering scenario has been invoked multiple times \citep{2010ApJ...713..816N,2009MNRAS.399..385B,2014MNRAS.441.2380B} to try to explain
the presence of hot dust around many stars \citep[see][for a review]{2017AstRv..13...69K}. More recently, \citet{Marino17} studied the effect of scattering by a chain of planets for a large parameter space so as to understand which planetary systems are more suited to create
large hot dust levels or maximise impacts on the chain of planets. They ran simulations for 1Gyr for different chains of planets with semi-major axes ranging from 1 to 50au and planet masses ranging from a few to 100M$_\oplus$.
In our case, we consider that interactions with the innermost planet of the chain dominate the comets' dynamical evolution as they reach to the very small pericentres considered here. However, we also have to
consider that some fraction of comets would have been ejected or accreted by the other planets
before reaching the innermost planet of the chain. \citet{Marino17} show that for a wide range of planet chain architectures, $f_{\rm in}=1-7$\% of comets originating in an outer belt end up reaching the inner system. For a close-in belt similar to the debated \citep[see][]{2018ApJ...855L...2M} belt recently invoked around the M-dwarf 
Proxima Centauri \citep{2017arXiv171100578A}, a single planet at 1au could be enough to scatter comets into the inner regions (but we note that we showed in Sec.~\ref{mmsn} that such a close-in belt is not likely to have survived around TRAPPIST-1 unless it formed recently). In that case, no comets are lost on the way through the chain and $f_{\rm in}=1$. 
We only consider the case of a planet coplanar to the 7 planets as this system seems well-aligned. A non-coplanar configuration would lead to an increased inclination distribution, which effect could be quantified using the analytics from Sec.~\ref{incl}.

Conservation of the Tisserand parameter\footnote{This parameter is a constant in the circular three-body problem, which constrains the orbit of a comet after being scattered by a planet.} \citep{1999ssd..book.....M} means that comets being scattered by an innermost 
planet that is on a circular orbit can only reach down to a minimum pericentre \citep[see][]{2012A&A...548A.104B}. Thus to scatter comets to small enough pericentres to reach the TRAPPIST-1 planets' locations, we consider that the innermost planet must be on an eccentric orbit, since in that case there is no minimum pericentre constraint \citep[see also][]{2014MNRAS.439.2442F}. We also consider that this planet should not be too massive, so as not to eject the comets before they can reach the innermost parts of the system. 

Guided by the results of \citet{2014MNRAS.439.2442F} we consider 1 and 10M$_\oplus$ planets with a 0.4 eccentricity orbiting at 1au to be representative of the kind of planets 
that are able to put comets on orbits that are capable of colliding with the seven known TRAPPIST-1 planets. We note that such planets are not massive enough and not close enough to gravitationally 
disturb the orbits of the seven currently known planets as can be checked
directly from \citet{2016MNRAS.457..465R}, so that the system of the seven inner planets stays stable even in the presence of such an additional planet \citep[see also][]{2017ApJ...842L...5Q}. We also note that these planet masses agree with current mass upper limits by \citet{2017arXiv170802200B} (i.e $< 4.6 M_{\rm Jup}$ within a 1 yr period, and $< 1.6 M_{\rm Jup}$ within a 5 yr period).
While an eccentricity of 0.4 is above the median eccentricity found for Earth to Super-Earth mass planets, such eccentric planets are observed. We also note that our scenario would still work for lower eccentricities as described in \citet{2014MNRAS.439.2442F} but the $\dot{q}$ value would vary.

Moreover, we find that the outermost planet interacting with the belt and causing the scattering  
could migrate outwards if it has a mass $\lesssim$ 10M$_\oplus$ \citep[see Eq.~58 in][where we used the surface density of the potential surviving belt shown in Fig.~\ref{fig1b}]{2012ApJ...758...80O} and stall if it is more massive. 
Such a migration is beneficial to sending more comets inwards as shown in \citet{2014MNRAS.441.2380B} because more time is available for the scattering process and it can access more material to scatter from. 
However, for a planet mass $\lesssim$0.1M$_\oplus$, the scattering would not be efficient anymore as shown by Eq.~52 of \citet{2012ApJ...758...80O}. Too massive an outer planet would also
prevent material from being scattered inwards as it would be more likely ejected but disc evolution models find that having another Jupiter in the TRAPPIST-1 system is unlikely \citep{2018MNRAS.475.5460H}. \citet{Marino17} show that for planet masses $\lesssim$100M$_\oplus$, $\gtrsim 2\%$ of the scattered comets still reach the inner region. More massive planets such as a Jupiter would 
more likely eject most of the material \citep[see][]{2017MNRAS.464.3385W}.

\begin{figure}
   \centering
  \includegraphics[width=9cm]{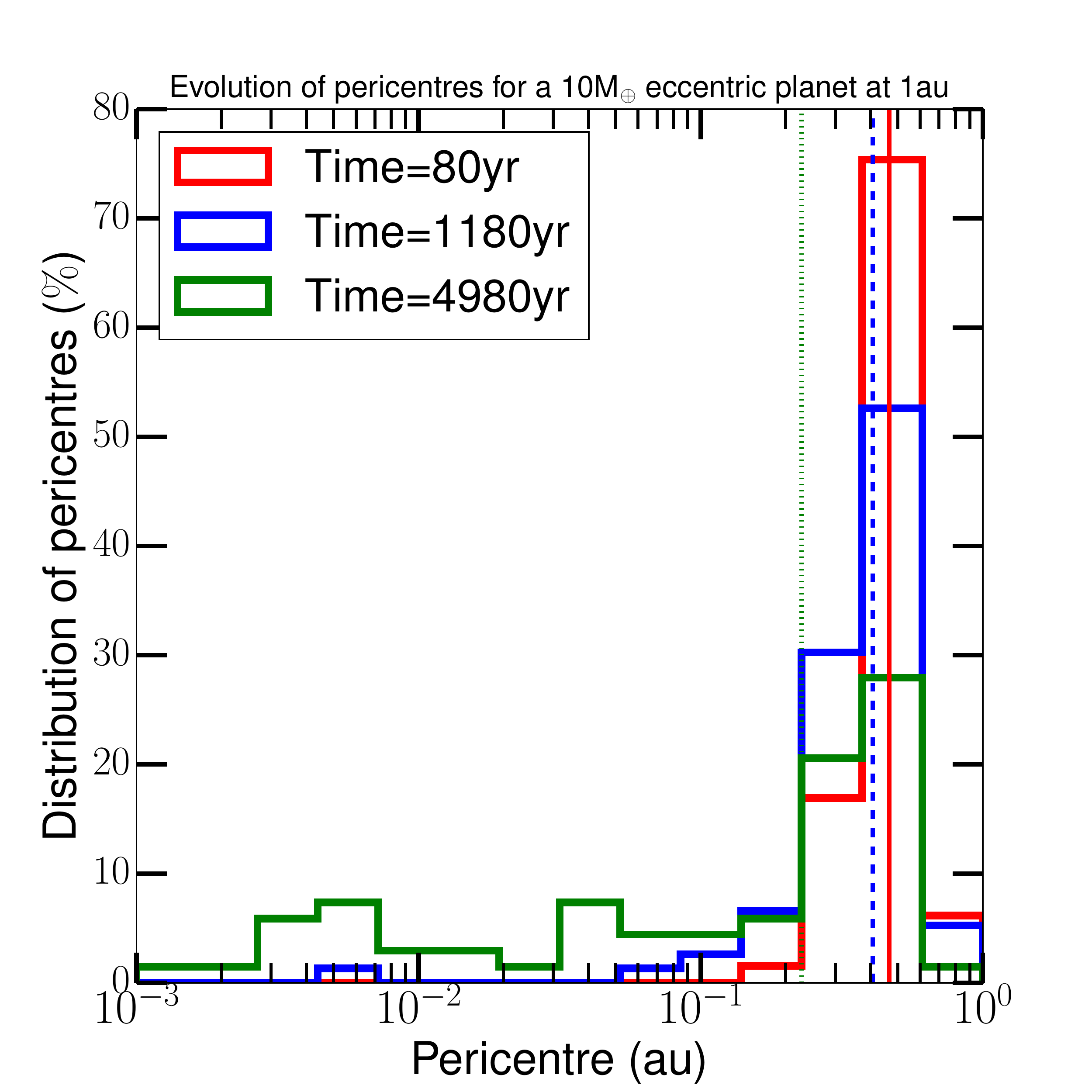}
   \caption{\label{fig5} Distribution of pericentres at 3 epochs (80, 1180 and 4980yr) due to perturbations of an eccentric (e=0.4), 10M$_\oplus$ planet located at 1au. Particles are initially placed in the planet's chaotic zone and the ones with apocentres at the planet get periodic kicks
that push their pericentres $q$ inwards. The median of distributions are shown as vertical lines. The median of $q$ goes inward with time and we find $\dot{q}=5 \times 10^{-5}$au/yr for this case.}
\end{figure}

We ran N-body simulations to follow the evolution of test particles initially randomly located
in the chaotic zone \citep[where resonances overlap, as classically defined in][]{1980AJ.....85.1122W} of such an eccentric planet. The planet is located at 1au and simulations are run until the planet runs out of material to scatter among the initial 5000 particles in 
the chaotic zone. Fig.~\ref{fig5} shows the evolution of the distribution of pericentres of particles that have their apocentres at the 10M$_\oplus$ planet location, which decreases steadily with time. Quantifying the rate of this decrease by looking at the evolution 
of the median of the distribution, we find that 

\begin{equation}\label{qp}
\dot{q}_P \sim 5 \times 10^{-5} {\rm au/yr}, 
\end{equation}

\noindent over many orbits for the 10M$_\oplus$ case. Running another simulation for an Earth mass planet with similar eccentricity at the same location, we find that $\dot{q}_P \sim 10^{-5}$au/yr.

While the path of an individual comet could be somewhat stochastic through the parameter space (jumping in individual scattering events), the effect for an ensemble of comets is that of a slow inward migration of $q$.
Therefore, we model this population, and how it is depleted due to interactions with the seven inner planets, by assuming that comets have an apocentre $Q$ that is fixed at the position of the innermost planet of the chain (1au), and considering the various depletion pathways
as the comets cross the parameter space in Figs.~\ref{fig1} and \ref{fig2} at a constant rate $\dot{q}$. That rate depends on the mass of the planet, so we keep this as a free parameter, noting that Eq.~\ref{qp} gives realistic values. That rate has a strong influence on the outcome (see Sec.~\ref{results}).

\subsection{Impacts from comets undergoing Kozai-Lidov oscillations due to an outer companion}\label{koz}

The incidence of binaries around M-dwarfs is around 27\% \citep{2012ApJ...754...44J}. Comets located at tens of hundreds of au (either in a disc or a more spherical Oort-cloud like distribution) could be perturbed by such distant companion stars.  
If the mutual inclination $i_0$ of some comets with this companion is greater than 39.23 degrees, the so-called Kozai-Lidov cycle can start and 
the mutual inclination starts decreasing while the eccentricity of the comet increases to reach a maximum \citep{1962AJ.....67..591K,1962P&SS....9..719L}. 

For the case of a circular outer companion, the maximum eccentricity reached by the comets is given by $\sqrt{1-(5/3) \cos^2(i_0)}$ \citep{1997AJ....113.1915I}. This means that to reach a pericentre $q < q_1=10^{-2}$au (to be able to reach the seven planets), the initial 
mutual inclination should be greater than $i_0 > \arccos \left (\sqrt{\frac{3}{5} \left (1-(1-q_1/a)^2 \right )} \right )$, where $a$ is the semi-major axes of the comets. 
If the belt is really close-in, i.e $a=0.1$au, this corresponds to $i_0>70.3$ degrees, and at 100au, it gives $i_0>89.4$ degrees (i.e., an almost perpendicular orbit is necessary in this latter case). We note that this inclination is between
the perturber and comets and the latter can be inclined compared to the planets. Therefore, finely tuned companions would be needed to send comets to the right location.

However, for an eccentric outer companion, the comets' eccentricity can reach values arbitrarily close to 1 \citep[e.g.][]{2011ApJ...742...94L,2013ApJ...779..166T}. While the periastron precession (due to GR) can 
dominate the dynamics \citep{2015MNRAS.447..747L} and stop the Kozai mechanism from working when the eccentricity comes close to 1, this occurs only for pericentres interior to the known TRAPPIST-1 planets \citep[see Eq.~50 and Fig.~6 of][]{2015MNRAS.447..747L}.
% show the maximum eccentricity reached owing to GR depending on the system's parameters
%such as the mass of the outer companion, its eccentricity, its semi-major axis and that of the test particle from which they derive a dimensionless parameter called $\epsilon_{\rm GR}$ (see their Eq.~33). They find that for $\epsilon_{\rm GR} \ll 1$, the maximum eccentricity is not affected by GR precession
%and test particles can still reach very small pericentres with $1-e \sim 10^{-6}$ \citep{2015MNRAS.447..747L}. Using our system parameters and a range of plausible values for the position, mass and eccentricity of the outer companion and disc position (see later in this section), we find that $\epsilon_{\rm GR} < 10^{-8}$, meaning that the GR 
%precession will not affect the Kozai mechanism in our problem. 

The Kozai oscillations will occur 
even if the perturbing companion is very distant and/or not very massive, only the timescale to achieve the eccentricity change will be longer in that case. Assuming the initial eccentricities $e_0$ of 
comets in a disc are small then the timescale $T_K$ to reach the maximum eccentricity given an eccentric perturber is \citep{2015MNRAS.452.3610A,2016ARA&A..54..441N}%\citep{1997AJ....113.1915I}Innanen97

\begin{equation}
 T_K \sim \frac{2.7}{\sqrt{\epsilon_{\rm oct}}} \left ( 1+ \frac{M_\star}{M_c} \right) \left( \frac{P_c^2}{P} \right) \left( 1-e_c^2 \right)^{3/2},
\end{equation}

\noindent where $a_c$, $e_c$, $M_c$ and $P_c$ are, respectively, the semi-major axis, eccentricity, mass and orbital period of the companion and $P$ is the orbital period of the comet being perturbed (i.e $P=2\pi \sqrt{a^3/(GM_\star)}$). The parameter $\epsilon_{\rm oct}=e_c(a/a_c)/(1-e_c^2)$ quantifies
the relative size of the octupole term of the Hamiltonian compared to the quadrupole term and is not equal to zero for an eccentric perturber \citep[the timescale for a circular orbit can be found by substituting $\epsilon_{\rm oct} \sim 259$, see][]{2015MNRAS.452.3610A}.

Then, we can determine
an order of magnitude for the rate of pericentre evolution given by
$\dot{q}_K \sim (a (1-e_0) -q_1)/T_K \sim a(1-e_0)/T_K$, so that

\begin{multline}
 \dot{q}_K \sim 2 \times 10^{-4} \left( \frac{a}{100{\rm au}} \right)^{3}  \left( \frac{1-e_0}{1-0.1} \right) \left( \frac{a_c}{150{\rm au}} \right)^{-7/2} \\ \left( \frac{e_c}{0.4} \right)^{1/2} \left( \frac{1-e_c^2}{1-0.4^2} \right)^{-2}   \left( \frac{1+M_\star/M_c}{1+0.08/0.01} \right)^{-1} \\ \left( \frac{M_\star}{0.08 M_\odot} \right)^{1/2}  \mathrm{au/yr}.
\end{multline}

\noindent Therefore, considering a belt at 100au perturbed by an eccentric companion of mass 0.01$M_\odot$ at 150au\footnote{We note that such a low mass companion at large distances is not yet ruled out \citep{2016ApJ...829L...2H,2017arXiv170802200B}.}, we find that $\dot{q}_K \sim 2 \times 10^{-4}$au/yr.
While a much farther companion could decrease that value and a farther exo-Kuiper belt would increase $\dot{q}_K$, we consider in Sec.~\ref{fraction} how evolution at typical $\dot{q}$ might affect the planetary atmospheres of TRAPPIST-1 planets. 

We have also checked that for such a configuration the Kozai mechanism cannot be suppressed by the precession induced by unknown planets in the system. Imagining the worst case scenario of the presence of a Jupiter-mass planet at 10au in this system, the Kozai dynamics remains dominated by the outer companion if the belt is located further 
than $\sim$30au \citep{2017ApJ...834..116P}, which is assumed here. 
We also looked at the effect of the seven known TRAPPIST-1 planets on the Kozai mechanism. The bodies that can reach these planets must be very eccentric and to take that into account properly, we model the effect of these planets as being an effective $J_2$ (quadrupole moment) and check
whether the precession rate due to $J_2$, i.e. $\omega_{J_2}$ is able to counteract the precession due to Kozai. Using Eq. 35 in \citet{2007ApJ...669.1298F}, we find that the effective $J_2$ of the 7 TRAPPIST-1 planets starts contributing and reduce the maximum Kozai eccentricity for
 
\begin{multline}\label{J2}
J_2 > 10^5 \left( \frac{a}{100{\rm au}} \right)^{5} \left( \frac{a_c}{150{\rm au}} \right)^{-3} \left( \frac{1-e_c^2}{1-0.4^2} \right)^{-3/2} \\  \left( \frac{M_c}{0.01M_\odot} \right) \left( \frac{M_\star}{0.08 M_\odot} \right)^{-1} \left( \frac{a_7}{0.06 {\rm au}} \right)^{-2} \\ \left ( (\sqrt{1-e^2}+\frac{1}{2})^2-\frac{1}{4} \right) \left( 1-e^2\right)^2,
\end{multline}

\noindent where $a_7$ is the semi-major axis of the outermost TRAPPIST-1 planet, and $e$ is the eccentricity of the comet, which for a belt of semi-major axis $a$ should be $1-0.01/a$ to be able to reach the innermost planet or $1-0.06/a$ to reach the outermost one. Using \citet{2017AJ....154..229B}, we find that
the $J_2$ due to the 7 TRAPPIST-1 planets would be $\sim 2 \times 10^{-5}$. Therefore, from Eq.~\ref{J2}, we estimate that for a 0.01$M_\odot$ at 150au, the belt of planetesimals should be at $\gtrsim 10$au to be able to reach the outermost planet or at  $\gtrsim 70$au to reach the innermost one.

We acknowledge that the change in inclination while undergoing Kozai oscillations is not taken into account in our previous general simulations shown in Sec.~\ref{nbody}. However, depending on the exact inclination of the companion compared to the belt, we can quantify using the equations
given in Sec.~\ref{incl} how it will affect the probability to be accreted onto the planets, which scales as $I_{\rm max}^{-1}$.

\subsection{Impacts from Oort-cloud comets perturbed by Galactic tides}\label{gal}

TRAPPIST-1 may have an Oort cloud, either because comets were captured from their neighbouring stars' belts at the cluster stage \citep{2010Sci...329..187L}, or because comets were scattered out by its planetary system \citep{1993ASPC...36..335T}. In our Solar System, \citet{1987AJ.....94.1330D}
propose that leftover comets between Uranus and Neptune would be thrown onto more extended orbits by the two planets until they reach a semi-major axis of $\sim 5000$au where Galactic tides change their angular momentum, therefore moving their periastron from reach of the planets.
While planets that are efficient at forming Oort clouds need to have the right ranges of mass and semi-major axis, which does not include the known TRAPPIST-1 planets \citep[e.g.][]{2017MNRAS.464.3385W}, other (as yet unseen) planets in the system could have scattered material into such an Oort cloud. 
Moreover, an Oort-cloud forming planet does not necessarily need to be at this exact location now as it could have migrated.

The same mechanism, i.e. Galactic tides, which increased angular momentum of leftover comets pumped up by Uranus and Neptune and thus detaching the comet orbits from the planets can also decrease angular momentum and bring back an outer Oort-cloud comet to
the planetary system.
For an Oort cloud, the Galactic tidal force (due to the Galactic disc potential) will slowly make the comets lose angular momentum resulting in a slow drift inwards of pericentre (because $e$ increases, $a$ is constant) 
at a rate $\dot{q}_G$ \citep[e.g.][]{1986Icar...65...13H,1992CeMDA..54...13M,2013MNRAS.430..403V}. The eccentricity reaches a maximum that is given by \citet{2005MNRAS.364.1222B}, which is greater for comets perpendicular to the orbital plane. It
is usually assumed that when a comet reaches a few au, it is lost from the Oort cloud due to planetary perturbations \citep[e.g.][]{1986Icar...65...13H,2006CeMDA..95..299F}.

The value of $\dot{q}_G$ can be estimated from the mean square change in angular momentum per orbit $\langle \Delta J^2 \rangle =1.2 \times 10^{-29} \rho_0^2  \, a^7 /M_\star$ \citep[in au$^4$/yr$^2$, Eq.~A4 of][]{2017MNRAS.464.3385W},
with $\rho_0$ the stellar mass density in units of 0.1M$_\odot$/pc$^3$ (local stellar mass density), a in au and $M_\star$ in M$_\odot$. Thus, since 
$\dot{q}_G= \sqrt{\langle \Delta J^2 \rangle \, q /(2 a^3)}/\pi$, we estimate that \citep{1987AJ.....94.1330D}
%\footnote{To check Eq.~\ref{qg}, we consider its application to the Solar System, for which $\dot{q}_G \sim 10^{-9} (a/10^4{\rm au})^{5.5}$ au/yr. This explains why new
%Oort-cloud comets have orbits with $a>10^4$au, otherwise they would not have had time to come within 5au over the age of the Solar System \citep{2014M&PS...49....8R}. 
%Similarly, we find that the pericentre of a Solar System comet with $a\sim30,000$au would be expected to vary by a few au over one orbit, which agrees with numerical simulations by \citet{2009Sci...325.1234K}.}

\begin{multline}\label{qg}
 \dot{q}_G \sim 7 \times 10^{-8} \left( \frac{\rho_0}{ 0.1{\rm M}_\odot{\rm/pc}^3} \right) \left( \frac{a}{10^4{\rm au}} \right)^{2} \\ \left( \frac{q}{0.06{\rm au}}\right)^{1/2} \left( \frac{M_\star}{0.08{\rm M}_\odot} \right)^{-1/2} \, \mathrm{au/yr}.
\end{multline}

\noindent We note that this $\dot{q}_G$ value is very small but it varies strongly with $a$.

%However,
%it could be that $\dot{q}_G$ was higher when TRAPPIST-1 was born. For instance, there are strong evidence that our Solar System was born in a denser stellar cluster with a mean density $1.5 \times 10^{4}$ M$_\odot$/pc$^3$ \citep{2010ARA&A..48...47A}. Also, for stars
%that would be discovered closer to the Galactic centre, the tidal forces would be stronger, scaling inversely with radius in the Galaxy and $\dot{q}_G$ would be much higher \citep{2013MNRAS.430..403V}. 

Therefore, for the case of the TRAPPIST-1 planets, it means that if the location of the Oort cloud is closer than a few $10^3$au, the time for moving the bulk of Oort-cloud bodies down to small pericentres close to the planet positions (i.e., $<$ 0.06au) would be greater than the age 
of the system. %Indeed, the mean eccentricity for comets in a typical classical Oort-cloud is $\sim$0.7 \citep[see Fig.~6 from numerical simulations in][]{1987AJ.....94.1330D}, with a semi-major axis
%typically at a few $10^4$au. A scaled down version of that would mean that the bulk of typical Oort-cloud comets will have pericentres $q_G$ at $>10^2$au and the time for the pericentres to reach the inner regions at {\bf the rate $\dot{q}_G$ would be longer than the age of TRAPPIST-1.
However, for an Oort-cloud at $10^5$au, it only takes $\sim$5Myr to reach the inner region but an origin at such a large distance becomes unlikely given that such comets would have been stripped by passing stars \citep{1993ASPC...36..335T}. Given the age and low-mass of TRAPPIST-1,
comets with a semi-major axis beyond 2000au should be strongly depleted by passing stars but some may still remain.
We also note that the presence of massive Jupiter-like planets in the outer regions of the TRAPPIST-1 system would have strong effects on the dynamics of the system \citep[e.g.][]{2009Sci...325.1234K} and our prescription would need to be revised if this type of planet is discovered. 
%We, therefore, require that $\dot{q}_G > q_G/7Gyr \sim 10^{-6}$au/yr for the comets to have time to move in before being accreted on to the planets.

\section{Results}\label{results}
\subsection{Fraction of comets that impact onto each planet}\label{fraction}

\subsubsection{For a general $\dot{q}$}\label{genfra}

\begin{figure*}
   \centering
  \includegraphics[width=19cm]{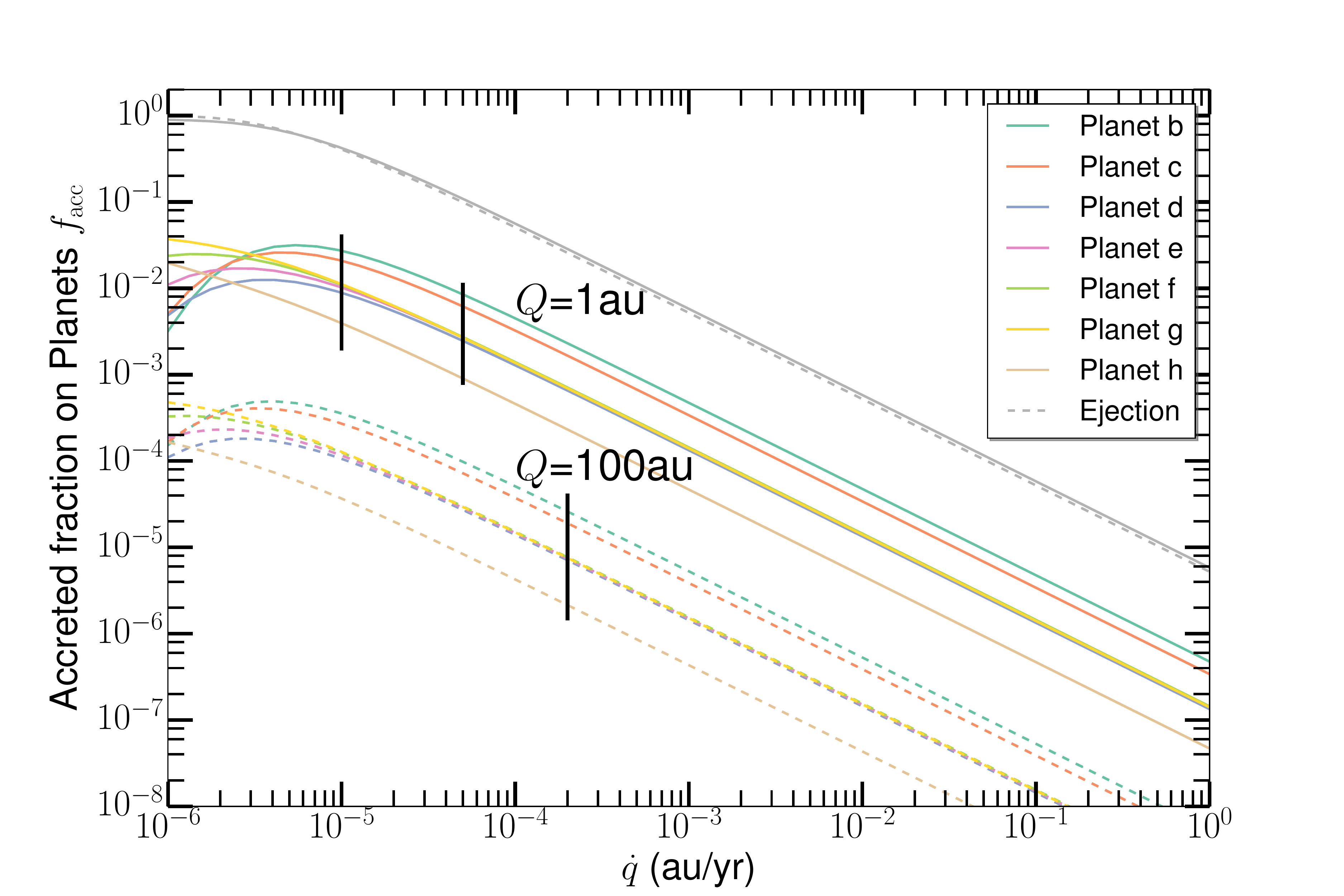}
   \caption{\label{fig6} Fraction of comets accreted $f_{\rm acc}$ onto each planet as a function of $\dot{q}$ for $Q=1$au (solid lines) and $Q=100$au (dashed). For large $\dot{q}$, $f_{\rm acc}$ scales approximately as $\propto Q^{-1}$ so this plot can be used for all $Q$ values. The Y-axis shows the fraction of comets that hits the different planets (i.e., accretion regime). 
The grey lines show the ejected fraction of comets.
The vertical black lines are typical $\dot{q}$ values for two scenarios: planet scattering (with $Q=1$au) and Kozai oscillations from a 100au belt (see the text in Sec.~\ref{fraction}).}
\end{figure*}

Here we use the results from sections \ref{nbod} and \ref{incl} to determine how much material will be accreted on the different planets depending on how fast comets move inwards, which is assumed to be set by a constant rate of change of pericentre $\dot{q}$. 
For a given $\dot{q}$ and apocentre $Q$, each (pericentre) $q$ cell of our parameter space is progressively crossed as the comet moves inwards to smaller $q$ values. Taking into account the timescales shown in Fig.~\ref{fig3} and 
scalings from Sec.~\ref{incl} (as we consider realistic inclined comets), we can use Figs. \ref{fig1} and \ref{fig2} (or their counterparts Figs. \ref{fig3d} and \ref{fig3e}) to work out the fraction of comets that are accreted onto the different planets or ejected along the way. 
Hereafter we only consider the inclined case using the results from Sec.~\ref{incl}.
%We can therefore compute the fraction of comets $f_{\rm acc}$ that will hit
%each planet using this procedure and can test the effect of changing $\dot{q}$. 

Fig.~\ref{fig6} shows the fraction accreted on the different planets $f_{\rm acc}$ for $Q=1$ and 100au. For small $\dot{q}$ (i.e $<10^{-5}$au/yr for the $Q=1$au case) 
most comets end up on planets g (yellow curve) and h (brown), while for large $\dot{q}$ each planet gets a fraction of comets accreted. We also show the fraction ejected as grey lines for $Q=1$ and 100au, which is close to 1 for very small $\dot{q}$ and decreases for larger values, meaning that
comets can go past the planets without being ejected nor accreted onto the planets for $\dot{q}>10^{-5}$au/yr. These comets may end up on the star or collisionally deplete before reaching it.

We see that the fraction of comets accreted\footnote{Here, we use accreted to say that a comet was not ejected but instead impacted onto a planet. It does not mean that the entirety of the accreted comet's material will stay in the atmosphere as we will show later.} $f_{\rm acc}$ onto the different planets varies significantly from 0.05 to $<10^{-8}$ for $10^{-6}<\dot{q}<1$au/yr. 
For small $\dot{q}$, the fraction accreted is dominated by planets h and g because $q$ decreases so slowly that these outermost planets catch all impacting comets before they reach further in.
On the other hand, for large $\dot{q}$,
the comets cannot efficiently accrete on the planets (as the loss timescale is long compared to $q/\dot{q}$, see Fig.~\ref{fig3}) and end up at small radii (where they either accrete onto the star or deplete collisionally). 
In between these two regimes, each planet accretes a fraction of the scattered comets. The fraction of comets accreted $f_{\rm acc}$ is also higher for smaller apocentres, as expected from Fig.~\ref{fig1}.
We find that for large $\dot{q}$, $f_{\rm acc} \propto \dot{q}^{-1} Q^{-1}$. The fraction accreted by the different planets vary by one order of magnitude in this regime (with b and h representing the extremes). This is due to both the difference in collisional cross sections and positions (since $t_{\rm acc} \propto a^2_{\rm pla}$, see Fig.~\ref{fig3c}).
In subsection~\ref{fracs}, we assess the outcome for the specific values of $\dot{q}$ that have been derived in Sec.~\ref{appl} for the different scenarios.

\subsubsection{For $\dot{q}$ derived from specific scenarios}\label{fracs}

%Now, we use some realistic values of $\dot{q}$ that we derived for the planet scattering scenario ($\dot{q}_P=10^{-5}$ and $5 \times 10^{-5}$ au/yr)
%and for the Kozai scenario ($\dot{q}_K=2 \times 10^{-4}$ au/yr) to assess the accreted fraction on the different planets for these scenarios. These $\dot{q}$ values are shown as vertical black lines in Fig.~\ref{fig6}.  

%For a belt close to 1au with a single eccentric planet scattering material in the inner regions, $Q$ is at 1au.
%For the planet chain scenario, $Q$ is the location of the innermost planet of the chain (also at 1au here) from which
%particles are being scattered to smaller $q$. In this case, the Kuiper-belt like disc location from which comets are scattered initially depends on the position of the outermost planet of the chain, which can be at 10s of au. 

%For the Kozai mechanism, . We note that in the case of an Oort-cloud, this distance can reach two to three orders of magnitude higher values but the ejection rate would then be close to 1 (see Fig.~\ref{fig2}) and
%$f_{\rm acc}$ that is $\propto \dot{q}^{-1} Q^{-1}$ would drop by more than one order of magnitude (because $\dot{q}$ will decrease $\propto Q^{-0.5}$ for a farther companion and belt) so that the mechanism would be much less efficient at accreting onto the planets (see Fig.~\ref{fig6}).

In Fig.~\ref{fig6}, we see that for a planet scattering scenario (both for a single planet or a chain) in which $Q=1$au and $\dot{q}=10^{-5}$au/yr (i.e. corresponding to a 1M$_\oplus$ planet in Sec.~\ref{sca}), we end up in the regime where a fraction of comets is accreted onto each planet. 
%Thus, there is a shielding effect in that comets are prevented from reaching the innermost planets as they are caught by planets g, h on their way in.
The fraction accreted is rather high in this case (between 0.01 and 0.03 for planets b to g) because the probability to be accreted for comets with $Q=1$au is rather high (as expected from Fig.~\ref{fig2}). 
This fraction accreted is valid for a single eccentric planet scattering material from a close-in belt at $\sim$ 1au similar to the debated belt potentially found around Proxima Cen \citep{2017arXiv171100578A}. However,
comets coming from tens of au belts would have to be scattered through a planet chain before making it to the innermost planet of the chain and some will be lost on the way. Here we assume that $f_{\rm in} \sim 5\%$ of the comets will make it to the innermost planet (see Sec.~\ref{sca}), 
which reduces the fraction accreted on the different planets from the initial reservoir (the Kuiper-belt like disc) to $\sim 5 \times 10^{-4}$. 

For the Kozai scenario, we consider that $Q$ represents the disc location from which 
the comets are perturbed by an outer companion and we take a typical distance of 100au as being representative. Considering the typical $\dot{q}_K$ value derived ($2 \times 10^{-4}$ au/yr), Fig.~\ref{fig6} shows that we are in the second regime where a fraction of comets ($\sim 10^{-5}$) is accreted onto each of the seven planets.
This is close to two orders of magnitude smaller than the chain of planets case. Here, we do not have to reduce the number of comets that arrives on to the seven planets (i.e $f_{\rm in}=1$) as Kozai oscillations operate directly from the outer belt. %The main difference with the previous case is the fraction of comets
%that ends up on the star. In the planet case, almost every comet that makes it to the 7 planets is accreted but in the Kozai case, a fraction of $\sim 3 \times 10^{-2}$ of comets make it to the star (taking into account ejection). This difference may be noticed in the star abundance ratio (see discussion).
%Even for the high 0.8au/yr $\dot{q}$ case, the fraction that end up on the star will still be of the same order, i.e most of the comets make it to the star anyway (even though the accreted fraction on planets goes down by a factor $10^3$).

For the Galactic tide scenario, from Fig.~\ref{fig6}, we evaluate that an Oort cloud at a few $10^4$au (that has a fast enough $\dot{q}_G$ to send comets to small pericentres within a fraction of the age of the system), i.e with large apocentres, the probability to be accreted is always $\lesssim 10^{-6}$ 
(for all the plotted $\dot{q}$, i.e more than order of magnitude smaller than for the Kozai mechanism). Therefore, the fraction of comets accreted is very low, which will not have any impacts on the atmospheres.
Therefore, we rule out Galactic tides as being an efficient mechanism\footnote{We note that we could fine tune the position of the Oort cloud to be in a narrow range in between $10^4$ and $10^5$au to maximise the fraction accreted while allowing enough time for the comets to reach the planetary system but this
would always result in an order of magnitude less efficient mechanism than Kozai. Moreover, it is not likely that an Oort cloud around a low mass star such as TRAPPIST-1 forms farther out than in our Solar System \citep{2017MNRAS.464.3385W}
as required here for maximising Galactic tide effects. And as shown in Sec.~\ref{gal}, such distant belts should be depleted owing to passing stars.} to modify the atmospheres of the TRAPPIST-1 planets. We also note that the same forces driving particles with high pericentres to low pericentres could also drive back these low pericentre orbits to high values, sometimes before they had time to 
reach the 7 inner planets \citep[e.g.][]{2007MNRAS.381..779E,2008CeMDA.102..111R}, and thus makes this scenario even more unlikely.
We are, thus, left with two plausible mechanisms to throw comets on the seven planets, namely, scattering by planets and Kozai oscillations due to an outer companion.  
%{\bf We also note that additional planets further out may prevent potential Oort cloud comets from reaching low pericentres \citep[e.g.][]{2007MNRAS.381..779E} and thus makes this scenario even more unlikely.}

\subsection{The relative effect of different impactor sizes on the atmospheres of the different planets}\label{rela}

\begin{figure*}
  % \centering
  \includegraphics[width=8.5cm]{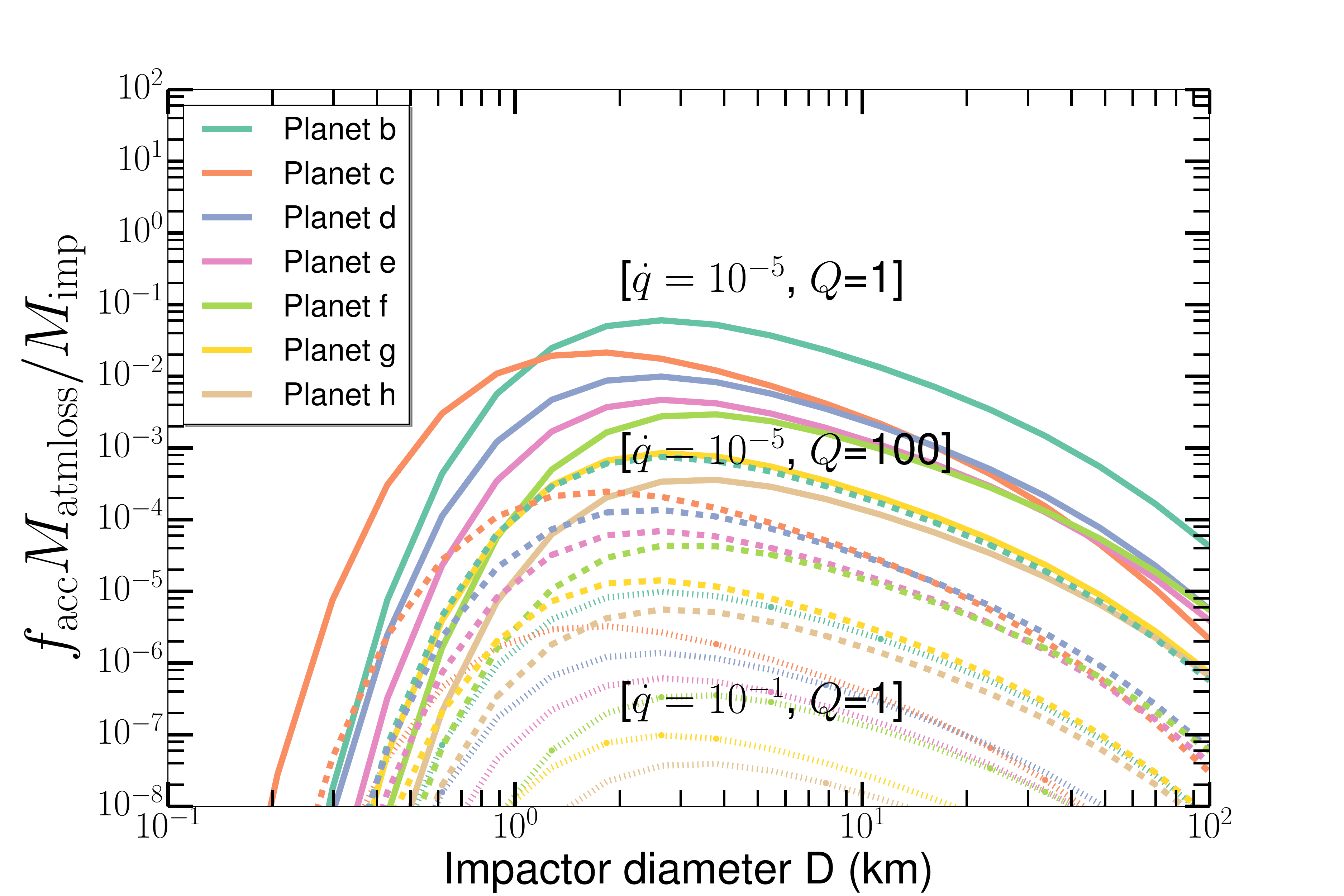}
   %\caption{\label{fig7} Atmospheric mass loss to impactor mass ratio $(M_{\rm atmloss}/M_{\rm imp}) f_{\rm acc}$ for a comet scattered from an outer belt for the different planets as a function of impactor sizes $D$ for $Q=1$ (solid lines) and 100au (dashed lines) with $\dot{q}=10^{-5}$au/yr and for $Q=1$au and 
%$\dot{q}=10^{-1}$au/yr (dotted line). It shows how much atmospheric mass is lost after a given comet is thrown in, taking into account that the fraction accreted is not equal to 1 (some incoming comets can be ejected or go further inwards without
%hitting the planets) as already seen in Fig.~\ref{fig6}.}
%\end{figure}
\hfill
%\begin{figure}
   %\centering
  \includegraphics[width=8.5cm]{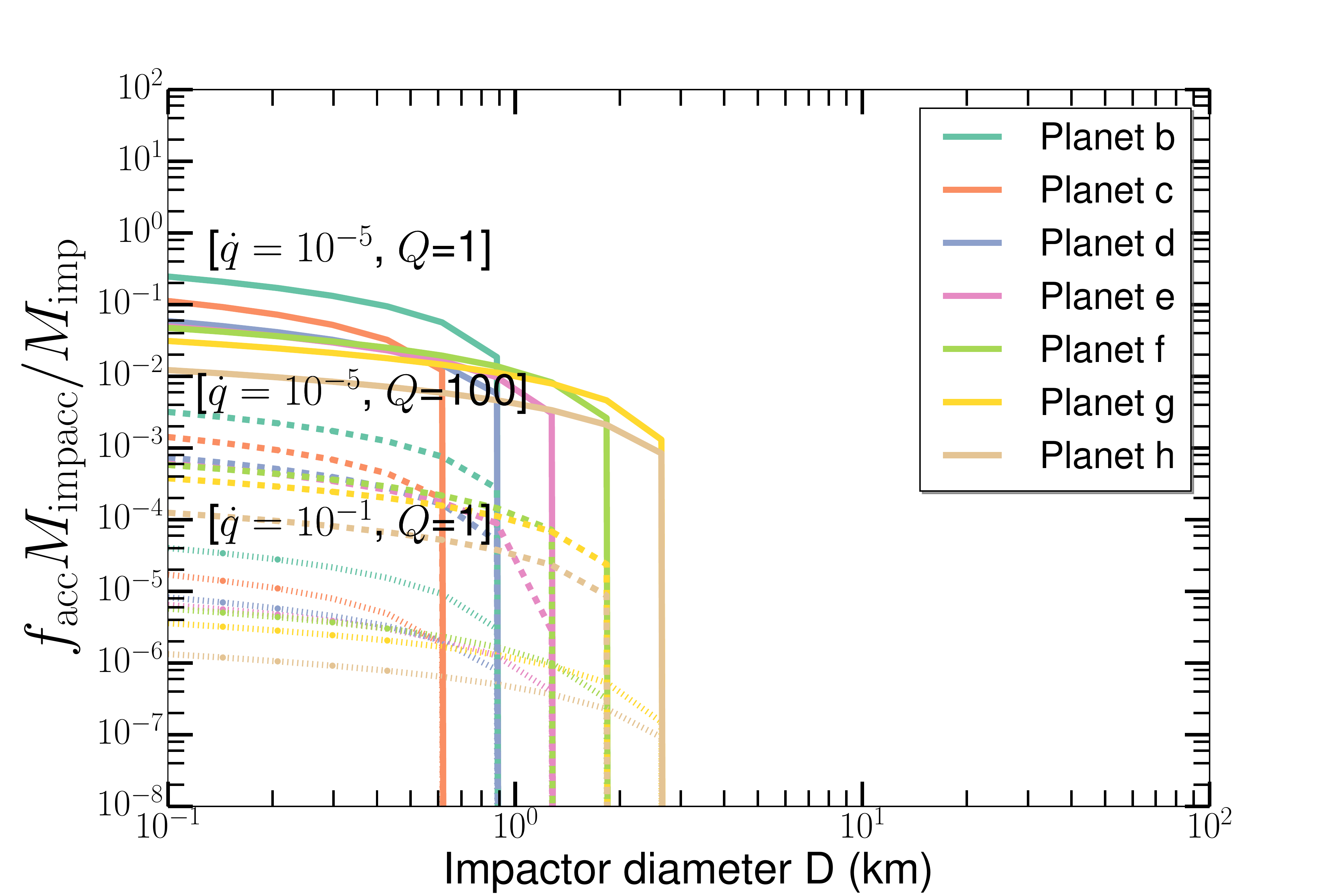}
   \caption{\label{fig7b} {\it Left}: Atmospheric mass loss to impactor mass ratio $(M_{\rm atmloss}/M_{\rm imp}) f_{\rm acc}$ for a comet scattered from an outer belt for the different planets as a function of impactor sizes $D$. 
It shows how much atmospheric mass is lost after a given comet is thrown in, taking into account that the fraction of comets that hit the different planets is not equal to 1 as already seen in Fig.~\ref{fig6}. {\it Right}: Accreted projectile mass to impactor mass ratio $(M_{\rm impacc}/M_{\rm imp}) f_{\rm acc}$ as a function of $D$. 
It shows how much projectile mass is accreted after a given comet is thrown in.
The different lines are for $Q=1$ (solid lines) and 100au (dashed lines) with $\dot{q}=10^{-5}$au/yr and for $Q=1$au and $\dot{q}=10^{-1}$au/yr (dotted line).}
\end{figure*}

In the previous subsection, we analysed the fraction of comets accreted $f_{\rm acc}$ by each planet. However, we want to quantify the effect of these impacts on the atmospheres of the seven planets. For example, we have seen in Fig.~\ref{fig4b} that impact velocities are much higher for planet d compared to further planets, so even if the 
fraction accreted is the same as that of the more distant planets in the planet scattering scenario, the effect on atmospheric mass loss may still be more important. Here, we quantify the atmospheric mass loss, and projectile mass accreted in the atmosphere (relative to impactor masses), i.e., the impactor
mass that does not escape the atmosphere after impact, for the different planets and
for different impactor sizes.

We use the numerical study of the effect of impacts on atmospheres by \citet{2009M&PS...44.1095S} to derive some conclusions for the TRAPPIST-1 planets. We first present the set of equations from \citet{2009M&PS...44.1095S} that we use to derive the atmospheric mass loss and projectile mass accreted after a given impact.
The outcome depends on the dimensionless variable $\eta$ \citep{2009M&PS...44.1095S}

\begin{equation}\label{eta}
\eta=\left( \frac{D}{H} \right)^3 \frac{ \rho_{\rm pr} \rho_{\rm t}}{\rho_{\rm atm0} (\rho_{\rm pr} + \rho_{\rm t})} \frac{ (V^2_{\rm imp}-V^2_{\rm esc})}{  V^2_{\rm esc}},
\end{equation}

\noindent where $D$ is the impactor diameter, $H$ the atmosphere scale height ($H=kT/(\mu m_H g)$ for an isothermal atmosphere with $g=GM_p/R^2_p$) and $\rho_{\rm t}$, $\rho_{\rm pr}$, $\rho_{\rm atm0}$ are the densities of the target (planet), projectile (exocomet), and
atmosphere at the surface, respectively. We assumed $\rho_{\rm t}=5000$ (terrestrial planet-like), $\rho_{\rm pr}=1200$ (comet-like), $\rho_{\rm atm0}=1.2$kg/m$^3$, $\mu=28.97$ (we assume an Earth-like atmosphere for now) and $T$ is taken to be the equilibrium temperature of the planets
\citep[assuming a null Bond albedo as calculated in][]{2016Natur.533..221G}. We note that recent observations suggest that some of the TRAPPIST-1 planet densities may be slighlty lower because of the potential presence of ice layers. \citet{2018arXiv180201377G} find that water mass fractions $<5\%$ 
can largely explain the observed mass-radius relationship of the less dense planets. Therefore, we can expect densities that are 10s of percent lower than assumed here, which would translate as a small uncertainty on $\eta$, which is however much lower than the 
uncertainties on the dynamics (see Sec.~\ref{appl}), and is thus not considered here in details.
$V_{\rm imp}$ is the impact velocity and $V_{\rm esc}=\sqrt{2 G M_{\rm pla}/R_{\rm pla}}$ is the escape velocity for the different planets. We have seen
in Sec.~\ref{vel} that $V_{\rm imp}$ is much greater than $V_{\rm esc}$, which simplifies the previous and following equations for most cases. 

The atmospheric mass loss per impactor mass is then defined as \citep{2009M&PS...44.1095S}

\begin{equation}\label{atmloss}
\frac{M_{\rm atmloss}}{M_{\rm imp}} = \frac{ (V^2_{\rm imp}-V^2_{\rm esc})}{  V^2_{\rm esc}} \rchi_a(\eta),
\end{equation}

\noindent where $M_{\rm imp}$ is the impactor mass and $\log_{10}{\rchi_a}=-6.375+5.239 \log_{10}{\eta} - 2.121  (\log_{10}{\eta})^2 + 0.397 (\log_{10}{\eta})^3 -0.037 (\log_{10}{\eta})^4 +0.0013 (\log_{10}{\eta})^5$. 

To get meaningful results, we compare the atmospheric mass loss $M_{\rm atmloss}$ to the impactor mass (of size $D$) that makes it to the inner regions and that is accreted on to the planet. Therefore, using the previous notations, we are interested in
$(M_{\rm atmloss}/M_{\rm imp}) f_{\rm in} f_{\rm acc}$ where we recall that $f_{\rm in}$ is the proportion of comets that are scattered from an outer belt and make it to the inner regions and $f_{\rm acc}$ is the accreted fraction onto a given planet. 
This ratio can therefore be understood as the atmospheric mass that is removed by one comet scattered from an outer belt, where only ``a fraction'' of the comet makes it to the inner regions and a fraction of that is accreted onto a specific planet.
In Fig.~\ref{fig7b} (left), we plot $(M_{\rm atmloss}/M_{\rm imp}) f_{\rm acc}$ keeping in mind that we should multiply that value by $f_{\rm in}$ (if it is different than 1, see Table~\ref{tab2}) to get the real value of accreted comets that make it to the inner regions.
  
Fig.~\ref{fig7b} (left) shows $(M_{\rm atmloss}/M_{\rm imp}) f_{\rm acc}$ as a function of impactor diameter $D$ for $Q=1$ and 100au (and $\dot{q} \sim 10^{-5}$au/yr) and for a higher $\dot{q}$ ($10^{-1}$au/yr) and $Q=1$au, using the impact velocity distributions shown in Fig.~\ref{fig4}.
The overall shape of the curves in Fig.~\ref{fig7b} (left) is explained in \citet{2009M&PS...44.1095S}. 
Impactors of size a few kms are the most harmful at removing atmospheric mass. Impactors smaller than 100m do not create large impact plumes and cannot accelerate large atmospheric mass to high latitudes. For impactors larger than a few 10s of kms, atmospheric
erosion continues to grow very slowly but the mass an impact removes cannot be greater than the total local atmospheric mass available. Therefore, for large impactors the atmospheric mass removed per increasing impactor mass becomes smaller.  
%We also note that for a size distribution of impactors $N(D) \propto D^{-\gamma}$, more massive bodies would still not be able to remove more mass than the km impactors because $(M_{\rm atmloss}/M_{\rm imp}) f_{\rm acc}$ scales as $D^{-2}$ for large diameters where
%the mass distribution scales as $D^{1/2}$, thus not making up for their limited effect on atmospheric mass loss.

The most harmful impactor size shifts along the x-axis for the different planets mainly because of the change in impactor velocity and the different properties of the planets through $H$ (the atmosphere scale height) with the relative scalings given in Eq.~\ref{eta}. 
The variations along the y-axis are mainly due to the different fraction accreted $f_{\rm acc}$ for each different planet
(see Fig.~\ref{fig6}) and the different impact velocities (see Fig.~\ref{fig4}) and scale as shown in Eq.~\ref{atmloss}. 
For example, we see that even though planets d, e, f, g accrete at the same level (see Fig.~\ref{fig6}), 
the atmospheric mass loss is greater for planet d because impact velocities are higher for the closer in planets (see Fig.~\ref{fig4}). 

The effect of increasing $Q$ from 1 to 100au (solid to dashed lines) is to shift all the lines down by a factor 100 because $f_{\rm acc}$ decreases by a factor 100. Changing $\dot{q}$ from $10^{-5}$ (solid) to $10^{-1}$au/yr produces a shift downwards of four orders of magnitude since
$f_{\rm acc}$ decreases by a factor $10^4$ between these two cases. Fig.~\ref{fig7b} (left) can therefore be used to work out the relative effectiveness of comets at removing mass from the atmosphere of each 
planet for any given $\dot{q}$ and $Q$, even though we show the results for only two different $Q$ (i.e., it is a general plot, not tied to a specific scenario from Sec.~\ref{appl}, and only $f_{\rm acc} \propto \dot{q}^{-1} Q^{-1}$ changes for different values of $\dot{q}$ and $Q$, making it easy
to compute results for different $\dot{q}$ and $Q$). 

The simulations of \citet{2009M&PS...44.1095S} also showed that the projectile mass accreted per impactor is given by 

\begin{equation}\label{imploss}
\frac{M_{\rm impacc}}{M_{\rm imp}} = 1 - \rchi_{\rm pr}(\eta),
\end{equation}

\noindent where  $\rchi_{\rm pr}={\rm min}\{1,0.07(\rho_{\rm t} / \rho_{\rm pr})(V_{\rm imp}/V_{\rm esc})( \log_{10}{\eta} - 1)\}$.
Similarly to atmospheric mass loss, Fig.~\ref{fig7b} (right) shows the accreted projectile mass per comet $(M_{\rm impacc}/M_{\rm imp}) f_{\rm acc}$ as a function of impactor diameter $D$ for $Q=1$ and 100au (and $\dot{q} \sim 10^{-5}$au/yr) and for a higher $\dot{q}$ ($10^{-1}$au/yr) and $Q=1$au.

The shape of the curves in Fig.~\ref{fig7b} (right) is already
known from \citet{2009M&PS...44.1095S}. The ejecta from impacting bodies that are $\lesssim 1$km does not have enough energy to escape after impact and is stranded in the atmosphere (though some material may condense on the planet surface at a later point, see Sec.~\ref{compo}). For more massive bodies, the ejecta after impact
is increasingly more energetic until the airless limit is reached (i.e., when atmospheric drag can be neglected before the after-impact plume expansion) where all the projectile material escapes. This cut-off happens for bodies larger than a few km.  

In Fig.~\ref{fig7b} (right), the variations along the x-axis (e.g. of the cut-off position) are due to different impact velocities (for instance a larger planetesimal can deliver material onto planet h because impacts happen at lower velocities) and it can also vary with the planets' properties through $H$
and the atmospheric density (assumed constant for now) with the scalings given by Eq.~\ref{eta}. Planets g and h can therefore get volatiles delivered from larger comets than further in planets.
The variations along the y-axis are mainly due to the fraction of comets accreted onto the planets and the different impact velocities and scale as depicted by Eq.~\ref{imploss}. 

The effect of increasing $Q$ from 1 to 100au (solid to dashed lines) or increasing $\dot{q}$ from $10^{-5}$ to $10^{-1}$au/yr is the same as explained when describing Fig.~\ref{fig7b} (left), i.e., due to the change in $f_{\rm acc}$.
This plot is therefore also general and can be used to compute the outcome of an impact for any values of $\dot{q}$ and $Q$, and is not tied to any of the specific scenarios explained in Sec.~\ref{appl}.

The volatile mass that ends up in the atmospheres is a fraction $f_{\rm vol}$ of the mass delivered. 
We assume that volatiles are delivered to the atmospheres in proportion to their fraction of the mass of the parent body. 
For a comet-like body, we assume a rock-to-ice mass ratio of 4 based on recent measurements in the 67P comet \citep{2015Sci...347a3905R}, i.e 20\% of ice by mass. For an asteroid-like body, the water mass fraction is lower and is found to vary between $10^{-3}$ and 0.1
\citep{2000orem.book..413A}. We will assume an intermediate value of 1\% for asteroid-like bodies\footnote{We assume that the bulk of the volatile mass is in water so that this value is representative of the total volatile mass, though a lower limit.}, which is typical of ordinary chondrites in our Solar System \citep[but we note that carbonaceous chondrites can reach 10\% of water by mass,][]{2004Icar..168....1R}. This gives us two 
extreme volatile delivery scenarios to consider with our model.

%{\bf We can also compute the planet mass loss after each impact... I don't know if I put it in???}
The CO or H$_2$O content of exocomets can be probed for the most massive belts and are found to be similar to Solar System comets \citep[e.g.][]{2016MNRAS.461..845K,2016MNRAS.460.2933M,2017ApJ...842....9M}. The potential to detect gas in debris disc systems will improve with new missions \citep[see][]{2017arXiv170308560K} and the assumptions used in this study could 
be refined with future estimates of the volatile content of exocomets in the TRAPPIST-1 system to get a better handle on the final atmospheric composition.

\begin{table*}
  \centering
  \caption{Table describing the parameters used for the different scenarios we tested. We list the rate of change of pericentre $\dot{q}$, the apocentre $Q$, the mean fraction accreted ${f_{\rm acc}}$ on each planet, the fraction of comets that makes it to the inner regions $f_{\rm in}$, 
 the mass fraction of volatiles on the exocomets/exoasteroids $f_{\rm vol}$, the minimum scattered mass $M_{\rm scadestroy}$ to destroy all 7 primordial atmospheres ($M_{\rm sca}=M_{\rm inc}/f_{\rm in}$), the mass of delivered volatiles $M_{\rm volmin}$ and water $M_{watmin}$ (assuming Solar-System comet-like compositions) for a belt scattering 
at the low scattering rate of the current Kuiper belt (i.e $M_{\rm inc} \sim 10^{-2} {\rm M}_\oplus \, f_{\rm in}$) for each of the planets f, g, h, and $M_{\rm volT}$, $M_{\rm watT}$
for a belt of 20M$_\oplus$ (close to the expected mass for a potential leftover belt around TRAPPIST-1, see section \ref{mmsn}) scattering 5\% (i.e $M_{\rm inc} \sim 1{\rm M}_\oplus \, f_{\rm in}$)
of its mass over 7Gyr. For the case of exoasteroids, $f_{\rm vol}=0.01$, and $M_{\rm volmin}$ as well as $M_{\rm volT}$ should be divided by 20, and $M_{watmin}$, $M_{\rm watT}$ by 10. M$_{\rm eo}$ means 1 Earth ocean (i.e $2.5 \times 10^{-4}$M$_\oplus$).}

 % \begin{threeparttable}
  \label{tab2}
  \begin{tabular}{|l|c|c|c|c|c|c|c|c|}
   \toprule
    Scenarios & $\dot{q}$ & $Q$ & ${f_{\rm acc}}$ & $f_{\rm in}$ & $f_{\rm vol}$ & $M_{\rm scadestroy}$ & $M_{\rm volmin}$,$M_{\rm watmin}$ & $M_{\rm volT}$,$M_{\rm watT}$ \\ 
		    & au/yr & au &  & & & M$_\oplus$ & M$_\oplus$,M$_{\rm eo}$ & M$_\oplus$,M$_{\rm eo}$ \\
    %\toprule
    %Star mass ($M_\odot$) &\multicolumn{8}{c}{0.0802} \\
    %Star radius ($R_\odot$) &\multicolumn{8}{c}{0.117} \\
    \midrule
    \midrule
    Single planet (1M$_\oplus$) & $10^{-5}$ & 1  & $7 \times 10^{-2}$ & 1 & 0.2/0.01 & $5 \times 10^{-4}$ & $4 \times 10^{-6}$,$8 \times 10^{-3}$ & $4 \times 10^{-4}$,$0.8$\\
    Single planet (10M$_\oplus$) & $5 \times 10^{-5}$ & 1 & $4 \times 10^{-2}$ & 1 & 0.2/0.01  & $3 \times 10^{-3}$ & $8 \times 10^{-7}$,$2 \times 10^{-3}$ & $8 \times 10^{-5}$,$0.2$ \\
    Planet chain (1M$_\oplus$) & $10^{-5}$ & 1 & $7 \times 10^{-2}$ & 0.05 & 0.2/0.01  & $10^{-2}$ & $2 \times 10^{-7}$,$4 \times 10^{-4}$ & $2 \times 10^{-5}$,$4 \times 10^{-2}$ \\
    Planet chain (10M$_\oplus$) & $5 \times 10^{-5}$  & 1 & $4 \times 10^{-2}$ & 0.05 & 0.2/0.01  & $5 \times 10^{-2}$ & $4 \times 10^{-8}$,$4 \times 10^{-5}$ & $4 \times 10^{-6}$,$4 \times 10^{-3}$ \\
    Kozai mechanism & $2 \times 10^{-4}$ & 100 & $1.5 \times 10^{-4}$ & 1 & 0.2 & $1$ & $2 \times 10^{-9}$,$4 \times 10^{-6}$ & $2 \times 10^{-7}$,$4 \times 10^{-4}$ \\

   \bottomrule
  \end{tabular}
%\begin{tablenotes}
     % \small
     % \item $^1$ mean $f_{\rm acc}$ value averaged over the seven planets
    %\end{tablenotes}
  %\end{threeparttable}
\end{table*}

\subsection{The integrated effect of these impacts over the age of the system}\label{integr}

\subsubsection{Total incoming mass over the system's age}\label{totinc}
We now work out the effect of impacts on the TRAPPIST-1 planets over the age of the system and more specifically, how much atmospheric mass is lost and how much projectile/volatile mass is accreted for a given total incoming mass of comets. 
To do so, we assume a typical $N(D) \propto D^\gamma$ size distribution with $\gamma=-3.5$ for the comets that are expelled from the belt \citep[e.g.][]{1969JGR....74.2531D,2007A&A...472..169T} up to a maximum size of 10km\footnote{We note that the size 
distribution of the Kuiper belt for the largest bodies is complicated and best-fitted by two shallow power laws and a knee or a divot between the two \citep{2018AJ....155..197L}, which would imply that most of the cross section would be in the biggest bodies. This is not 
representative of what is observed in general for the debris disc population, for which a -3.5 slope all the way through the largest bodies is able to explain the observations.}. Indeed,
integrating over the assumed size distribution for the total atmospheric mass loss (or accreted material) shows that $>10$km impactors are unimportant \citep[as already concluded by][]{2015Icar..247...81S} because $M_{\rm atmloss} \propto D^{-2}$ for large bodies as seen from Fig.~\ref{fig7b} (left), which decreases faster than the gain in mass 
of these larger bodies ($\propto D^{0.5}$). 
Very massive giant impacts \citep[e.g.][]{2015A&A...573A..39K} of bodies with radius $>1000$km (i.e Pluto-sized or greater) can have a devastating effect on the atmosphere of a planet \citep{2015Icar..247...81S}, which is not modelled in \citet{2009M&PS...44.1095S}, but these impacts are rare and thus neglected in this study.

%Slightly smaller objects between 100 and 1000km are also rare and not capable of destroying the entire atmosphere. Rather, they remove all the atmosphere above the tangent plane at impact, thus only $H/(2R_{\rm pla})$ can be removed for a given impact. Because of that, \citet{2015Icar..247...81S}
%show that the most effective impactors to remove mass from the atmospheres are bodies of a few kms. Therefore, we decide not to model larger impactors in our study and focus on the 0.1 to 100km range.

\begin{figure}
   \centering
  \includegraphics[width=9cm]{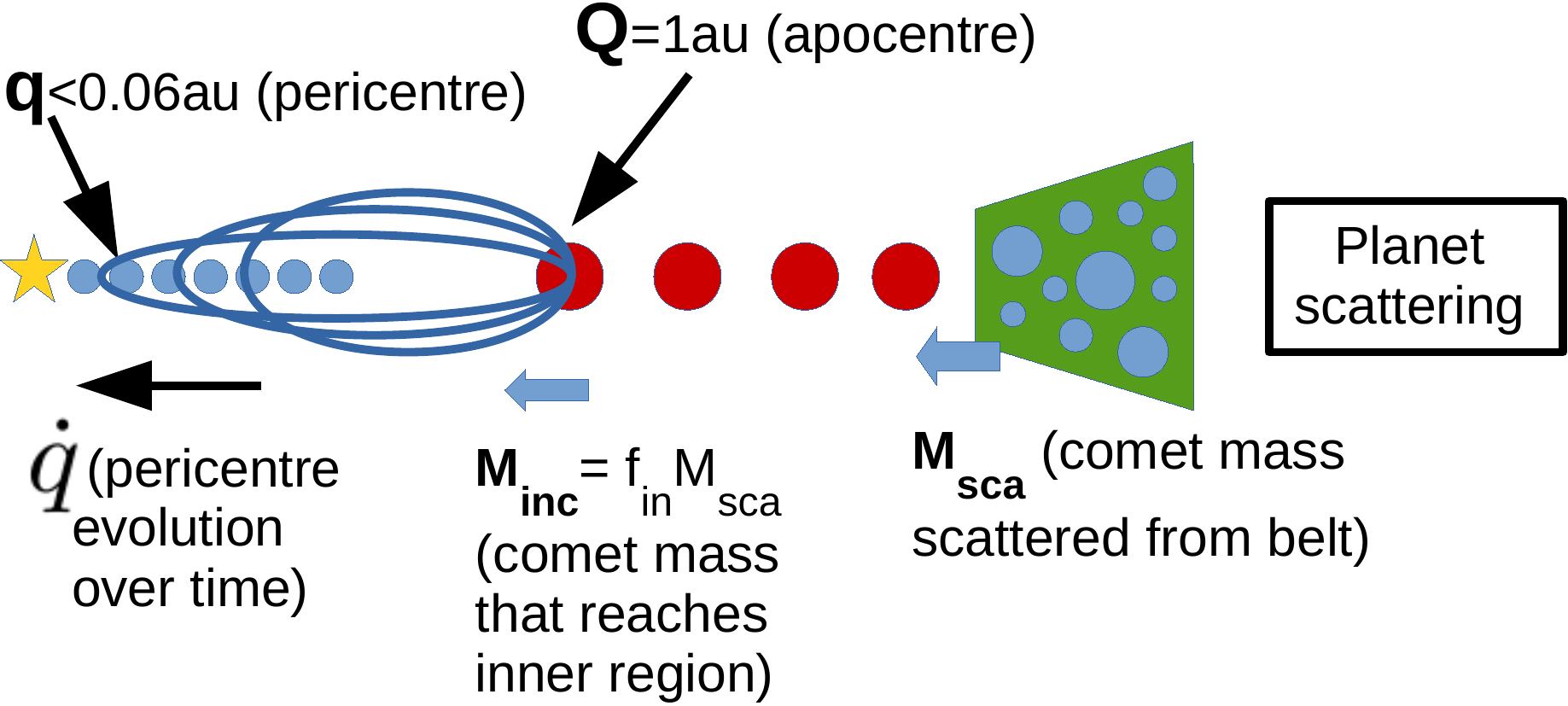}
   \caption{\label{diag2} Schematic showing the main variables used to parametrise comet scattering (here, for the planet chain case but this is general) in the inner regions of the TRAPPIST-1 system.}
\end{figure}

We consider an incoming mass of comets $M_{\rm inc}$ that reaches and can potentially hit the TRAPPIST-1 planets 
after a mass $M_{\rm sca}$ of comets has been scattered from this outer belt over the system's age.
Taking into account the efficiency to reach inner regions, $M_{\rm inc}=M_{\rm sca} f_{\rm in}$ (see Fig.~\ref{diag2}). 
%We will also vary the assumed size distribution to check the effect it has on the delivery of material to the planets.

%The value of $f_{\rm in}$ is scenario dependent. It is equal to 1 if all the bodies scattered from the belt make it to the TRAPPIST-1 planets' region (Kozai scenario and single planet scattering bodies from a close-by belt) and is smaller otherwise (e.g. scattering by a chain of planets where
%comets are lost on the way in before reaching the seven inner planets).

The integrated amount of mass scattered from a belt $M_{\rm sca}$ over the system's age can be evaluated.
In Sec.~\ref{mmsn}, we predicted that a potential planetesimal belt of 20M$_\oplus$ could potentially have survived around TRAPPIST-1 at tens of au. By the action of a nearby planet, many planetesimals may have been scattered inwards over the lifetime of the system. Assuming that 5\% of the belt mass is scattered over 7 Gyrs \citep[using results by][]{Marino17}, we get that $M_{\rm sca} \sim 1$M$_\oplus$ leading to $M_{\rm inc} \sim f_{\rm in}$ M$_\oplus$.

In our Solar System, $\sim 0.27$ comet/yr leave the Kuiper belt towards the inner regions \citep{1997Icar..127...13L}. The typical mass of comets in \citet{1997Icar..127...13L}'s study is $\sim 4 \times 10^{13}$kg so that the rate of scattered incoming comets is $\dot{M}_{\rm sca} \sim 2 \times 10^{-3}$M$_\oplus$/Gyr.
Therefore, a similar Kuiper belt around TRAPPIST-1 would give $M_{\rm sca} \sim 10^{-2}$M$_\oplus$ over 7Gyr leading to $M_{\rm inc} \sim 10^{-2} f_{\rm in}$ M$_\oplus$. 

However, the Kuiper belt is thought to have been a lot more massive in its youth \citep[e.g.][and see Sec.~\ref{mmsn}]{2011AJ....142..152L} and in general, debris discs that are observed can have fractional luminosities of up to $10^4$ greater than this 
low-mass belt \citep{2008ARA&A..46..339W}, which is an indicator of them being more massive.  We note that the Kuiper belt is so light \citep[$\sim 0.1$M$_\oplus$,][]{2009AJ....137...72F,2010A&A...520A..32V} that current instruments could not even detect it around another star \citep{2012A&A...540A..30V,2017arXiv170308560K}. 
From an MMSN-like calculation, the initial Kuiper belt mass may have been of several 10s of Earth masses \citep{1981PThPS..70...35H,2003EM&P...92....1M}, meaning that $M_{\rm sca}$ could have been of the order of a few $10$M$_\oplus$ owing to the depletion of the belt to reach its current mass. 
In other words, we expect

\begin{equation}\label{minc}
10^{-2} f_{\rm in} \, {\rm M}_\oplus \lesssim M_{\rm inc} \lesssim 30 f_{\rm in} \, {\rm M}_\oplus.
\end{equation}

\subsubsection{Atmospheric mass loss}\label{atml}

\begin{figure*}
   \centering
  \includegraphics[width=19cm]{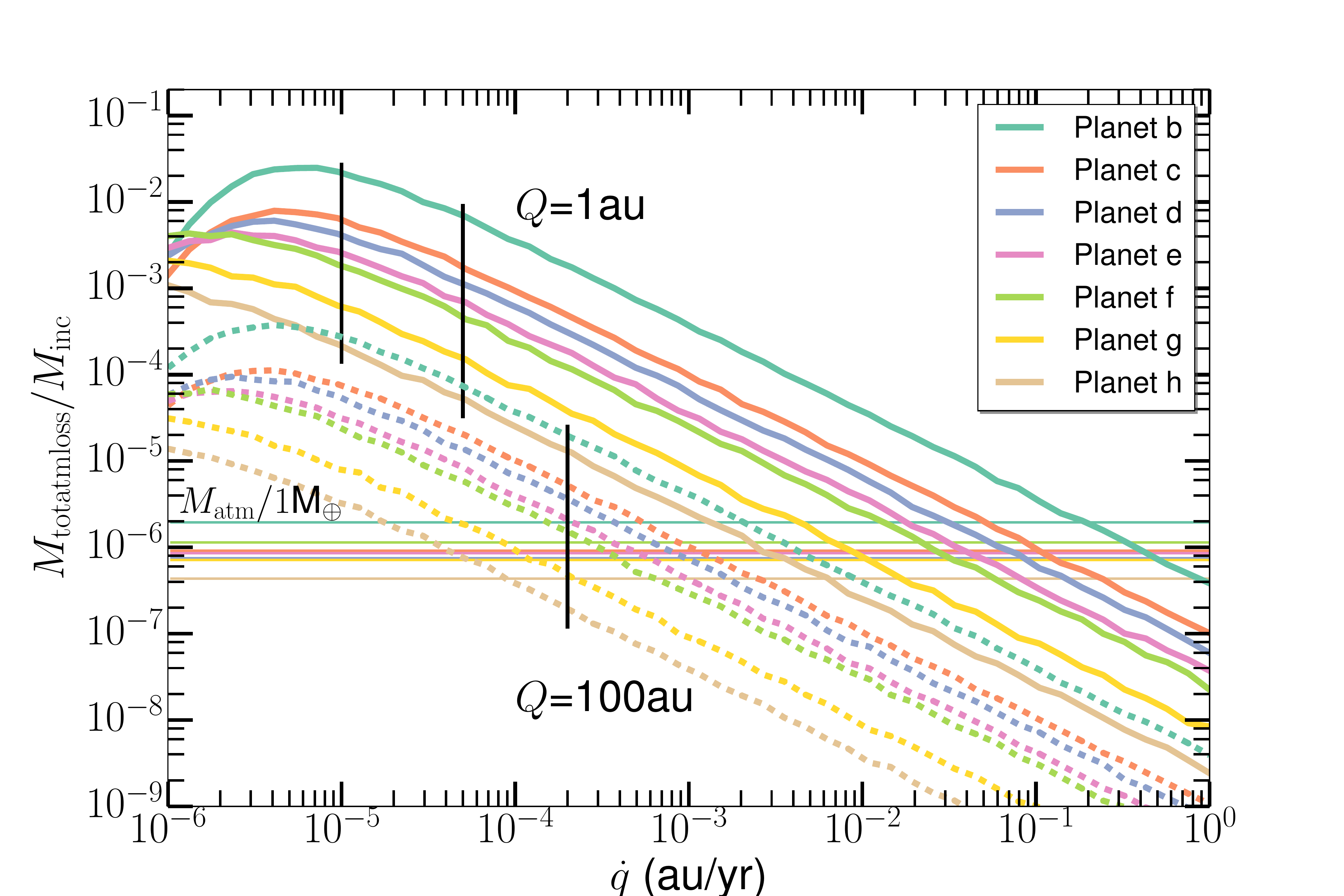}
   \caption{\label{fig8} Total atmospheric mass loss for a given incoming mass of comets that reach the inner region, i.e $M_{\rm totatmloss}/M_{\rm inc}$ (assuming a -3.5 size distribution) as a function of $\dot{q}$ for $Q=1$ (solid lines) and 100au (dashed).
The vertical black lines are typical $\dot{q}$ values for each scenario (see the text in Sec.~\ref{fraction}). The primordial atmospheric masses of each planet are plotted as horizontal lines for an incoming mass of 1$M_\oplus$, assuming an Earth-like atmospheric density and composition.}
\end{figure*}

The total atmospheric mass loss for a given planet over the system's age is

\begin{equation}\label{min}
M_{\rm totatmloss}=\int_{D_{\rm min}}^{D_{\rm max}} N(D) M_{\rm atmloss}(D) f_{\rm acc} \, {\rm d}D,
\end{equation}

\noindent where $N(D)$ is the number of bodies in each impactor diameter bin $D$ that make it to the inner regions.

% of size $D_i$ that have been scattered to the inner regions within 7Gyr, which 
%is given by $(M_{\rm inc}/M_{\rm imp})(D_{i+1}^{\gamma+4}-D_i^{\gamma+4})/(D_{\rm max}^{\gamma+4}-D_{\rm min}^{\gamma+4})$. 
%We can see that $M_{\rm imp}$ cancels out and that $M_{\rm totatmloss}/M_{\rm inc}$ is independent of $M_{\rm inc}$, and we can use Eq.~\ref{minc} to get $M_{\rm inc}$ from a belt mass $M_{\rm belt}$.

Fig.~\ref{fig8} shows $M_{\rm totatmloss}/M_{\rm inc}$, i.e the total atmospheric mass loss compared to the incoming mass $M_{\rm inc}$ of comets injected into the inner regions over the lifetime of the star.
Once again, this figure is general (and can be used for any $\dot{q}$ and $Q$) and is not tied to a specific scenario (only the black vertical lines are scenario dependent). We show the atmospheric mass removed for specific values of apocentres $Q=1$ and 100au
but values for other $Q$ can also be estimated (as $M_{\rm totatmloss} \propto f_{\rm acc} \propto Q^{-1}$). 
Atmospheric mass loss remains lower for planets g and h because impacts happen at lower velocities (see also Fig.~\ref{fig7b} left).
The mean total atmospheric mass loss for the seven planets can be approximated as

\begin{equation}\label{gen}
\frac{M_{\rm totatmloss}}{M_{\rm inc}} \sim 2 \times 10^{-3} \left( \frac{\dot{q}}{10^{-5} {\rm au/yr}} \right)^{-1} \left( \frac{Q}{1{\rm au}} \right)^{-1},
\end{equation}

\noindent where we note that this ratio is accurate for planets d, e and f but can be a factor 10 more or less for a specific planet (e.g. 10 times higher for planet b and 10 times lower for planet h), and Fig.~\ref{fig8} should be used to get more accurate values.

To assess whether the impact process is capable of 
destroying an entire primordial atmosphere, we first estimate the primordial atmospheric masses of the different planets. These primordial atmopsheric masses are not known and so for reference we asssume an Earth-like composition and density. Computing the scale height for each planet (as in Eq.~\ref{eta}) and assuming an isothermal atmosphere of temperature $T$ 
(the equilibrium temperature of the planets), we integrate over the height of the
planet atmospheres to get their masses $M_{\rm atm}=4 \pi \rho_{\rm atm 0} H (R_{\rm pla}^2+2 H R_{\rm pla} + 2 H^2)$. This gives primordial atmospheric masses of $2, 0.9, 0.7, 0.9, 1.1, 0.7, 0.4 \times 10^{-6}$ $M_\oplus$ for planets b to h. This is shown on Fig.~\ref{fig8} as horizontal lines, where this mass has been divided by 1M$_\oplus$ to show the effect of an incoming mass of 1M$_\oplus$.

Therefore, a primordial Earth-like density atmosphere on the TRAPPIST-1 planets could be destroyed if

 \begin{equation}\label{gen2}
M_{\rm inc} > 5 \times 10^{-4} \left( \frac{\dot{q}}{10^{-5} {\rm au/yr}} \right) \left( \frac{Q}{1{\rm au}} \right) {\rm M}_\oplus.
\end{equation}

\noindent For the specific physical scenarios from Sec.~\ref{appl} (see vertical black lines on Fig.~\ref{fig8} for the planet scattering and Kozai scenarios), Table~\ref{tab2} shows the minimum scattered mass needed $M_{\rm scadestroy}$ from an outer belt (the minimum incoming mass would be $f_{\rm in} M_{\rm scadestroy}$) to destroy the primordial atmospheres
of the seven planets.

For example, for the planet scattering scenario with a single Earth-mass planet at 1au (i.e $\dot{q} \sim 10^{-5}$au/yr, $Q=1$au and $f_{\rm in}=1$, see Table~\ref{tab2}), using Eq.~\ref{gen2} we find that $M_{\rm inc} \gtrsim 5 \times 10^{-4}$M$_\oplus$ can destroy the primordial atmospheres of the seven planets. This corresponds
to a belt that is being depleted for 7Gyr at a rate ten times lower than that at which the current Kuiper belt is being depleted. If the comets had to be passed in through a planetary system before reaching the planet at 1au, the inefficiency in the inward scattering process results in an additional factor $f_{\rm in}=0.05$. This means that even with this factor, the current Kuiper belt 
scattering rate is enough to destroy the atmospheres of the seven TRAPPIST-1 planets.

For the Kozai scenario $\dot{q}$ values are higher ($\dot{q} \sim 2 \times 10^{-4}$au/yr) and $Q$ is at larger distances (100au), meaning that interactions with planets are much more likely to result in ejections rather than accretions (see Fig.~\ref{fig2}). We find that
$M_{\rm inc} > 1$M$_\oplus$ is needed to destroy the primordial atmospheres, i.e two orders of magnitude larger than in the planet chain case.
For a 1M$_\oplus$ incoming mass (i.e. 100 times the current Kuiper-belt like incoming mass rate), Fig.~\ref{fig8} shows that the atmospheric mass loss is $\sim 2 \times 10^{-5} M_\oplus$ for planet b and a factor 10 less for planets c, d, e, and f, and about another factor 5-10 less for planets g, h (all of which are higher than the primordial Earth-like atmospheric masses assumed here
except for planets g and h that are a factor 2 too small).

%Therefore, we conclude that a Kuiper-belt like incoming mass is enough to destroy the primordial atmospheres of the seven planets in the case of the planet scattering scenario (both for a single planet and a planet chain). As for the Kozai scenario, the primordial atmospheres of 
%all planets would not entirely be destroyed (at least for a low Kuiper-belt like belt mass) but we note that if $\dot{q}$ were to be slightly smaller, the atmospheres of planets b, c and d may have been fully destroyed.

%However, $M_{\rm belt}$ could be much higher than for the Kuiper belt as explained in Sec.~\ref{totinc}. For a $10^{3}$ more luminous belt than the Kuiper belt (commonly observed), $M_{\rm inc}$ reaches $\sim 10$M$_\oplus f_{\rm in}$.
%In this case, both the planet scattering and Kozai scenarios lead
%to total destruction of the primordial atmospheres of the seven planets, even in the case where the initial atmospheres are ten times denser than Earth's atmosphere (and not even accounting for UV photodestruction processes) and even if the impact process only happened
%for a fraction of the system's age (e.g for 1Gyr instead of 7Gyr). 

Given that the exo-Kuiper belts detected around F, G, K stars are much more massive than the Kuiper belt, and that the possible belt mass we derive for the TRAPPIST-1 belt in Sec.~\ref{mmsn} is $\sim 20$M$_\oplus$), the scattering may be even higher than assumed here (i.e., up to a factor of a few $10^3$
the Kuiper-belt incoming mass), and we conclude that if a scattering belt is around TRAPPIST-1, {\it the primordial atmospheres would not survive impacts over the system's lifetime} for both a planet scattering and a Kozai scenario.

%Matm2 1.95506331957e-06 0 1 0 16099.3417839 6918906.0 6371000.0
%Matm2 9.17869438607e-07 1 1 0 8012.10615988 6727776.0 6371000.0
%Matm2 7.45234029979e-07 2 1 0 12140.6538253 4918412.0 6371000.0
%Matm2 8.584469571e-07 3 1 0 9905.69623107 5848578.0 6371000.0
%Matm2 1.14499031275e-06 4 1 0 10199.2121921 6657695.0 6371000.0
%Matm2 7.11707145653e-07 5 1 0 5459.10090916 7180117.0 6371000.0
%Matm2 4.33826554751e-07 6 1 0 7082.04806473 4918412.0 6371000.0

\subsubsection{Water mass loss}\label{atmwat}

In Table~\ref{tabwat}, we also quantify the maximum water mass loss for the single and planet chain scenarios. The water mass loss $M_{\rm watLossT}$ is given for each planet 
for a belt of 20M$_\oplus$, which is close to the expected mass for a potential leftover belt around TRAPPIST-1 (see section \ref{mmsn}) scattering 5\% (i.e., $M_{\rm inc} \sim 1{\rm M}_\oplus \, f_{\rm in}$)
of its mass over 7Gyr. For the planet chain scenario, the planets can lose up to 4, 1.2, 0.8, 0.6, 0.4, 0.12, 0.06 $M_{\rm eo}$ (Earth ocean mass), for b, c, d, e, f, g, h, respectively, and 20 times more for the single planet case. These values can be compared to the water mass loss from hydrodynamic escape
due to XUV irradiation during the runaway greenhouse phase, for which they found upper limits of \citep{2017A&A...599L...3B}, 80, 40, 2.9, 1.5, 0.9, 0.4, 0.1 $M_{\rm eo}$, for b, c, d, e, f, g, h, respectively. These values are, however, to be taken as strict upper limits because it is uncertain that hydrogen can reach the very top 
layers at the base of the hydrodynamic wind, which is needed for it to escape \citep{2017MNRAS.464.3728B}. Also, this hydrodynamic escape works well to eject hydrogen but other atoms are difficult to drag along \citep{2017MNRAS.464.3728B}. For the impact case,
not only hydrogen would escape but the whole fluid in the ejected plume. Bearing these caveats in mind, we can now compare the water mass loss from hydrodynamic escape to the impact scenario. For the planet chain case, the water mass loss due to impacts seems to be less efficient than
hydrodynamic escape for planets b and c and both scenarios are within a factor of a few for the other planets. For the most optimistic case of the single planet case, impacts could produce the same water loss as hydrodynamic escape for planets b and c and be an order of magnitude higher for
planets d to h.

%Matm2[i]=4.e0*np.pi*rhoatm*H*(Rearth**2+2.e0*H*Rearth+2.e0*H**2)/Mearth#in kg assuming an isothermal atmosphere
\begin{table*}
  \centering
  \caption{Amount of water lost due to impacts for the planet scenario (single and chain). The water mass loss $M_{\rm watLossT}$ is given for each planet 
for a belt of 20M$_\oplus$ (close to the expected mass for a potential leftover belt around TRAPPIST-1, see section \ref{mmsn}) scattering 5\% (i.e $M_{\rm inc} \sim 1{\rm M}_\oplus \, f_{\rm in}$)
of its mass over 7Gyr. M$_{\rm eo}$ means 1 Earth ocean (i.e $2.5 \times 10^{-4}$M$_\oplus$).}
  \label{tabwat}
  \begin{tabular}{|l|c|c|c|c|c|c|c|c|c|c|}
   \toprule
    Scenarios & $\dot{q}$ & $Q$ & \multicolumn{7}{c}{$M_{\rm watLossT}$} \\ 
		    & (au/yr) & (au) & \multicolumn{7}{c}{(M$_{\rm eo}$)} \\
     \cmidrule(l){4-10}%\cmidrule(l){7-10}
     & & & b & c & d & e & f & g & h \\

     %& & & \multicolumn{7}{c}{yo} \\

    %Star mass ($M_\odot$) &\multicolumn{8}{c}{0.0802} \\
    %Star radius ($R_\odot$) &\multicolumn{8}{c}{0.117} \\
    \midrule
    \midrule
    Single planet (10M$_\oplus$) & $5 \times 10^{-5}$ & 1 & 80 & 24 & 16 & 12 & 8 & 2.4 & 0.12 \\
    Planet chain (10M$_\oplus$) & $5 \times 10^{-5}$  & 1 & 4 & 1.2 & 0.8 & 0.6 & 0.4 & 0.12 & 0.06 \\
   \bottomrule
  \end{tabular}
%\begin{tablenotes}
     % \small
     % \item $^1$ mean $f_{\rm acc}$ value averaged over the seven planets
    %\end{tablenotes}
  %\end{threeparttable}
\end{table*}

\subsubsection{Delivery of volatiles}\label{volsec}

\begin{figure*}
   \centering
  \includegraphics[width=19cm]{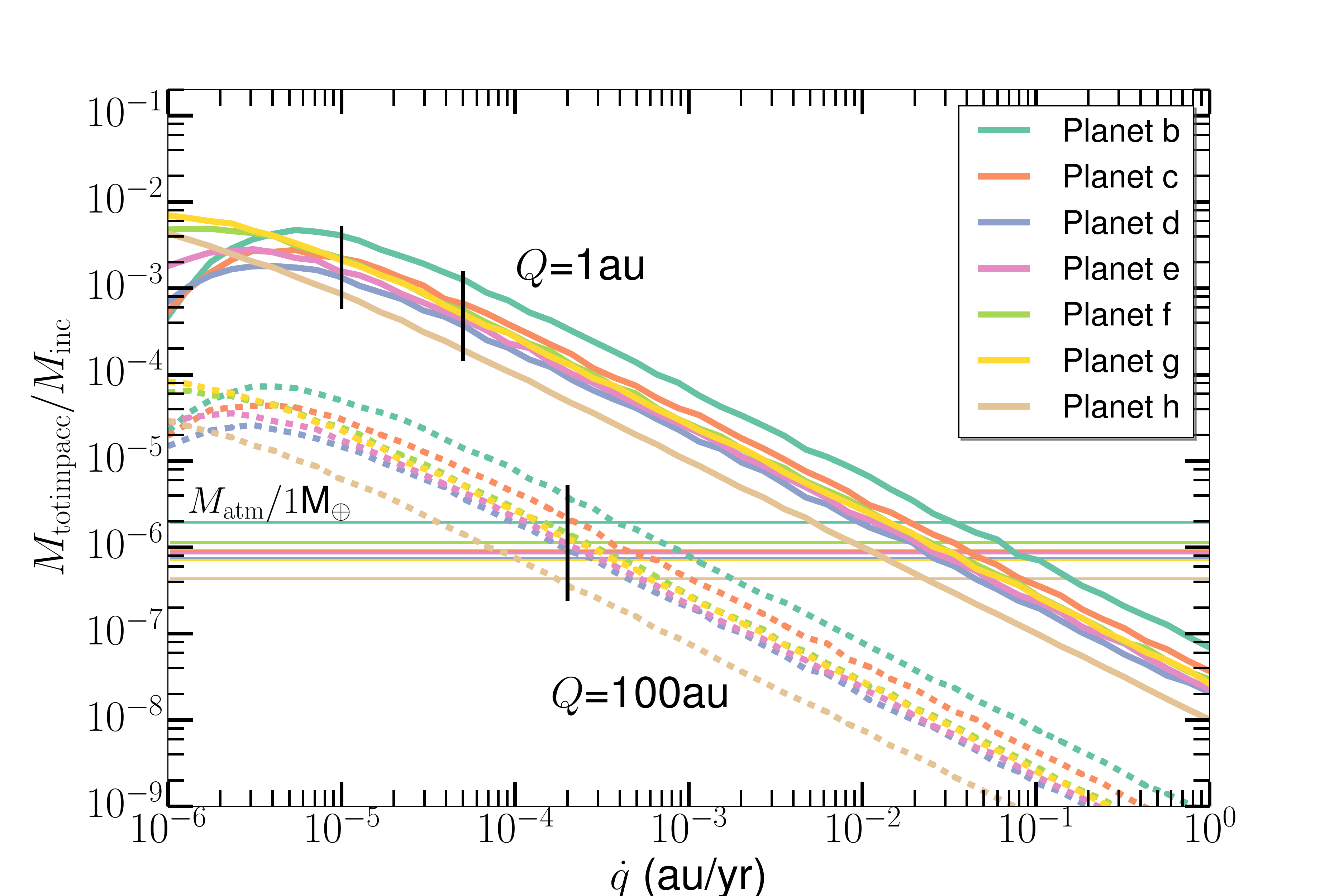}
   \caption{\label{fig9} Total accreted projectile mass for a given incoming mass of comets sent to inner regions, i.e $M_{\rm totimpacc}/M_{\rm inc}$ (assuming a -3.5 size distribution) as a function of $\dot{q}$ for $Q=1$ (solid lines) and 100au (dashed).
The vertical black lines are typical $\dot{q}$ values for each scenario (see the text in Sec.~\ref{fraction}). The primordial atmospheric masses of each planet are plotted as horizontal lines for an incoming mass of 1$M_\oplus$, assuming an Earth-like density and composition.}
\end{figure*}

We now evaluate the total mass of material and volatiles that can be delivered from the impactors over the system's lifetime.
We derive the total accreted projectile mass $M_{\rm totimpacc}$ by integrating the mass accreted per impactor (Fig.~\ref{fig7b} right) over the assumed size distribution.
This accreted mass is assumed to be deposited in the planets' atmospheres.
 
Fig.~\ref{fig9} shows $M_{\rm totimpacc}/M_{\rm inc}$, the total accreted projectile mass compared to mass of comets injected into the inner regions over the lifetime of the star. 
The overall shape is similar to Fig.~\ref{fig8}, but note that planet h is far better for delivery of mass into its atmosphere than having its atmosphere depleted because impacts are
at lower velocities (which means material from larger planetesimals can be accreted, see Fig.~\ref{fig7b} right). This means that the mass delivered on planet h (and planets with similar impact velocities) may be greater than that lost after each impact. 

\begin{figure}
   \centering
  \includegraphics[width=9cm]{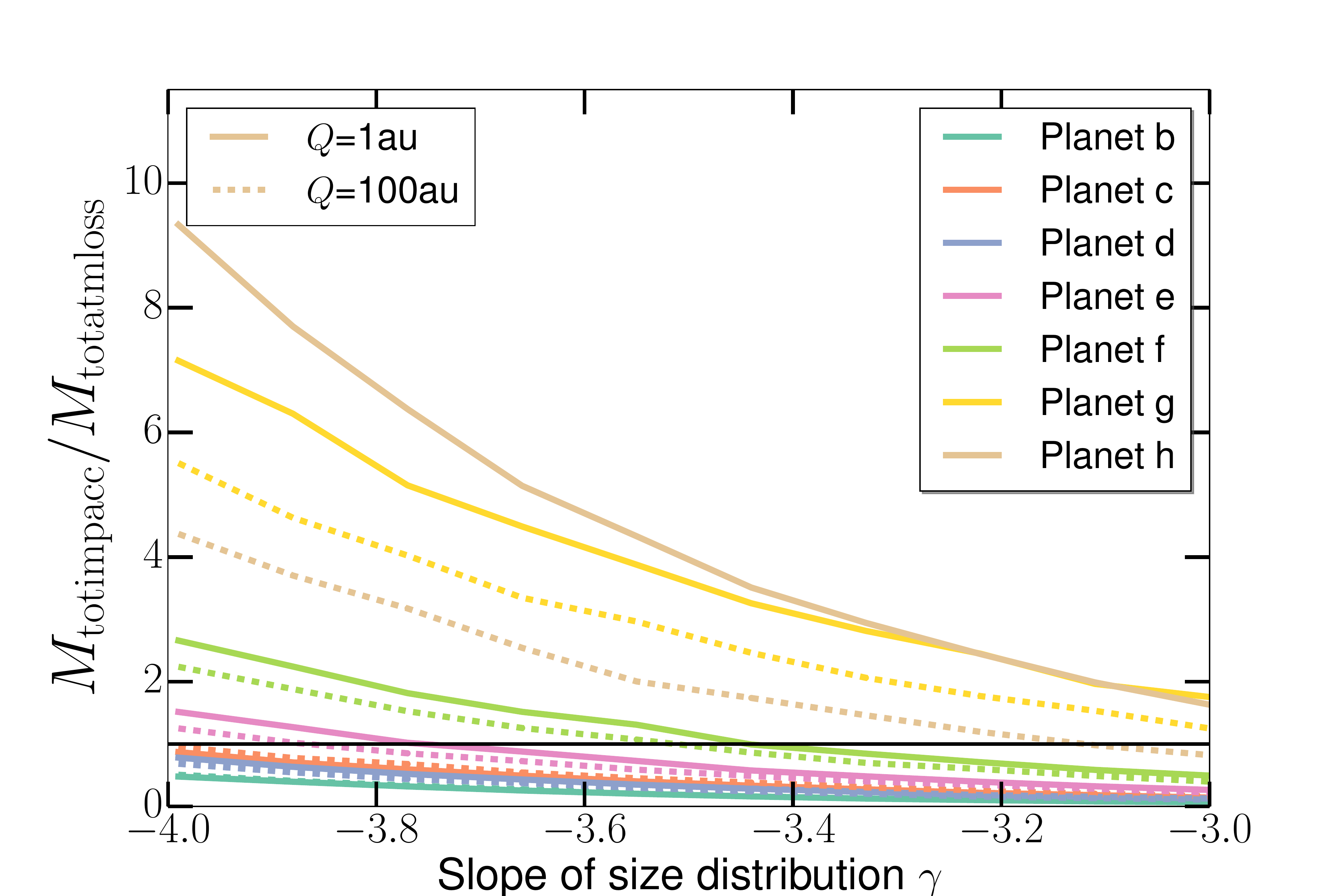}
   \caption{\label{fig10} Ratio of accreted projectile mass over atmospheric mass lost for $Q=1$ (solid) and 100au (dashed) for the different planets as a function of the slope of the size distribution.}
\end{figure}

To quantify this, Fig.~\ref{fig10} shows the ratio of the accreted projectile mass and atmospheric mass lost, which does not significantly depend on $\dot{q}$, instead only depending on the size distribution of comets and slightly on $Q$.
Thus, this ratio is plotted as a function of the slope in the size distribution $\gamma$ for two different values of $Q$ (1 and 100au). 
For $\gamma=-3.5$, this ratio is greater than
one for planets g and h and close to 1 for planet f but lower for the other planets.
This means that even if all of the accreted mass ends up in the atmosphere, the total atmospheric mass must be decreasing for the inner planets and can only increase for the outer three planets if all mass ends up in the atmospheres. Regardless, all planets
will have their atmospheres enriched by the planetesimals' composition and the situation is similar for all $Q$. 
%This means that subsequent volatile delivery will only be possible for the outermost planets. For a ratio of the accreted projectile mass to atmospheric mass lost $<1$, some volatiles
%can be delivered but they will not accumulate as further impacts will remove the new created atmosphere in a greater quantity than what was delivered. The conclusion for planets g and h
%is the same for all $\gamma$ values and for all $Q$ values $<100$au. The only difference is that bodies delivered with a very steep size
%distribution may be able to retain more projectile material (and then more volatiles) but the effect is modest for the innermost planets. 
%However, the ratio can increase/decrease by a factor two for planets g and h for a steeper/shallower size distribution than -3.5 but it will always be greater than 1, meaning that the atmospheres of these planets would be mostly made of cometary material after 7Gyr of 
%impacts (note that some additional sources of volatiles may be present, see Secs.~\ref{add} and \ref{compo} for a discussion on that). 

Consider now the fraction of this delivered projectile mass that will be in volatiles, i.e $M_{\rm totvolacc}=M_{\rm totimpacc} f_{\rm vol}$, which could be delivered to planets from comets.
To assess the amount of volatiles that are delivered to the planets we consider two types of material that impact on to these planets (presented in Sec.~\ref{rela}); 1) Cometary-like material with 20\% of ice by mass ($f_{\rm vol}=0.2$),
and 2) Asteroid-like material with $\sim$1\% of volatiles by mass ($f_{\rm vol}=0.01$). 

One important question is whether the icy material will have dissappeared through sublimation before impacting the planets.
\citet{2016P&SS..133...47M} show that a 1km comet survives sublimation for $\sim560$ orbits around a 0.1$L_\odot$ star. TRAPPIST-1 is 200 times less luminous and so the comets will survive much longer. Extrapolating \citet{2016P&SS..133...47M}'s formula to TRAPPIST-1
luminosity, we get that a 1km comet passing at small pericentres (0.1au) would need more than $10^5$ orbits to sublimate. As it only takes a few 100s of orbits for the comets to be accreted
on the planets (see Fig.~\ref{fig3}), we assume that most of the icy content of the comets will not have sublimated and so will be available to be delivered at impact. We note that during the $\dot{q}$ evolution, the sublimation will start happening only in the very last phase, i.e when the pericentre 
is close to the planets already \citep[because for larger pericentres the mass loss from comet sublimation is very slow and the timescale of evolution of the pericentre is much faster,][]{2016P&SS..133...47M}. Thus, the impact timescale of 100s of orbits is a good indicator of 
the number of orbits before impact during which sublimation could happen. 

Therefore, we estimate the mean of the total volatile mass delivered on each of the seven planets as

 \begin{equation}\label{gen3}
\frac{M_{\rm totvolacc}}{M_{\rm inc}} \sim 2 \times 10^{-3} f_{\rm vol} \left( \frac{\dot{q}}{10^{-5} {\rm au/yr}} \right)^{-1} \left( \frac{Q}{1{\rm au}} \right)^{-1},
\end{equation}

\noindent which is, for all planets, within a factor 3 of that from Fig.~\ref{fig9}. For all planets, we can also estimate the incoming mass needed to deliver more volatiles than the primordial atmospheric mass

 \begin{equation}\label{gen4}
M_{\rm inc} > 5 \times 10^{-4}  f_{\rm vol}^{-1} \left( \frac{\dot{q}}{10^{-5} {\rm au/yr}} \right) \left( \frac{Q}{1{\rm au}} \right) {\rm M}_\oplus,
\end{equation}

\noindent where we assumed primordial atmospheres of Earth-like densities.

Thus, for the planet scattering scenario, we find that only a small incoming mass is needed to deliver enough volatiles to potentially replenish an atmosphere with an Earth-like density
(e.g., $M_{\rm inc} > 3 \times 10^{-3}$M$_\oplus$ for comet-like bodies scattered from an outer belt to a 1M$_\oplus$ planet at 1 au). The incoming mass needed for the Kozai scenario is larger, 5M$_\oplus$, but not implausible to reach as shown by Eq.~\ref{minc}. 

\begin{figure}
   \centering
  \includegraphics[width=9cm]{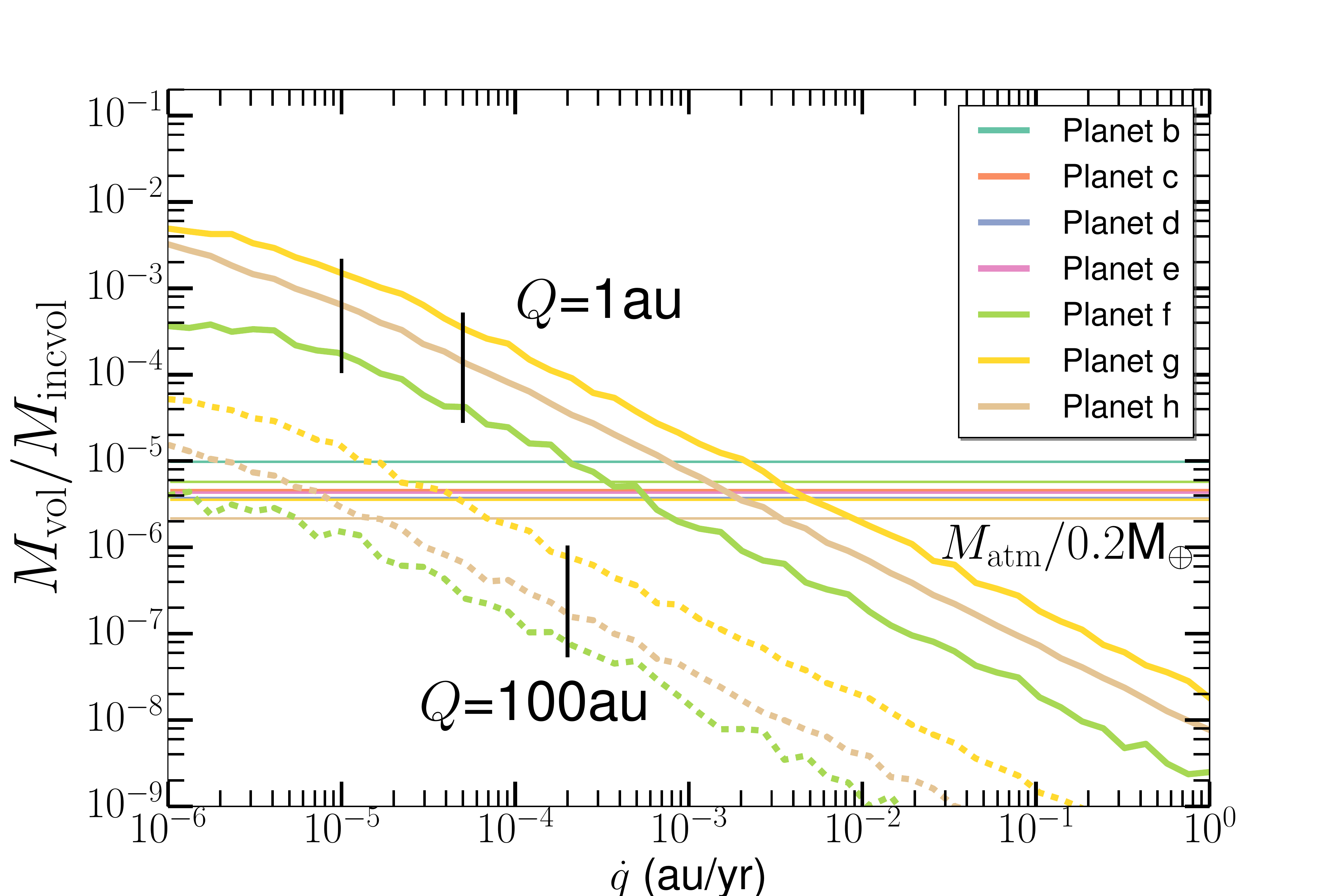}
   \caption{\label{fig10b} Volatile mass that can accumulate $M_{\rm vol}$ in the atmosphere impact after impact for a given incoming mass of cometary volatiles $M_{\rm incvol}$ with $M_{\rm incvol}=M_{\rm inc} f_{\rm vol}$ as a function of $\dot{q}$ for $Q=1$ (solid lines) and 100au (dashed).
The vertical black lines are typical $\dot{q}$ values for each scenario (see the text in Sec.~\ref{fraction}). The primordial atmospheric masses of each planet are plotted as horizontal lines for an incoming mass in volatiles of 0.2$M_\oplus$ (i.e. $M_{\rm inc}=1$M$_\oplus$ for $f_{\rm vol}=0.2$), assuming an Earth-like density and composition.}
\end{figure}

From Fig.~\ref{fig10}, we have shown that only planets g and h (and possibly f) would be able to retain the
largest part of the delivered volatiles.
This means that for the planet-scattering and Kozai scenarios, the new atmospheric compositions of planets f, g and h could be entirely set by the comet volatile content, which would replenish the atmospheres over the system's age.
However, the absolute level of the volatile content that will remain in the atmosphere is difficult to constrain as some fraction of the volatile mass will be ejected by later impacts or end up on the planet's surface and some other sources of volatiles could be present (see Sec.~\ref{add}).
We can, however, estimate the amount of volatiles that will survive after each impact assuming that a fraction $f_r$ of the accreted material remains in the atmosphere rather than condensing on the planet.
Therefore, after a given impact $f_r M_{\rm totimpacc}$ of material will be added to the atmosphere and the next impact could remove a maximum of $M_{\rm totatmloss}$ from this added material.
Assuming that $f_r=1$ (if impacts are frequent enough, e.g. LHB-like, material does not have time to condense back on the surface), we compute the fraction of volatiles that would accumulate from subsequent impacts in Fig.~\ref{fig10b}. 
We note that some additional volatiles could be added by degassing of the planets' interiors but that $f_r$ may also be smaller so that the exact volatile mass that can accumulate depends on complex physics that cannot be modelled in this paper. 
We see that indeed, only planets f, g, and h have positive values (i.e. they gain volatiles over time) and 
therefore appear\footnote{However, for $f_r<0.2$, the atmospheric mass loss takes over for planets g and h (and for $f_r<0.8$ for planet f) so that no secondary atmospheres would accumulate in this case, but this neglects outgassing which would add more volatiles and would make it harder to not build up secondary atmospheres on these three planets.} in Fig.~\ref{fig10b} showing $M_{\rm vol}/M_{\rm incvol}$, where $M_{\rm incvol}=M_{\rm inc} f_{\rm vol}$ is the incoming mass of volatiles. We can also derive a general formula as a function of $\dot{q}$ and $Q$ (similar to Eq.~\ref{gen3}) that gives the mass of volatiles that can accumulate $M_{\rm vol}$ rather than the total volatile
mass delivered. We do that in Sec.~\ref{timesec} and give the temporal evolution (assuming a constant rate of impact) of the build up of the secondary atmospheres that are created for planets f, g, and h (see Eq.~\ref{voltime}). 

We thus conclude that the atmospheres of planets f, g and h might be more massive than that of the innermost planets of the TRAPPIST-1 system if cometary bombardement has happened, and that a fraction of their composition should reflect the cometary abundances in this system.
We note that the build-up of secondary atmospheres for planets f, g and h is mainly allowed by the impact velocities that are low enough on these outermost planets to both reduce
the atmospheric mass loss after each impact and allow to deliver more volatiles (from larger bodies).

\subsubsection{Delivery of water}
%Using the results from the previous subsection and focusing on water delivery on planets f, g, h (that can retain volatiles), we can retrieve the specific mass of water that would be delivered to the planets in general (i.e as a function of $\dot{q}$ and $Q$) and for the specific scenarios we study.

Water on Solar System comets makes up more than $f_{\rm wat}=50\%$ of the volatiles \citep{2011ARA&A..49..471M}. Depending on $f_{\rm wat}$ for exocomets, the amount of water $M_{\rm water}$ delivered on the seven planets can be approximated by

 \begin{multline}\label{gen5}
M_{\rm water} \sim 2 \times 10^{-4} \left( \frac{f_{\rm vol}}{0.2} \right) \left( \frac{f_{\rm wat}}{0.5} \right) \left( \frac{\dot{q}}{10^{-5} {\rm au/yr}} \right)^{-1} \\ \left( \frac{Q}{1{\rm au}} \right)^{-1} M_{\rm inc},
\end{multline}

\noindent where $f_{\rm vol}\sim0.2$ for exocomets ($\sim$0.01 for asteroids) and $f_{\rm wat} \sim 0.5$ ($\sim$1 for asteroids).
For example, for the single Earth-mass planet scattering scenario (i.e $\dot{q} \sim 10^{-5}$au/yr, $Q=1$au, and $f_{\rm in}=1$), we find that a belt scattering at the same low rate as the current Kuiper-belt would result in the planets accreting 
$\sim 8 \times 10^{-3}$ Earth oceans of water (or 10 times less for asteroid-like bodies), assuming that one Earth ocean equals $1.5 \times 10^{21}$kg (see $M_{\rm watmin}$ in Table~\ref{tab2}). 
We note that for the planet chain case (where $f_{\rm in}=0.05$), these values would be a factor 20 smaller and for a larger incoming mass $M_{\rm inc}$ these values could go up by a factor more than $10^3$ (see Eq.~\ref{minc}). We find that a belt of 20M$_\oplus$ (similar to
the plausible belt mass we predict around TRAPPIST-1 in Sec.~\ref{mmsn}) that would scatter 5\% of its mass over 7Gyr (i.e., $M_{\rm inc} \sim 1{\rm M}_\oplus \, f_{\rm in}$) would deliver $\sim 1$ Earth ocean of water to the planets for the single planet case and $\sim 0.04$ Earth ocean for a planet 
chain (see $M_{\rm watT}$ in Table~\ref{tab2}).

For the Kozai scenario, we find that between $\sim 10^{-5}$ (pessimistic case with a Kuiper belt scattering rate) and $\sim 0.01$ (optimist case with $M_{\rm inc} \sim 20{\rm M}_\oplus$) Earth oceans of water could be delivered to the planets. 

This delivered water will presumably recondense as ice on the surface of planet h (but when the star was younger this planet was in the HZ and water could have been in liquid form for a long period, see Sec.~\ref{timesec}), but for warmer planets such as planets f and g, we expect that a rain cycle would create liquid water on these planets that would then be reinjected into the atmospheres cyclically (see Sec.~\ref{compo}). 
The temporal evolution of the build up of the amount of water in these secondary atmospheres can be obtained from Fig.~\ref{fig10b} or from the coming Eq.~\ref{voltime} for planets f, g, and h.
%This is a lower limit because some water can rain quickly onto the surface of the planet and be protected for the next impact.

\section{Discussion}\label{discu}

\subsection{Comparison between timescales of the different processes}
The consideration of timescales is important because it constrains the duration over which atmosphere loss/gain occurs compared with other processes which may be taking place, but which are beyond the scope of this manuscript to consider in detail.

\subsubsection{Timescale to lose primordial atmospheres from impacts}

Assuming a constant rate of scattering $\dot{M}_{\rm sca}$ over 7Gyr, we compute the atmospheric mass lost as a function of time

 \begin{multline}\label{atmtime}
M_{\rm atmlossc}(t) \sim  2 \times 10^{-3}  f_{\rm in} \left( \frac{\dot{q}}{10^{-5} {\rm au/yr}} \right)^{-1} \left( \frac{Q}{1{\rm au}} \right)^{-1}  \\ \left( \frac{t}{7{\rm Gyr}} \right) \left (\frac{\dot{M}_{\rm sca}}{0.1{\rm M}_\oplus/{\rm Gyr}} \right) {\rm M}_\oplus,
\end{multline}

\noindent where we note that $\dot{M}_{\rm sca}=0.1{\rm M}_\oplus/{\rm Gyr}$ corresponds to a belt with a total incoming mass of $\sim$1M$_\oplus$ $f_{\rm in}$, i.e similar to what would be expected for a 20M$_\oplus$ belt scattering 5\% of its material over the age of the star. 
$M_{\rm atmlossc}(t)$ becomes greater than an atmospheric mass of $10^{-6}$M$_\oplus$ for 

 \begin{multline}\label{destroytime}
t_{\rm destroy} \sim  4{\rm Myr} \, f_{\rm in}^{-1} \left( \frac{\dot{q}}{10^{-5} {\rm au/yr}} \right) \left( \frac{Q}{1{\rm au}} \right) \\ \left (\frac{\dot{M}_{\rm sca}}{0.1{\rm M}_\oplus/{\rm Gyr}} \right)^{-1}.
\end{multline}

\begin{figure}
   \centering
  \includegraphics[width=9cm]{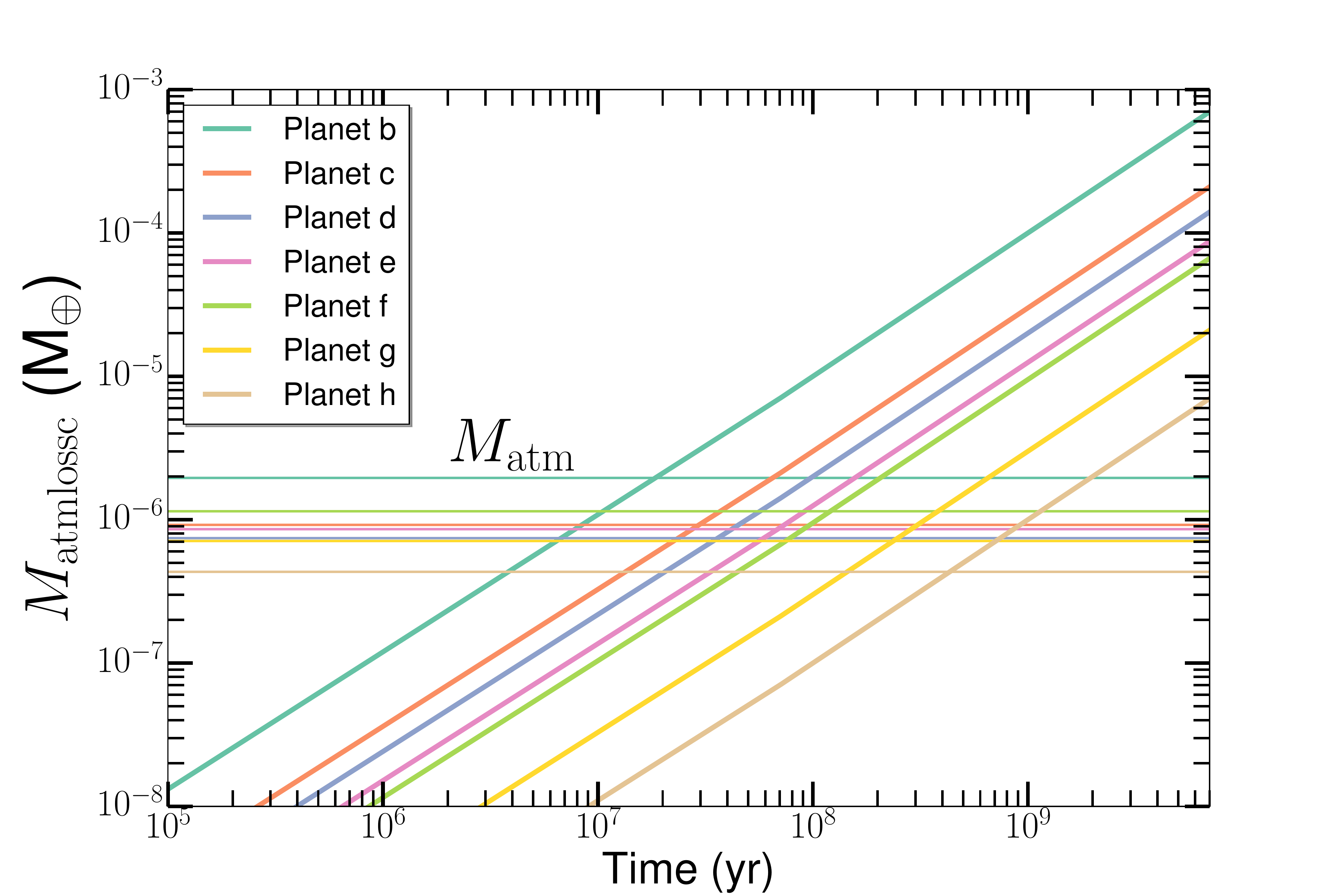}
   \caption{\label{fig11} Atmospheric mass loss $M_{\rm atmlossc}$ as a function of time for a belt scattering at a constant rate $\dot{M}_{\rm sca}=0.1{\rm M}_\oplus/{\rm Gyr}$ for 7Gyr. 
We assume the planet chain scenario with $\dot{q}=10^{-5}$au/yr, $Q=1$au, $f_{\rm in}=0.05$. The primordial atmospheric mass of each planet is plotted as a horizontal line assuming an Earth-like density.}
\end{figure}

Now, we consider the planet chain scenario (i.e with $\dot{q}=10^{-5}$au/yr, $Q=1$au, and $f_{\rm in}=0.05$) with a scattering rate $\dot{M}_{\rm sca}=0.1{\rm M}_\oplus/{\rm Gyr}$ and look at the temporal evolution of the atmospheric mass loss $M_{\rm atmlossc}(t)$ due to
the series of impacts over the system's age as shown by Fig.~\ref{fig11}. By comparing to the primordial atmospheric masses of the planets (horizontal lines in Fig.~\ref{fig11}), we see that for this scenario, it takes between 10 and 400Myr to destroy the primordial atmospheres of all seven planets 
(assuming an Earth-like atmospheric density). This is very fast compared to the age of the system.

\noindent This shows that the timescales over which the primordial atmospheres can be destroyed are much shorter than the age of the system. Therefore, we confirm the previous conclusion (see Sec.~\ref{atml}) that cometary impacts may have entirely stripped 
all planets of their primordial atmospheres by 7Gyr, even if the scattering rate is smaller by a factor more than 10 than assumed here (i.e., close to the Kuiper-belt scattering rate level).

\subsubsection{Timescale to regenerate secondary atmospheres from impacts for planets f, g, and h}\label{timesec}

We also compute the temporal evolution of the volatiles $M_{\rm vol}$ that are deposited and accumulate after each impact (i.e we take into account that subsequent impacts remove part of the volatiles delivered by the preceding impact as in Fig.~\ref{fig10b}). 
For planets g and h, $M_{\rm vol}$ is given by (for any $\dot{q}$ and $Q$)

 \begin{multline}\label{voltime}
M_{\rm vol}(t) \sim  10^{-3} f_{\rm in} f_{\rm vol} \left( \frac{\dot{q}}{10^{-5} {\rm au/yr}} \right)^{-1} \left( \frac{Q}{1{\rm au}} \right)^{-1}  \\ \left( \frac{t}{7{\rm Gyr}} \right) \left (\frac{\dot{M}_{\rm sca}}{0.1{\rm M}_\oplus/{\rm Gyr}} \right) {\rm M}_\oplus,
\end{multline}

\noindent and is a factor 5 smaller for planet f. The amount of water delivered is simply $M_{\rm water}=f_{\rm wat} M_{\rm vol}$. Now, we work out the timescale to replenish the secondary atmospheres of planets g and h in cometary volatiles at the level of a $10^{-6}$M$_\oplus$ atmospheric mass

 \begin{multline}\label{volrepltime}
t_{\rm replenish} \sim  7{\rm Myr} \, f_{\rm in}^{-1} f_{\rm vol}^{-1} \left( \frac{\dot{q}}{10^{-5} {\rm au/yr}} \right) \left( \frac{Q}{1{\rm au}} \right)  \\ \left (\frac{\dot{M}_{\rm sca}}{0.1{\rm M}_\oplus/{\rm Gyr}} \right)^{-1},
\end{multline}

\noindent and a factor 5 longer for planet f. The replenishment timescale shows that in most physically motivated cases planets f, g, and h will have had time (over the age of the system) to rebuild secondary atmospheres with masses of at least $10^{-6}$M$_\oplus$, i.e., equal or greater than an Earth-like primordial atmosphere. 
We note that most of the volatiles delivered by the comets have low condensation temperatures and thus would remain in the atmosphere rather than go on the planet's surface but water could condense as ice on planet h 
and cycle from the surface to the atmosphere on planets f, g owing to rain (see Sec.~\ref{compo}). Therefore, we expect $M_{\rm vol}$ to be a good estimate of the amounts of volatiles that can accumulate for planets f and g and note that up to 50\% of the volatiles (to account for water)
could transform into ice on planet g and thus reduce $M_{\rm vol}$ by a factor 2 (but this ice could outgas at a later stage because of the planet activity).

\begin{figure}
   \centering
  \includegraphics[width=9cm]{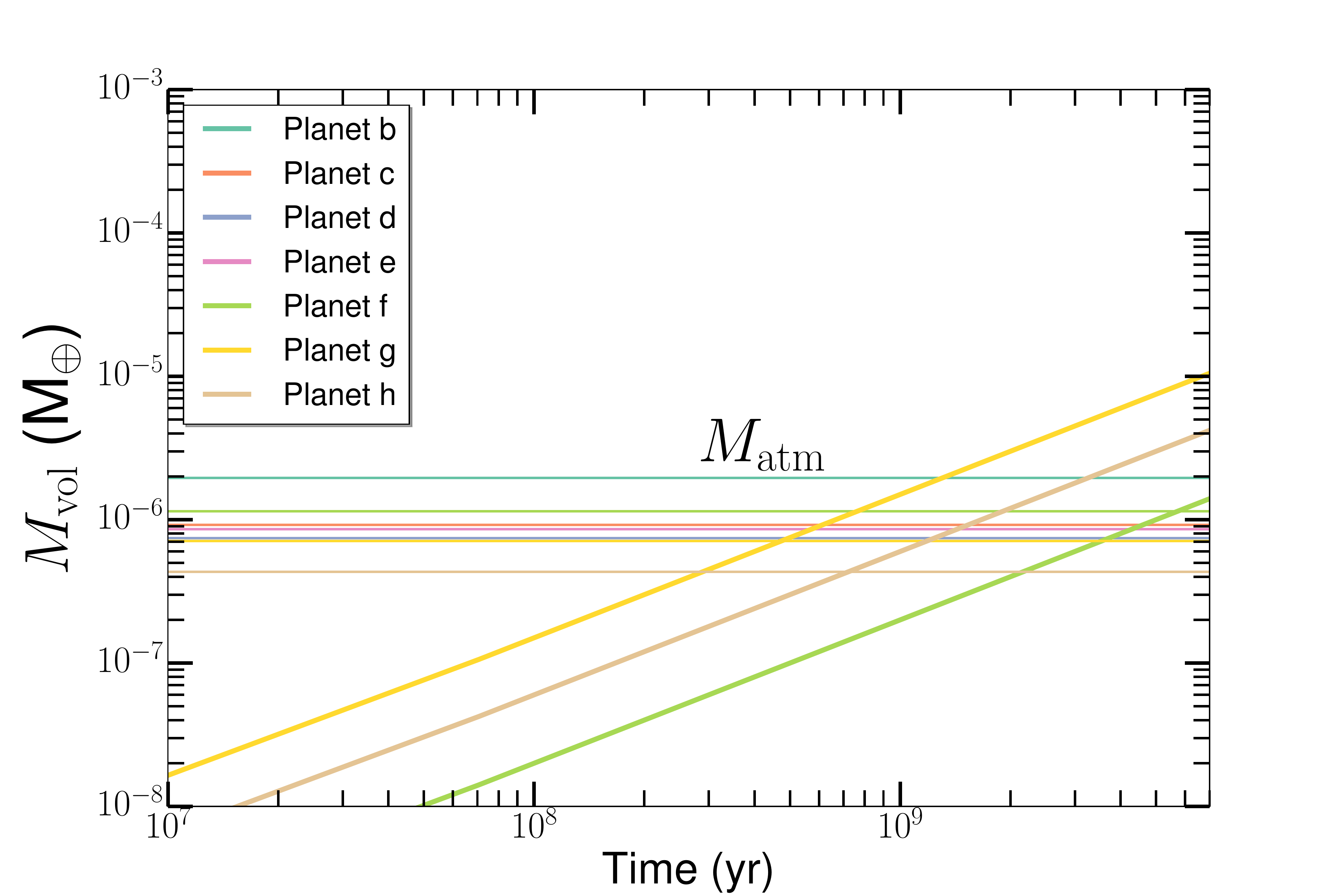}
   \caption{\label{fig12} Volatile mass that accumulates impact after impact $M_{\rm vol}$ as a function of time. We assume the planet chain scenario with $\dot{q}=10^{-5}$au/yr, $Q=1$au, $f_{\rm in}=0.05$ and $f_{\rm vol}=0.2$ and a constant comet input rate over 7Gyr of $\dot{M}_{\rm sca}=0.1{\rm M}_\oplus/{\rm Gyr}$. The primordial atmospheric mass of each planet is plotted as a horizontal line assuming an Earth-like density.}
\end{figure}

Fig.~\ref{fig12} shows an example for the planet chain scenario with $\dot{q}=10^{-5}$au/yr, $Q=1$au, $f_{\rm in}=0.05$ and $f_{\rm vol}=0.2$ assuming $\dot{M}_{\rm sca}=0.1{\rm M}_\oplus/{\rm Gyr}$. We see that only planets f, g and h can accumulate volatiles over time (see section \ref{volsec}). For this planet chain scenario,
planets g and h can get their atmospheres replenished by cometary volatiles at the level of their primordial atmospheres over a timescale of 500-700Myr and $\sim$6Gyr for planet f.

We now compare the replenishment timescale to the evolution of the position of the HZ. \citet{2015AsBio..15..119L} show that for a 0.08M$_\odot$ star, the HZ location moves inwards to its present-day position after $\sim 1$Gyr. 
This means that planet h will be the first to enter the liquid water HZ, which it will do at a point when the closer-in planets are still in a runaway greenhouse state (assuming they have retained any atmospheres).  According to 
the \citet{2015AsBio..15..119L} model, planet h crosses into the empirical habitable zone at $\sim$30Myr.  Coupled with our results, this scenario indicates that planet h could have received significant volatile delivery at a point in its history (i.e., between 30Myr and 1Gyr) when liquid water was stable 
at its surface (Fig.~\ref{fig12}). This raises the prospect for an early carbon cycle being established on this planet, stabilising climate through water-rock interaction as is inferred for Earth \citep{1981JGR....86.9776W}.

\subsection{Additional sources of volatiles}\label{add}
\subsubsection{Volatiles created by vapourised material from the planet's surface during impact}\label{source}
The volatile fraction that ends up in the atmospheres of the TRAPPIST-1 planets does not only build up from the impactor material but also from the vapourised material from the planet surface, as was probably the case for the Chicxulub impact that may have released large 
quantities of gas and dust contributing to the environmental stress that led to the demise of dinosaurs on Earth \citep{1997JGR...10221645P}.

From \citet{1977Sci...198.1249O}, we can estimate the volume of material $V_{\rm vap}$ vapourised from a given meteoritic impact (with a volume $V_{\rm pr}$). They find that $V_{\rm vap} = 0.4 S V_{\rm pr}$, where
$S=(\rho_{\rm pr}/\rho_{\rm t}) (V_{\rm imp}/C_p)^2$, using the same notations as in previous sections and $C_p$ being the bulk sound speed of planetary surface, which varies depending on the planet ground composition \citep{1989icgp.book.....M}. We assume an Earth-like composition for which
$C_p \sim 7$km/s. We thus find that the vapour mass $M_{\rm vap}$ produced for a given impactor of mass $M_{\rm imp}$ is $M_{\rm vap}=0.4 M_{\rm imp} (V_{\rm imp}/(7{\rm km/s}))^2$. 

However, some of the vapour ejecta will escape and only a fraction will have a low enough velocity to be retained
in the atmosphere. Once again, using results from \citet{2009M&PS...44.1095S}, we get that the maximum ejected fraction of target material after impact is $M_{\rm taresc} \sim 0.02 M_{\rm imp} (V_{\rm imp}/V_{\rm esc})^2$. 
This maximum is reached for bodies that are larger than $\sim$1km and for smaller bodies, the planet retains almost all of the target material created at impact. Of course, above a certain threshold it means that the whole target mass escapes (as $M_{\rm taresc}$ becomes greater than the total atmospheric mass), 
which is similar to the projectile mass behaviour (where volatiles from bodies larger than $\sim$10km cannot be retained in the atmosphere). We thus find that $M_{\rm vap}$ is a good indicator of the vapourised mass that will remain in the atmosphere (as $M_{\rm taresc} \ll M_{\rm vap}$).

From Eq.~\ref{imploss}, we notice that the mass delivered from the projectile quickly tends to $M_{\rm imp}$ for bodies smaller than about 10km.
Thus, for planets g and h that have median impact velocities of $\sim 25$ and 20km/s, $M_{\rm vap}$ will be slightly higher but of the same order of magnitude as $M_{\rm impacc}$. This means that some volatiles such as SO$_2$, CO$_2$ or water could also be formed
from the vapourised planets' crust \citep[see][]{1997JGR...10221645P}. However, we note that the typically low concentration of volatiles in planetary basalts that would form the bulk of a crust would not release as many volatiles as for the Chicxulub impact \citep[e.g.][]{1987Icar...71..225D,2002Natur.419..451S}.

\subsubsection{Outgassing on the planets}\label{source3}

Degassing may happen early during accretion when forming the planets but this is not a concern in our study as we expect the primordial atmospheres to be totally destroyed. Degassing from tectonic activity may also happen at a later stage that could affect the amount of
volatiles in the atmospheres. Another way of producing degassing is from stellar induction heating. A recent study that focused on the effect of this mechanism on the TRAPPIST-1 planets 
finds that induction heating could create strong outgassing on planets b, c, d that are very close to their host star but it should not affect the outermost planets e, f, g, and h \citep{2017arXiv171008761K}. 

For the plate-tectonic degassing, we take the degassing on Earth as an upper bound because plate tectonics is very active on Earth and may be less efficient/active on other planets\footnote{We note that a recent study shows that even
for planets in the stagnant lid regime (i.e., without plate-tectonics), volcanic outgassing rates suitable for habitability could possibly be maintained \citep{2017arXiv171203614F}.}. Earth produces $\sim 22$km$^3$ of basaltic magmas each year \citep{1984JVGR...20..177C}.  Given a magma density of 2600kg/m$^3$, 
we estimate a total degassing rate of $\sim 6 \times 10^{13}$kg/yr. Assuming a typical water content of 0.3wt\% and the extreme case of perfectly efficient degassing with no subduction recycling of water to the planet's mantle, we find that 
an upper bound on the tectonically driven water degassing rate is $\sim 3 \times 10^{-5}$ M$_\oplus$/Gyr (0.11 Earth oceans per Gyr). Therefore,
if the tectonic activity on planets f, g and h were as active as on Earth, degassing of water could occur at a similar rate to the water delivered from impacting comets (see Table~\ref{tab2}), thus enhancing the amount of water on planets f, g and h. 

%A study by \citet{2008Icar..195..447P} that assumes that plate tectonics is active on Earth-like exoplanets finds that the the bulk of the degassing (see their table~4) operates within $<10^6$yr for planets of similar masses to the TRAPPIST-1 planets. This outgassing process
%can therefore also be neglected here to contribute significantly to the secondary atmospheres. Another way of producing degassing is from stellar induction heating. A recent study that focused on the effect of this mechanism on the TRAPPIST-1 planets 
%finds that induction heating could create strong outgassing on planets b, c, d that are very close to their host star but it should not affect the outermost planets e, f, g, and h \citet{2017arXiv171008761K}. 

\subsubsection{Volatiles that are ejected of the atmosphere and reaccreted later}\label{source2}
The material that escapes the planetary atmospheres after each impact because they have velocities greater than the escape velocity will end up in an eccentric torus around the star close to the given planet location \citep[e.g.][]{2014MNRAS.440.3757J,2017AsBio..17..721C}. The eccentricity
will vary depending on the ejection velocity of the material. While we expect that high-velocity ejecta may reach neighbouring planets \citep[e.g. in a Panspermia-manner,][]{2017ApJ...839L..21K,2017PNAS..114.6689L}, most of the material in the torus would interact with the planet it has been ejected from.
We note that for an Earth-like planet on a slightly wider orbit than planet h, the escape velocity (of about 10km/s) could become greater than the planet's Keplerian velocity and thus the material would not form a torus but rather be ejected on unbound orbits.

The fate of the material in the torus is not straightforward to model. The material could deplete collisionally due to high-velocity collisions in the elliptic torus and be ground down to dust, which would be blown out from the system by stellar wind radiation pressure \citep{2008ARA&A..46..339W} and at the same time
eject the ices or volatiles present on the grains.
While the ejecta is also partly made up of gas, one could also expect that the gas material (at least the fraction that is not blown out by radiation pressure) in the torus will viscously spread (maybe dragging dust with it) and end up on more distant planets. The fate of the material that would be able to interact with a planet for a long enough timescale is to 
be reaccreted onto the progenitor planet \citep{2017MNRAS.464.3385W}. The exact outcome depends on the exact chemico-physical conditions in the TRAPPIST-1 planets environment, which is not known, and thus goes beyond the scope of this paper.

\subsection{Composition of the atmospheres at the end of the impact process}\label{compo}
Thanks to our model, we are able to retrieve the amount of volatiles that is delivered to the different planets as well as the atmospheric mass removed by a long-term series of impacts. 
For the outermost planets, we find that the volatiles delivered by impacts may accumulate and be abundant, which could give us a way
to constrain the atmospheric composition of planets f, g, h, the former two being in the HZ. However, we need to understand how these delivered volatiles would evolve in their new atmospheres to predict the current atmospheric compositions of these planets. 
They could chemically react to form new species, condense on the surface as ice and some additional volatiles may be produced as seen in the previous Sec.~\ref{add}.

For instance, the delivered water will presumably condense as ice on the surface of the colder planet h (when it finishes being in the HZ, see Sec.~\ref{timesec}) but for warmer planets in the HZ (e.g. planets f and g), a rain cycle could create liquid water on the planets that is then reinjected into the atmospheres cyclically. 
Volatiles such as CO, CO$_2$, or CH$_4$ have a low condensation temperature and will remain in the atmosphere along with other similar volatiles delivered by the comets. 
However, when liquid water is on the planet, this can draw the CO$_2$ content down by silicate weathering that fixes CO$_2$ in the planet's surface (forming carbonates) as shown in \citet{1968Sedim..11....5S}.

Over longer timescales, these volatiles can further chemically react to form new molecules. 
However, the exact composition of the delivered volatiles depends on the composition of exocomets in our scenario. The latter has been found to be consistent with the composition of comets in our Solar System \citep[e.g.][]{2017ApJ...842....9M}
but there is still a wide range of observed compositions amongst the Solar System's comets \citep[e.g.][]{2011ARA&A..49..471M}.

Another complication is that, as discussed in the previous subsection \ref{add}, volatiles may also be formed from the vapourised planet's crust during impact, from outgassing and even by reaccretion of previously ejected material, which would mix with the volatiles delivered by impacts.

All of these factors (active chemistry, potential additional volatiles, exocomet composition) makes it hard to predict the exact final compositions of the atmospheres after a few Gyr of evolution.
An atmosphere model that would make assumptions about what happens without impacts could be fed by our impact predictions to come up with a plausible likely composition, but this goes beyond the scope of the present paper. We note however that these extra sources of volatiles do not change our conclusion
that in the presence of a belt scattering comets, the atmospheres of the outermost planets f, g, h should be more massive. %We also emphasise that the relative compositions of the atmospheres do not depend on the timescale on which impacts are assumed to have happened. 

\subsection{Impacts in very dense Venus-like atmospheres}
We note that our model is not valid for very massive atmospheres. If the atmospheres of planets f, g, h become massive enough (Venus-like, i.e., 200bars) due to impacts (or if the primordial atmospheres were Venus-like), 1-10km impactors do not create craters anymore but rather decelerate and get fragmented before touching the ground 
and create big aerial bursts that are very effective at removing atmospheric mass \citep{2014P&SS...98..120S}. The amount of accreted projectile material is also
very high (close to 100\%) for these aerial burst type of impacts \citep{2014P&SS...98..120S}.
Therefore, for very dense atmospheres, we expect an increased delivery of volatiles from the impactors and less from the vapourised crust. We also expect that Venus-like primordial atmospheres would still be destroyed, since those impacts are more effective at removing mass, therefore not changing 
our conclusions.

\subsection{Implications for life on these planets}\label{life}

One of the prime motives in searching for planets orbiting very low mass stars is to study the chemical composition of their atmospheres, and discover whether they contain large quantities of gas of a likely biological origin \citep[e.g.][]{2016AsBio..16..465S}. Here, we consider
the implications of our results concerning impacts towards creating the first forms of life.

Many elements can affect the emergence of life, most of which currently remain unconstrained empirically.  We chose to apply our study to the TRAPPIST-1 system because its seven planets mark an important milestone. In addition to the multiple advantages of having a 
very low-mass host star for atmospheric characterisation \citep[e.g.][]{2017MNRAS.464.2687H}, these seven worlds allow us to compare each to one another. All seven have followed a similar history in terms of UV irradiation for instance (modulo their distance to the star). Here we 
have tried to quantify whether all planets would receive a similar impact history, which may be important to kick start life as explained further. 

UV irradiation has often been seen as prejudicial to habitability. Its main disadvantages are: 1) to photodissociate water molecules, of which the hydrogen is then lost to the space, depleting its oceans \citep[e.g.][]{2017A&A...599L...3B}, and 2) to break complex molecules on the surface, 
and affect replication \citep[e.g.][]{2017MNRAS.469L..26O}. The situation is particularly sensitive for planets orbiting very low-mass stars like TRAPPIST-1, since these spend a long time contracting onto the main-sequence, in a 1 Gyr stage of particularly heightened far UV
activity \citep[e.g.][]{2015ApJ...806..137R}.

However, these issues might be mitigated by several effects: 1) Ocean loss depends on the initial water reservoir \citep[e.g.][]{2016A&A...596A.111R,2017MNRAS.464.3728B}, and the TRAPPIST-1 planets might have been initially rich in water, having possibly assembled beyond 
the snow-line \citep{2017A&A...598L...5A,2017A&A...604A...1O} and/or accreted water at a later stage owing to impacts (as shown in this study);
 2) UV photons do not penetrate water well, and organisms can protect themselves under a few metres of water \citep[e.g.][]{2017arXiv170805400E}; 3) UV irradiation accelerates mutations, leading to Darwinian evolution; 4) the non-illuminated side of a tidally synchronised planet is 
protected; and 5) {\it UV irradiation, impacts, and a hard surface might be required to kick-start life (abiogenesis)}.

The literature contains much debate on many of the points above, except on the very last one, which we describe in more detail here as it is related to the outcome of this paper. 

Recent advances in biochemistry \citep[summarised in][]{Suth17} have shown a prebiotic chemical path leading from hydrogen 
cyanide (HCN) to formaldehyde (CH$_2$O), a known precursor to ribonucleotides (the building block to biologically relevant molecules such as ATP, RNA and DNA), amino acids (required for proteins) and lipids \citep{2015NatCh...7..301P}. Hydrogen cyanide, the initial molecule needed to inititate the
process, can be produced in the plasma created when impactors enter in contact with an atmosphere \citep{2015PNAS..112..657F}. In the presence of UV radiation, hydrogen cyanide can then react with other compounds that can be found concentrated on a planetary surface to create the building blocks of life. 
The impactor itself may have another role to play, which is to excavate underground material, and reveal chemically interesting strata \citep{2015NatCh...7..301P}, thereby acting as a chemical reactor. 

We show in this paper that, if a belt scattering comets is present in the system, numerous impacts with different energies will happen throughout the history of the TRAPPIST-1 planets. From these impacts, we expect to create a subsequent amount of HCN in 
the impactor plasma \citep{2015PNAS..112..657F}. We also note that as HCN is found in comets \citep[e.g.][]{2011ARA&A..49..471M}, it may also be present on the potential exocomets of TRAPPIST-1 and be delivered along with the other volatiles \citep[e.g. see][]{Matra}. 
We also emphasise that if the planets are tidally locked, it does not affect the emergence of life in this scenario as we predict that about half of the impacts would happen on the night side and the other half on the day side so that the UV photons from the star necessary for reactions to happen
will be able to play their role. 

Thus, our scenario
offers the seed to create the first building blocks of life and more detailed modelling is needed to quantify how many ribonucleotides, amino acids and lipids could be created from the impact properties (e.g. impact velocities, rate of impacts) we predict. This is beyond the scope of this paper
but should give birth to new interesting studies in the near future. Panspermia may also be viable to transport some potential life forms to other planets, which can enhance the probability of life spreading in the system \citep{2017ApJ...839L..21K,2017PNAS..114.6689L}.

To conclude, we cannot be certain yet that such a path is where biology originated, however, it provides a different narrative, one that requires UV irradiation, impacts and a limited amount of water. 
Ultraviolet, in this context, becomes beneficial by removing excess liquid water and transforming hydrogen cyanide into formaldehyde, whereas impacts would bring in energy to create hydrogen cyanide, and replenish 
the planet in volatiles such as water, much like what happened in the LHB \citep[e.g.][]{2014GeCoA.145..175C,2017AJ....153..103N} after a desiccating moon-forming impact \citep[e.g.][]{2014RSPTA.37230175C}.

\section{Conclusion}\label{ccl}
In this paper, we have studied the effects of impacts on the seven TRAPPIST-1 planets in terms of atmospheric mass loss and delivery of volatiles and water.
We derive general results for any scenario where the comet pericentres slowly migrate inwards at a rate $\dot{q}$. We also specifically test three scenarios 
for the delivery of comets from an outer belt to the inner planets (located within 0.1au): 1) Planet scattering by a single or a chain of planets,
2) An outer companion forcing Kozai oscillations on comets leading them to small pericentres, 3) Galactic tides on an exo-Oort cloud. We model these three scenarios by a steadily decreasing pericentre (constant $\dot{q}$) that
is quantified in Sec.~\ref{appl} for each of the scenarios.
The results can be summed up as follows:

\begin{itemize}
 \item We find that applying a minimum mass TRAPPIST-1 nebula approach lead to a surface density $\Sigma \sim 122 \, (r/1{\rm au})^{-1.97}$ kg/m$^2$. We show that a potential belt around TRAPPIST-1 could not survive within 10au because of collisional 
erosion (if it was created at the end of the protoplanetary disc phase). Assuming that such a belt is between 10 and 50au,
and extrapolating the derived minimum surface density, we infer that this belt would have mass of at least 20M$_\oplus$ and may be observable in the far-IR or sub-mm with ALMA (see Sec.~\ref{mmsn}).
 \item We ran a suite of N-body simulations to understand the dynamics of comets that impact onto the seven different planets. We find the impact and ejection probabilities for each comet's orbit (see Figs.~\ref{fig1} and \ref{fig2}). We also provide the accretion timescales for these different comet families (see Fig.~\ref{fig3}). We analytically explain the main
dependencies for these probabilities and timescales.
 \item We give the impact velocity distributions for each planet, and we find that they typically have double-peaked profiles (see Fig.~\ref{fig4}). The median impact velocity for planet b is close to 100km/s, whilst for planet h, it is close to 20km/s (see Fig.~\ref{fig4b}). These impact velocities are always much above the escape velocities of the planets and gravitational focusing is not important.
 \item We find that the fraction of comets accreted on each planet depends on the decreasing rate of pericentres ($\dot{q}$) and apocentre $Q$ (scaling as $\dot{q}^{-1} Q^{-1}$). We find two regimes, for small $\dot{q}$, most of the impacts end up on planets g and h and for higher $\dot{q}$, each planet gets a fraction of comets accreted (see Fig.~\ref{fig6}).
 \item The atmospheric removal is dominated by comets of a few km in diameter (see Fig.~\ref{fig7b} left).
 \item The delivery of volatiles is only possible for comets $\lesssim 3$km in size (see Fig~\ref{fig7b} right). For bigger comets, the projectile material escapes and no delivery is possible.
 \item We find that the higher impact velocities for the innermost planets lead to a higher atmospheric removal rate for a given cometary impact rate and a lower amount of volatile delivered.
 \item In general, we find that if the incoming mass of comets that reach the inner regions
$M_{\rm inc} > 5 \times 10^{-4} \left( \frac{\dot{q}}{10^{-5} {\rm au/yr}} \right) \left( \frac{Q}{1{\rm au}} \right) {\rm M}_\oplus$,
 the primordial atmospheres of the seven planets would be totally destroyed (see Fig.~\ref{fig8}), i.e {\it a belt with a low scattering rate similar to the current Kuiper belt is enough to destroy all primordial planetary atmospheres}.
 \item We quantify the amount of water lost owing to impacts and find that it is similar (possibly higher) to the amount of water lost through hydrodynamic escape (see Sec.~\ref{atmwat} and Table.~\ref{tabwat}).
 \item As for the delivery of volatiles to the comets (see Fig.~\ref{fig9}), we find that planets g and h (and most likely f) may retain volatiles from the impacting comets in their atmospheres and the 
conclusion holds for any size distribution of incoming comets between -3 and -4 (see Fig.~\ref{fig10}).
\item {\it We thus predict that if the planets were hit by comets, the atmospheres of planets f, g, and h would be more massive, which could be checked by future missions in the next decade}.
\item We also show that for an incoming mass of comets $M_{\rm inc} > 5 \times 10^{-4}  f_{\rm vol}^{-1} \left( \frac{\dot{q}}{10^{-5} {\rm au/yr}} \right) \left( \frac{Q}{1{\rm au}} \right) {\rm M}_\oplus$ (where $f_{\rm vol}$ is the volatile fraction on solids), the volatile mass delivered by comets is greater than Earth-like atmospheric masses (assuming Earth-like densities for the 7 planets). 
\item We provide a prescription for the amount of water or volatiles that can accumulate as a function of time (see Eq.~\ref{voltime}) that could be used to feed an atmospheric model to check the actual composition of atmospheres dominated by the delivery of comets.
\item We find that a large quantity of volatiles may have been delivered to planet h while it was still in the liquid water habitable zone.
 \item We find that a planet chain that would scatter comets from an exo-Kuiper belt or an outer companion that would force Kozai oscillations on a comet belt are two plausible mechanisms to throw an important number of comets on the seven planets over the system's lifetime (see Secs.~\ref{sca} and \ref{koz}).
 \item On the other hand, we rule out a potential Oort-cloud around TRAPPIST-1 as being a significant source of impacting comets (see Sec.~\ref{fracs}).
 \item For the planet-scattering scenario, we find that even a belt with a low scattering rate similar to the current Kuiper-belt is enough to destroy typical Earth-like primordial atmospheres for the seven planets. 
Taking into account that typically observed debris belts are much more massive than the Kuiper belt, we find
that the Kozai (slightly less efficient) scenario can also strip primordial atmospheres even if the impact process only lasts a fraction of the system's age.
 \item As for the volatile delivery, we find that for the planet-scattering scenario, planets f, g, and h can get (more than) an Earth ocean mass of water (and other volatiles) delivered, which can accumulate impact after impact. 
We find that the primordial atmospheres are gradually replaced by cometary material and may lead to subsequent build up of new secondary atmospheres with exocomet-like compositions. 
These new secondary atmospheres may become more massive than the initial primordial atmospheres.
\item Table~\ref{tab2} summarises the results for the different scenarios as for the minimum scattered (incoming) mass needed to destroy the primordial atmospheres and the volatile/water masses that can be delivered onto each planet. 
 \item We also discuss the implications of impacts to create the building blocks of life. We detail new emerging pathways that can lead to life showing that UV irradiation, impacts and a hard planetary surface might be enough to kick start biological reactions and form ATP, RNA, DNA, amino acids and lipids
that are essential to life (see Sec.~\ref{life}).
\end{itemize}

In brief, we find that the primordial atmospheres of the seven planets orbiting around TRAPPIST-1 would not survive over the lifetime of the system if a belt scattering comets at a similar low rate than the Kuiper belt (or faster) were around TRAPPIST-1. 
According to our calculations based on applying a minimum mass extrasolar nebula approach for the TRAPPIST-1 system, we expect a potential 20M$_\oplus$ belt may have survived around TRAPPIST-1 that would be observable with ALMA. 
We also show that a large fraction of the delivered cometary volatiles remains in the atmospheres of the outermost planets f, g and h, which gradually replace their primordial atmospheres. We predict that the new secondary atmospheres 
of planets f, g and h may be more massive than that of the innermost planets (which may soon be checkable/observable with the JWST) and their composition might be dominated by the composition of exocomets in this system (i.e., impacts leave an imprint). 
We also predict that more than an Earth ocean mass of water could be delivered to planets f, g, and h owing to impacts that may be in liquid form on planets f and g.

\section*{Acknowledgments}
This paper is dedicated to Mila. We thank the two referees for comments that greatly improved the quality of the paper.
QK and MCW acknowledge funding from STFC via the Institute of Astronomy, Cambridge Consolidated Grant. QK thanks J. Teyssandier for interesting discussions about the Kozai mechanism.
Simulations in this paper made use of the REBOUND code which can be downloaded freely at http://github.com/hannorein/rebound.

%\appendix
%\section{Accretion and ejection probability maps for the case of an inclined distribution of comets}

%\begin{figure}
 %  \centering
  %\includegraphics[width=9cm]{images/planettejee.pdf}
   %\caption{\label{fig3f} ejection timescale.}
%\end{figure}

%\section{Collision velocities}

%Fig.~\ref{fig10} shows the impact velocity map for each orbit.

%\begin{figure*}
%   \centering
%  \includegraphics[width=19cm]{images/planetVcol.pdf}
%   \caption{\label{fig10} bla.}
%\end{figure*}

\label{lastpage}

\end{document}